\documentclass[11pt,letterpaper]{article}
\usepackage{jheppub}
\usepackage{style}

\begin{document}

\preprint{FERMILAB-PUB-25-0772-SQMS-T}

\title{Quantum Calculations of the Cavity Shift in Electron Magnetic Moment Measurements}
\author[1]{Hannah Day\,\orcidlink{0000-0002-4496-5600},}
\affiliation[1]{Department of Physics, University of Illinois Urbana-Champaign, Urbana, IL 61801, USA}
\author[2,3]{Roni Harnik\,\orcidlink{0000-0001-7293-7175},}
\affiliation[2]{Theoretical Physics Division, Fermi National Accelerator Laboratory, Batavia, IL 60510, USA}
\affiliation[3]{Superconducting Quantum Materials and Systems Center (SQMS), Fermi National Accelerator Laboratory, Batavia, IL 60510, USA}
\author[1,3,4]{Yonatan~Kahn\,\orcidlink{0000-0002-9379-1838},}
\affiliation[4]{Department of Physics, University of Toronto, Toronto, ON M5S 1A7, Canada}
\author[1]{Shashin~Pavaskar\,\orcidlink{0000-0003-1886-1266},}
\author[5,6]{and~Kevin~Zhou\,\orcidlink{0000-0002-9810-3977}}
\affiliation[5]{Leinweber Institute for Theoretical Physics, University of California, Berkeley, CA 94720, USA}
\affiliation[6]{Theory Group, Lawrence Berkeley National Laboratory, Berkeley, CA 94720, USA}

\emailAdd{hjday2@illinois.edu}
\emailAdd{roni@fnal.gov}
\emailAdd{yf.kahn@utoronto.ca}
\emailAdd{pavaskar@illinois.edu}
\emailAdd{kzhou7@berkeley.edu}

\abstract{The measurement of the anomalous electron magnetic moment $g-2$ through quantum transitions of a single trapped electron is the most stringent test of quantum field theory. These experiments are now so precise that they must account for the effects of the cavity containing the electron. Classical calculations of this ``cavity shift'' must subtract the electron's divergent self-field, and thus require knowledge of the exact Green's function for the cavity's electromagnetic field. We perform the first fully quantum calculation of the cavity shift in a closed cavity, which instead involves subtracting linearly divergent cavity mode sums and integrals. Using contour integration methods, we find perfect agreement with existing classical results for both spherical and cylindrical cavities, justifying their current use. Moreover, our mode-based results can be naturally generalized to account for systematic effects, necessary to push future measurements to the next order of magnitude in precision.}

\maketitle
\setcounter{page}{2}

\vspace{-6mm}
\paragraph{Conventions and Notation.} We use a mostly-negative spacetime metric and natural units, $\hbar = c = 1$, with rationalized Heaviside--Lorentz units for electromagnetic fields (i.e., SI units with $\epsilon_0 = \mu_0 = 1$). The electron has mass $m$ and charge $-e$, and the fine-structure constant is $\alpha = e^2/(4 \pi)$. We use the Dirac representation for the gamma matrices, 
\begin{equation*}
\gamma^0 = \begin{pmatrix} I & 0 \\ 0 & -I \end{pmatrix}, \qquad \gamma^i = \begin{pmatrix} 0 & \sigma^i \\ -\sigma^i & 0 \end{pmatrix}, \qquad \gamma^5 = \begin{pmatrix} 0 & I \\ I & 0 \end{pmatrix}.
\end{equation*}

\section{Introduction}

Measurements of the electron magnetic moment $\mu \equiv \mu_B g/2$ have reached a fractional uncertainty of $1.3 \times 10^{-13}$, making it the single most precisely measured property of any fundamental particle~\cite{PhysRevLett.130.071801,Fan:2022oyb}. Such measurements directly probe five-loop corrections in quantum electrodynamics~\cite{PhysRevD.97.036001}.\footnote{This calculation involves diagrams with electron, muon, and tau loops, as well as hadronic and electroweak corrections. There was a $5 \sigma$ discrepancy in the calculation of the five-loop QED diagrams without lepton loops~\cite{PhysRevLett.109.111807,PhysRevD.100.096004}, which was recently resolved~\cite{Volkov:2024yzc,Aoyama:2024aly}, and found to arise from a bias in the numeric integration procedure. The use of electron $g-2$ to test the Standard Model is currently limited by the $5 \sigma$ discrepancy in the leading measurements of $\alpha$ by atom interferometry~\cite{Parker:2018vye,Morel:2020dww}.} Future refinements could improve the precision by over an order of magnitude~\cite{PhysRevLett.126.070402,PhysRevA.103.022824,PhysRevA.111.042806}, which would yield better sensitivity to generic physics beyond the Standard Model than that of the leading muon $g-2$ measurements~\cite{Muong-2:2025xyk}.

\begin{figure}[t]
\centering
\includegraphics[width=0.9\columnwidth]{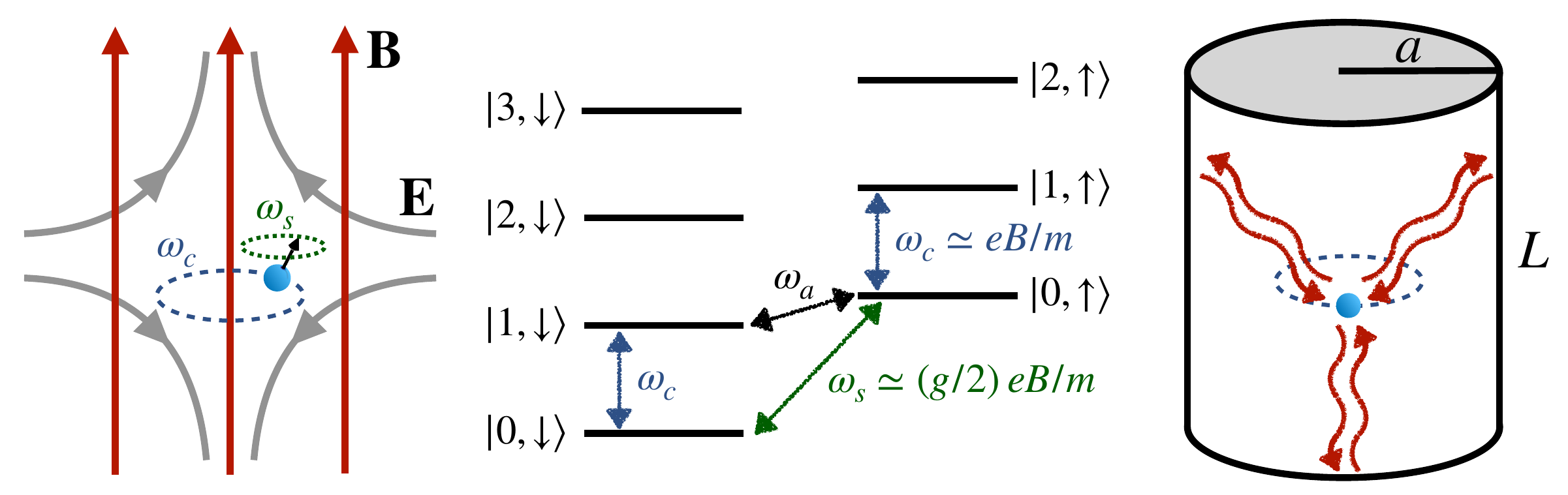}
\caption{Schematic depiction of the technique for current measurements of electron $g-2$. \textbf{Left:} an electron is placed in a uniform magnetic field, which causes cyclotron motion and spin precession. A weak quadrupolar electrostatic field traps the particle in the axial direction parallel to $B$. \textbf{Center:} measuring the transition energies $\omega_a$ and $\omega_c$ allows $g-2$ to be extracted, independent of the magnitude of the $B$-field to leading order. \textbf{Right:} the entire system is placed in a cylindrical cavity which reflects the electron's cyclotron radiation, extending the cyclotron motion's lifetime and shifting the cyclotron frequency. This ``cavity shift'' is the focus of this work.}
\label{fig:overview}
\end{figure}

The leading technique for measuring the electron $g$-factor involves making precise measurements of the energy levels of a single electron in a magnetic field $B$~\cite{Brown:1985rh,Gabrielse:2025jep}, as depicted in Fig.~\ref{fig:overview}. The energy eigenstates are two ladders of Landau levels $|n, \downarrow \rangle$ and $|n, \uparrow \rangle$, with the Landau levels and spin-flip transitions ideally separated by the cyclotron and spin precession angular frequencies respectively,
\begin{equation} \label{eq:omega_ideal}
\omega_c^0 = \frac{eB}{m}, \qquad \omega_s = \frac{g}{2} \frac{eB}{m}.
\end{equation}
The value of $g-2$ is inferred by measuring both $\omega_c \simeq \omega_c^0$ from the transition $|0, \uparrow \rangle \to |1, \uparrow \rangle$, and the energy $\omega_a \simeq \omega_s - \omega_c^0$ of the transition $|0, \uparrow \rangle \to |1, \downarrow \rangle$, and computing the ratio
\begin{equation} \label{eq:wa_wc_ratio}
\frac{\omega_a}{\omega_c} \simeq \frac{g-2}{2}
\end{equation}
which cancels out dependence on $B$, and greatly improves the relative precision on $g$ since $g \approx 2$. To achieve sufficient precision on the measurement of $\omega_c$, the electron is confined in a cavity of length scale $R$ to suppress spontaneous emission. 

The measured cyclotron angular frequency $\omega_c$ is shifted from the ideal value $\omega_c^0$ due to a variety of small effects, and these affect the determination of $g$ via $\Delta \omega_c / \omega_c \simeq - \Delta g / g$. Thus, at the current level of precision, all fractional shifts in $\omega_c$ of order $10^{-13}$ or larger must be accounted for. One of the most subtle is the (transverse) ``cavity shift.'' Classically, the electron's orbital motion produces radiation, which reflects off the cavity walls and acts back on the electron. This does not substantially affect $\omega_s$, but shifts $\omega_c$ by roughly
\begin{equation}
\label{eq:CyclotronShift}
\frac{\Delta \omega_{c}}{\omega_c} \sim \frac{\alpha}{m R} \sim 10^{-12} \left(\frac{4 \, {\rm mm}}{R}\right).
\end{equation}
This shift is so small that it cannot be measured by any other type of experiment, but it must be accounted to achieve the desired precision. Furthermore, the cavity shift has an intricate dependence on $\omega_c$, as it is resonantly enhanced when $\omega_c$ is near a cavity mode. 

A classical calculation may be sufficient, since the cavity size $R$ is macroscopic ($mR \gg 1$), but it is not obvious how to recover the cavity shift in a fully quantum calculation. Viewed quantum-mechanically, the electron does not have a well-defined classical trajectory, and the cavity walls determine the mode structure of the electromagnetic field, modifying the photon propagator from its free space value. In a quantum treatment, the leading cavity shift Eq.~\eqref{eq:CyclotronShift} should arise from a one-loop self-energy diagram. 

In the early 1980s, several groups attempted such a calculation in toy setups with one or two infinite conducting plates, and found comparable shifts of both $\omega_c$ and $\omega_s$, in disagreement with the classical calculation and with each other~\cite{Barton_1977,PhysRevD.30.2356,PhysRevLett.54.742,PhysRevD.34.1429}. By the late 1980s, consensus was reached that these conflicting results were due to errors such as computing gauge-dependent quantities, improperly using hard cutoffs, and failing to account for mass renormalization~\cite{PhysRevD.32.729,Barton_1988,PhysRevD.36.2181,Kreuzer_1988}. Ultimately, the quantum calculations agreed with the classical result. 

Following the infinite-plate calculations, the cavity shift was derived classically for an electron at the center of a cylindrical~\cite{PhysRevA.32.3204} or spherical~\cite{PhysRevA.34.2638} cavity. (See also Ref.~\cite{PhysRevA.83.052122} for a generalization to arbitrary points in a cylindrical cavity, and Ref.~\cite{kinoshita1990quantum} for a pedagogical review.) Here the parallel-plate results are not useful, since these cavity shapes confine the electron to a finite volume, changing the photon mode structure much more dramatically. Furthermore, the classical calculations require a careful subtraction of the electron's divergent self-field from the classical Green's function. This was the basis for Kramers' original attempts at renormalizing electromagnetism~\cite{Kramers1938,schweber1994qed}, and in the present context, it can be done exactly for an ideal sphere or cylinder. However, it is unclear how to generalize the subtraction to less ideal or symmetric cavity geometries. Finally, these calculations do not naturally accommodate cavity modes with different quality factors, which is important for future measurements. 

In this paper, we address these shortcomings by computing the renormalized cavity shift quantum-mechanically, writing it in terms of an explicit mode sum. Ultimately we find that the quantum-mechanical calculations exactly match classical calculations for perfectly-conducting spherical and cylindrical cavities, putting existing measurements which rely on these calculations on firmer theoretical footing. Furthermore, since we express the cavity shift in terms of a mode sum, our work enables a more precise understanding of the cavity shift which accounts for cavity imperfections.

We begin in Sec.~\ref{sec:general} by reviewing the setup of the $g-2$ experiment, and the many effects that shift the value of $\omega_c$. We review why many of them can be neglected or separately accounted for, simplifying our later treatment of the cavity shift, and also explain how the (transverse) cavity shift discussed in this work differs from the longitudinal ``image charge'' shift measured in trapped ion experiments. Finally, we illustrate why explicit renormalization is required, in either a classical or quantum calculation. 

In Sec.~\ref{sec:nrqm_shift}, we describe the electron with nonrelativistic quantum mechanics, and use second-order perturbation theory to compute the energy level shift $\Delta \omega_c^{\text{cav}}$, as a sum over cavity modes. We also compute the shift $\Delta \omega_c^{\text{free}}$ in free space, as an integral over plane waves. Both these quantities are linearly divergent, and the physical cavity shift is their difference. This derivation is similar in spirit to Bethe's nonrelativistic calculation of the Lamb shift, though in our case some matrix elements are easier to evaluate analytically. Famously, after performing such subtractions, Bethe showed that the Lamb shift is only logarithmically divergent in the nonrelativistic theory, allowing him to find a decent estimate of its magnitude~\cite{PhysRev.72.339}. However, finding the precise numerical value required a renormalized calculation in the relativistic theory, including a nontrivial matching of the high-energy and low-energy descriptions of the system at a scale comparable to the electron mass, as reviewed in Refs.~\cite{weinberg,schweber1994qed,itzykson2012quantum}.

Motivated by possible subtleties in matching low- and high-energy calculations, in Sec.~\ref{sec:qft_shift} we compute the cavity shift in the relativistic theory, using Schwinger's exact propagator for the electron in an external magnetic field, and the cavity-modified photon propagator. However, we find the result is essentially the same as in the nonrelativistic theory, and no matching step is required. This shows that the cavity shift is an inherently infrared effect.

We then compute the cavity shift in specific cavity geometries. In Sec.~\ref{sec:sphere_result}, we warm up with a spherical cavity, and show that the difference between the regulated sum $\Delta \omega_c^{\text{cav}}$ and integral $\Delta \omega_c^{\text{free}}$ can be computed in a regulator-independent way by evaluating an appropriate contour integral. This yields exactly the same analytic expression previously derived classically. We then consider a cylindrical cavity in Sec.~\ref{sec:cylinder_result}. We are able to use the same technique, though the result is significantly more complicated due to the presence of multiple mode sums and branch cuts, and it again matches the classical result.

Our approach thus provides a flexible new method applicable to more general cavities. In Sec.~\ref{sec:conclusion}, we discuss how the renormalized cavity shift can be computed with an explicit regulator; in particular, we show that even a hard cutoff with a small number of terms can yield a highly accurate result, as long as the cutoff is chosen appropriately. We conclude by discussing the relationship between our method and other calculations, and applications to future electron $g-2$ measurements. Technical details are relegated to the appendices, and referred to throughout the main text.

\section{General Considerations}
\label{sec:general}

Many effects can potentially shift $\omega_c$ at the $10^{-13}$ level, so we will begin by reviewing and estimating them. We will show that many are actually negligible, or can be accounted for independently of the cavity shift, simplifying the analysis of later sections. We largely follow Ref.~\cite{Brown:1985rh}, but also incorporate more recent theoretical results. Where specific choices of parameters such as magnetic field and cavity size are needed, we choose values relevant to the latest iteration of the experiment~\cite{PhysRevLett.130.071801,Fan:2022oyb}.

\paragraph{Frequency Scales in a Penning Trap.} In the $g-2$ experiment, an electron is placed in a uniform magnetic field $\mathbf{B} = B \, \hat{\mathbf{z}}$, in which the angular frequency of the cyclotron orbit is 
\begin{equation}
\omega_c^0 = \frac{eB}{m} = 6 \times 10^{-4} \, \mathrm{eV} \sim (0.3 \, {\rm mm})^{-1}.
\end{equation}
In a pure magnetic field, the electron would not be confined along the $z$-axis, so a weak quadrupole electrostatic field with potential $\phi \propto z^2 - \rho^2/2$ (where $\rho$ is the distance to the $z$-axis) is imposed. This combination of fields is called a Penning trap. The angular frequency of oscillations along the axial direction is 
\begin{equation}
\omega_z = 5 \times 10^{-7} \, \mathrm{eV} \sim 10^{-3} \, \omega_c.
\end{equation}
Furthermore, in the presence of both of these fields, the electron can perform slow, circular ``magnetron'' motion in the $xy$ plane, with angular frequency $\omega_m \simeq \omega_z^2/2 \omega_c$. 

Since the quadrupole potential also provides a radial force, the physical cyclotron frequency $\omega_c$ differs from $\omega_c^0$ by a fractional amount of order $\omega_z^2/(\omega_c^0)^2 \sim 10^{-6}$. However, by measuring $\omega_c$, $\omega_z$, and $\omega_m$ in the presence of the potential, one can infer the value $\omega_c^0$ that would occur in the absence of the potential using the ``invariance theorem''~\cite{Brown:1985rh}
\begin{equation} \label{eq:invariance}
\omega_c^0 = \sqrt{\omega_c^2 + \omega_z^2 + \omega_m^2},
\end{equation}
which holds even if the trapping potential is misaligned or otherwise imperfect. As discussed in Ref.~\cite{Brown:1985rh}, $\omega_c^0$ is essentially the quantity that should be used to determine $g$ via Eq.~\eqref{eq:wa_wc_ratio}. It will turn out that the dominant effect of the cavity shift is through its effect on $\omega_c$, but we will note when relevant how shifts of $\omega_z$ and $\omega_m$ affect the determination of $\omega_c^0$. 

\paragraph{Relativistic and Radiative Corrections.} The above discussion treats the electron's motion as nonrelativistic. The leading relativistic effects on the frequencies of the lowest cyclotron levels, computed in Ref.~\cite{Brown:1985rh}, have a fractional size $\Delta \omega_c / \omega_c \sim \omega_c / m \sim 10^{-9}$. A complementary perspective using classical equations of motion is given in Ref.~\cite{ketter_2014}.

Intuitively, the main effects at this order are the increased relativistic momentum $\mathbf{p} = \gamma m \mathbf{v}$, which slows down the cyclotron motion, and Thomas precession, which shifts the spin precession frequency. The effects of axial and magnetron motion on the measurement of $g$ are suppressed by multiple powers of $\omega_z/\omega_c$, and are thus negligible. Relativistic effects at the next order were computed in Ref.~\cite{Wienczek:2022cwa} using the Foldy--Wouthuysen transformation, but these contribute at most $(\omega_c / m)^2 \sim 10^{-18}$ and are also negligible. Thus, we may neglect all relativistic effects besides the few mentioned above, which are simply accounted for by shifting the value of $\omega_c$. 

Famously, one-loop corrections in QED affect the spin precession frequency at the level $\Delta \omega_s / \omega_s \sim \alpha / 2 \pi \sim 10^{-3}$. However, loop corrections generally have a much smaller effect on $\omega_c$. First, as discussed in Ref.~\cite{Jentschura:2023irs}, vacuum polarization effects are exponentially suppressed, because they only modify the potential near charges, and the trapped electron is macroscopically far away from other charges. Second, the one-loop correction to the electron self-energy yields an effect analogous to the Lamb shift. Since $\omega_c/m$ plays the role of $Z \alpha$ in an atom, one would expect $\Delta \omega_c / \omega_c \sim \alpha \, (\omega_c / m) \log(m / \omega_c) \sim 10^{-10}$. However, this energy level shift turns out to be independent of the cyclotron energy level to leading order, and thus does not affect measurements of $\omega_c$~\cite{Jentschura:2023irs,PhysRev.96.523,PhysRevD.8.3446}. The leading state-dependent shift appears only at the next order in the strength of the $B$-field, giving a negligible $\Delta \omega_c / \omega_c \sim \alpha \, (\omega_c / m)^2 \log (m/\omega_c) \sim 10^{-19}$.

Together, these parametric estimates indicate that nonrelativistic quantum mechanics can reliably be used to compute the cavity shift, though in Sec.~\ref{sec:qft_shift}, we perform a relativistic calculation and take the appropriate nonrelativistic limit to recover the same results.

\paragraph{Cavities and the Dipole Approximation.} In free space, cyclotron motion is damped due to the emission of radiation. Applying the Larmor formula gives a damping rate
\begin{equation}
\gamma = \frac{e^2 \omega_c^2}{3 \pi m} \sim \frac{\alpha \omega_c^2}{m} \sim 10^{-11} \, \omega_c.
\end{equation}
This yields a negligible fractional frequency shift $\Delta \omega_c / \omega_c \sim (\gamma / \omega_c)^2 \sim 10^{-22}$. However, the short lifetime of the cyclotron states makes it difficult to measure $\omega_c$ to the required precision. 

To address this, the Penning trap is placed inside a cavity, which modifies the free space photon modes to discrete cavity modes. When $\omega_c$ coincides with the angular frequency of a cavity mode, the damping rate is resonantly enhanced by the mode quality factor $Q \gtrsim 10^3$, and when $\omega_c$ is well away from a resonant mode, the contribution of that mode to the decay rate is suppressed by $1/Q$. By tuning $\omega_c$ to sit between cavity modes, the decay rate can be sufficiently reduced. The length scale of the cavity is $R \simeq 4 \, \mathrm{mm}$; in the cylindrical cavity used for the most recent measurement, $R$ is roughly both the radius and the half-length~\cite{Fan:2022oyb}. This value is chosen so that $\omega_c$ is comparable to that of fairly low-lying cavity modes, $\omega_c R \sim 10$, where the modes can be unambiguously identified, and their $Q$-factors may be reliably measured.

The electron is very well-localized relative to the cavity size: the typical radius of the lowest cyclotron orbit is the ``magnetic length'' $\ell \sim 1 / \sqrt{m \omega_c} = 1/\sqrt{eB} \sim 10^{-8} \, \mathrm{m}$, while the spread in the axial direction is $\sigma_z \sim \sqrt{q/m \omega_z} \sim 10^{-5} \, \mathrm{m}$, as the axial quantum number was $q \sim 100$ in the latest measurement. Since $\ell, \sigma_z \ll R$ the electron itself never ``sees'' the cavity walls; the cavity only affects the photon modes.

In the calculations below, it will be useful to make the ``dipole approximation'', which amounts to evaluating the photon mode functions at the mean position of the electron. For simplicity we assume the electron is centered in the cavity, and denote the center of the cavity by $\mathbf{r} = \bm{0}$. As we will see, the dipole approximation is reasonable because the renormalized cavity shift is dominantly due to low-lying cavity modes, with wavelengths of order $R \gg \ell, \sigma_z$.\footnote{A realistic cavity becomes transparent above the plasma frequency $\omega_p \sim 15 \, \mathrm{eV}$, corresponding to radiation of a wavelength $2 \pi / \omega_p \sim 10^{-7} \, \mathrm{m}$. Thus, the dipole approximation would hold, at least in the radial directions, for all cavity modes that even exist. However, this assumption will not be needed below.} 

Finally, we will assume the photon modes with nonzero values at the cavity's center can be separated into a set proportional to $\hat{\mathbf{z}}$ (``axial modes'') and a set proportional to $\hat{\mathbf{x}}\pm i \hat{\mathbf{y}}$ (``radial modes''). This holds for the spherical and cylindrical geometries we consider, and it is also generically a good assumption for any cavity geometry with axial symmetry. With this assumption, we can isolate the dependence on cavity geometry from the rest of the calculation.

\paragraph{Classical Estimates of the Cavity Shift.} The cavity suppresses the imaginary part of the electron's self energy, but by the general logic of the Kramers--Kronig relations, analyticity implies that a shift in the imaginary part is inevitably accompanied by a shift in the real part. Classically, the cyclotron frequency shifts because the radiation fields produced by the electron reflect off the cavity walls and act back on the electron~\cite{PhysRevD.32.729}. As we will see, there are several parts of this ``cavity shift'', though only one is currently relevant.

To build intuition, we replace the cavity with an infinite flat conducting plate a distance $R$ away from the electron; then the cavity shift is due to the electric field $\mathbf{E}'$ of the electron's image charge. This can be decomposed into a longitudinal (``electrostatic'') part with magnitude $E'_L \sim q / (4 \pi R^2)$ due to the image charge's Coulomb field, and a transverse part $E'_T \sim q a_c / (4 \pi R)$ due to radiation fields, where $a_c$ is the acceleration of the cyclotron orbit.

Numerically, $\mathbf{E}'_L$ is larger, but a constant $\mathbf{E}'_L$ would just be equivalent to shifting the center of the trapping potential. Furthermore, for symmetric cavities, $\mathbf{E}'_L$ vanishes at the center of the cavity. Thus, the leading effect of the longitudinal field comes from the linear term of the Taylor expansion, $E'_L \sim q r_c / (4 \pi R^3)$. Part of this term cannot be absorbed into a redefinition of the trapping potential, so it can indeed produce an observable cavity shift.

To estimate the effect of these fields on the cyclotron frequency, we compute the ratio of the image charge-induced forces $q E'$ to the Lorentz force $q v_c B$, which gives 
\begin{equation} \label{eq:shift_scaling}
\frac{\Delta \omega_c}{\omega_c} \sim \begin{cases} \alpha / (m R) \sim 10^{-12} & \text{transverse}, \\ \alpha / (m \omega_c^2 R^3) \sim 10^{-14} & \text{longitudinal}. \end{cases}
\end{equation}
Thus, the cavity shift from $\mathbf{E}_T'$ is important for the current generation of experiments, while that of $\mathbf{E}_L'$ is currently negligible, but may become relevant in the near future. 

We emphasize that the longitudinal and transverse cavity shifts are independent effects, which must be summed to yield the total cavity shift. As we will discuss in Sec.~\ref{sec:conclusion}, the longitudinal cavity shift dominates at low frequencies, and is therefore more important for trapped ion experiments~\cite{PhysRevA.40.6308,vandyck2006231,BASE:2014drs,BASE:2016yuo,Schneider:2017lff,PhysRevLett.119.033001}, where it is called the ``image charge'' shift. In this work, we will focus exclusively on the transverse cavity shift, which arises from the fields of cavity modes. As we will see, this effect is more subtle because the mode sum must be renormalized. 

The (transverse) cavity shift can be enhanced by as much as $Q \sim 10^3$ on resonance with a cavity mode, making it relevant even for measurements performed in the 1980s. In practice, $\omega_c$ is tuned to sit between cavity modes to avoid this further enhancement. However, the presence of many cavity modes implies that the cavity shift in a closed cavity has an intricate frequency dependence, which is not captured by the toy model of a conducting plate. 

There are several other components of the cavity shift, but they are negligible. First, we have neglected the magnetic field $B_T' \sim E_T'$, because the magnetic force is smaller by a factor of $v_c \sim \sqrt{\omega_c / m}$. Second, the axial and magnetron frequencies are also shifted, by~\cite{PhysRevA.32.3204,Jentschura:2023mbv}
\begin{equation}
\frac{\Delta \omega_z}{\omega_z} \sim \frac{\Delta \omega_m}{\omega_m} \sim \frac{\alpha}{mR} \frac{1}{(\omega_z R)^2} \sim 10^{-8}.
\end{equation}
This can affect the determination of $\omega_c^0$ through the invariance theorem, but by at most
\begin{equation}
\frac{\Delta \omega_c}{\omega_c} \sim \left( \frac{\omega_z}{\omega_c} \right)^2 \frac{\Delta \omega_z}{\omega_z} \sim \frac{\alpha}{mR} \frac{1}{(\omega_c R)^2} \sim 10^{-14}
\end{equation}
which may become relevant in the near future. Finally, the spin precession frequency can be modified by the field of the electron's image magnetic dipole moment $\mu \sim q/m$. As was first argued in Ref.~\cite{PhysRevD.32.729}, since the radiative magnetic field is $B_T' \sim \mu \omega_s^2 / (4 \pi R)$, we have 
\begin{equation}
\frac{\Delta \omega_s}{\omega_s} \sim \frac{B_T'}{B} \sim \frac{\alpha}{m R} \frac{\omega_s}{m} \sim 10^{-21}
\end{equation}
which is completely negligible. Thus, currently the only relevant part of the cavity shift is the direct shift of $\omega_c$ due to the transverse electric field of the electron's accelerating charge. 

\paragraph{Classical Calculations of the Cavity Shift.} The cavity shift has only been calculated for a closed cavity in cylindrical~\cite{PhysRevA.32.3204} and spherical~\cite{PhysRevA.34.2638} geometries. In both cases, $\mathbf{E}_T'$ was evaluated at the electron's position using the cavity's exact Green's function, with the electron's divergent self-field subtracted. In the spherical case, a simple analytic form for the Green's function exists, while for the cylindrical case, it can be written as an infinite series by first considering the Green's function due to two infinite parallel plates, then adding the contribution due to the cylinder's curved walls. The derivation relies on the symmetry of the cavity, which allows the use of the method of images. In both cases, it is possible to account for finite quality factors by shifting the frequency of the Fourier-space Green's function into the complex plane, but one must assign the same quality factor to all modes, or in the case of the cylinder, a uniform $Q$-factor to all TE modes and another $Q$-factor to all TM modes. 

In practice, the quality factors are not uniform, and the modes do not take their ideal forms, due to imperfections of the cavity geometry and the effects of the electrodes that generate the trapping potential. Thus, in practice the cavity is characterized by measuring properties of its modes. Given this information, one can compute the cavity shift by considering the classical excitation of each cavity mode $n$, giving a result of the form~\cite{PhysRevA.37.4163}
\begin{equation} \label{eq:w_cav_classical}
\frac{\Delta \omega_c^{\text{cav}}}{\omega_c} = \frac{\alpha}{m R} \sum_n \frac{\omega_n^2}{\omega_n^2 - \omega_c^2} \, c_n
\end{equation}
where $c_n$ is a mode-dependent order-one coefficient. One can then estimate the cavity shift by summing over a few modes with $\omega_n$ close to $\omega_c$, as was suggested in Refs.~\cite{dehmelt1984,dehmelt1992}.

However, while this method does capture the resonant enhancement of the cavity shift near a mode, where $\Delta \omega_c^{\text{cav}}/\omega_c \sim Q \alpha / (m R)$, it does not yield a quantitatively correct answer in general, because the sum in Eq.~\eqref{eq:w_cav_classical} is actually divergent. To renormalize the sum, we must subtract it against the corresponding divergent shift in free space. In the next section, we will evaluate both of these quantities in nonrelativistic quantum mechanics.

\section{Cavity Shifts in the Nonrelativistic Quantum Theory}
\label{sec:nrqm_shift}

The Hamiltonian for the trapped electron is 
\begin{equation}
H = \frac{(\mathbf{p} + e \mathbf{A}_c + e \mathbf{A}_q )^2}{2m} - e (\phi_c + \phi_q),
\end{equation}
where we have separated out the classical scalar potential $\phi_c$ and vector potential $\mathbf{A}_c$ due to the background electric and magnetic fields. We have neglected the electron spin since it contributes negligibly to the cavity shift, as described in Sec.~\ref{sec:general}. 

We will solve the Schrodinger equation with the classical backgrounds exactly, and define the mechanical momentum by $\bpi \equiv \mathbf{p} + e\mathbf{A}_c$. Then the perturbation
\begin{equation}
\delta H = \frac{e}{2m}\left(\bpi \cdot \mathbf{A}_q + \mathbf{A}_q \cdot \bpi \right) + \frac{e^2 \mathbf{A}_q \cdot \mathbf{A}_q}{2m} - e \phi_q
\end{equation}
gives rise to the cavity shift. We can simplify this further by working in Coulomb gauge, $\nabla \cdot \mathbf{A}_q = 0$, which implies $[\bpi, \mathbf{A}_q] = 0$. In this gauge, $\phi_q$ contains contributions from the electron's longitudinal field, which was shown in Sec.~\ref{sec:general} to be negligible. Dropping it gives 
\begin{equation}
\delta H = \frac{e}{m}\left(\mathbf{A}_q \cdot \bpi \right) + \frac{e^2 \mathbf{A}_q \cdot \mathbf{A}_q}{2m}.
\end{equation}
For an electron eigenstate $|N; 0\rangle$ with no photons, the energy level shift at first order in $\delta H$ is 
\begin{equation}
\langle N; 0 | \delta H | N; 0 \rangle = \frac{e^2}{2m} \langle 0 | \mathbf{A}_q \cdot \mathbf{A}_q |0 \rangle.
\end{equation}
This is a constant, independent of the electron state, so it does not affect the spacing between cyclotron levels; we can remove it by taking $\mathbf{A}_q \cdot \mathbf{A}_q$ to be normal-ordered. Then the leading contribution to the cavity shift arises at second order in $\delta H$. Keeping only the order $e^2$ term, 
\begin{equation} \label{eq:pert_expr}
\delta E_N \simeq \frac{e^2}{m^2} \sum_{\sigma s} \sum_{N'} \frac{|\langle N; 0 | \mathbf{A}_q \cdot \bpi | N'; 1_{\sigma s} \rangle|^2}{E_N - (E_{N'} + \omega_{\sigma s})+i\epsilon},
\end{equation}
where $\sigma s$ indexes the polarizations and mode indices of one-photon states. (In free space, we would instead index by the polarization $\lambda$ and momentum $\mathbf{k}$.) An analogous expression occurs in the computation of the Lamb shift in nonrelativistic quantum mechanics, where the unperturbed Hamiltonian is the Coulomb potential of the hydrogen atom.

The cavity shift comes from the real part of $\delta E_N$, while the $i \epsilon$ ensures that $\delta E_N$ acquires a negative imaginary part when the denominator vanishes, $E_N = E_N' + \omega_{\sigma s}$, corresponding to a positive decay rate $\Gamma_{N} = -2 \, {\rm Im}(\delta E_N)$. The imaginary part is finite and does not require renormalization, so we will always implicitly take the real part. Since $\text{Re}(1 / (x + i \epsilon)) = \mathcal{P}(1/x)$, integrals that arise from Eq.~\eqref{eq:pert_expr} will always be Cauchy principal values. 

\subsection{Electron and Photon States}

To compute the matrix elements in Eq.~\eqref{eq:pert_expr}, we need to specify the electron and photon eigenstates and their associated quantum numbers.

\paragraph{Electron Eigenstates.} The Hamiltonian of a nonrelativistic electron in a Penning trap can be solved exactly~\cite{Brown:1985rh}, and we briefly review the results here. To warm up, first suppose there is only a uniform magnetic field, and work in symmetric gauge, where $\mathbf{A}_c = \frac{1}{2}(\mathbf{B} \times \bm{\rho})$ with $\bm{\rho}$ the radial vector in cylindrical coordinates. The classical equations of motion are linear, which implies that the quantum states can be expressed as a combination of harmonic oscillators at angular frequency $\omega_c^0$. To do this, we define the operator
\begin{equation} \label{eq:a_op_def}
a = \frac{\pi_x - i \pi_y}{\sqrt{2 m \omega_c^0}}
\end{equation}
which satisfies the harmonic oscillator algebra, $[a, a^\dagger] = 1$. Then the Hamiltonian becomes
\begin{equation}
H = \frac{\bpi \cdot \bpi}{2m} = \omega_c^0 \left( a^\dagger a + \frac{1}{2}\right) + \frac{p_z^2}{2m}
\end{equation}
which describes the cyclotron motion and the axial motion. To describe the third degree of freedom, which will correspond to the magnetron motion, we define the ladder operator
\begin{equation}
b = \frac{\widetilde{\pi}_x + i \widetilde{\pi}_y}{\sqrt{2 m \omega_c^0}}
\end{equation}
in terms of the non-gauge-invariant quantity $\widetilde{\bpi} = \mathbf{p} - e \mathbf{A}_c$. Then in symmetric gauge, we have $[\bpi, \widetilde{\bpi}] = 0$, which implies $[a, b] = 0$, so that we can write the electronic states as
\begin{equation}
|N \rangle \equiv |n l \rangle \otimes |q_z\rangle, \qquad |n l \rangle = (a^\dagger)^n (b^\dagger)^l |0 \rangle
\end{equation}
where $|q_z \rangle$ is a plane wave in the $z$-direction, and $|nl \rangle$ are the ``radial'' states. The Landau levels are indexed by $n$, and each Landau level has an infinite degeneracy in $l$, which determines the angular momentum of the state. For $n = 0$ and $l = 0$, the radial wavefunction is a Gaussian centered at the origin with radial spread $\ell = 1/\sqrt{m \omega_c^0} = 1/\sqrt{eB}$.

With a quadrupole field, the motion in the $z$-direction is now harmonic with axial angular frequency $\omega_z$, corresponding to harmonic oscillator states $|q \rangle$ with spatial extent $\sigma_z \sim \sqrt{q/m \omega_z}$. The magnetron motion is also harmonic at angular frequency $\omega_m$, with $l$ the quantum number for magnetron oscillations. Finally, the cyclotron frequency $\omega_c$ is shifted slightly, satisfying $\omega_c^0 = \omega_c + \omega_m$, where $\omega_m = \omega_z^2/(2 \omega_c)$, which satisfies the invariance theorem~(\ref{eq:invariance}). The energy eigenstates and eigenvalues are thus
\begin{equation}
|n l q \rangle = (a^\dagger)^n (b^\dagger)^l (a_z^\dagger)^q|0 \rangle
\end{equation}
and
\begin{equation}
E_{nlq} = \left(n + \frac{1}{2}\right) \omega_c + \left(q + \frac{1}{2}\right) \omega_z - \left(l + \frac{1}{2}\right) \omega_m.
\end{equation}
Note the magnetron term is negative, so the Hamiltonian is technically unbounded from below; in practice, the system is metastable, with a lifetime of years.

Including the quadrupole potential, the components of the mechanical momentum may be expanded in cyclotron, magnetron, and axial ladder operators as~\cite{Brown:1985rh}
\begin{align}
\pi_x & = \sqrt{\frac{m}{2(\omega_c - \omega_m)}}\Big(\omega_c(a + a^\dagger) - \omega_m (b + b^\dagger) \Big), \\
\pi_y & = i\sqrt{\frac{m}{2(\omega_c - \omega_m)}}\Big(\omega_c(a - a^\dagger) +\omega_m (b - b^\dagger) \Big), \\
\pi_z &= -i\sqrt{\frac{m \omega_z}{2}}(a_z - a_z^\dagger),
\end{align}
which will be sufficient to evaluate all the required matrix elements. 

\paragraph{Photon Mode Expansion.} We canonically quantize $\mathbf{A}_q$ in a generic cavity as
\begin{equation} \label{eq:cavity_A_expansion}
\mathbf{A}_q(\mathbf{x}) = \sum_{\sigma s}\frac{1}{\sqrt{2\omega_{\sigma s}}}\left[\mathbf{u}_{\sigma s}(\mathbf{x})a_{\sigma s}+ {\rm h.c.} \right]
\end{equation}
where $\sigma \in \{\rm TE, TM\}$, $s$ is a multi-index for the integer mode numbers with associated angular frequencies $\omega_{\sigma s}$, and the mode functions obey the normalization condition
\begin{equation} \label{eq:mode_norm}
\int_V d^3\mathbf{x} \, \mathbf{u}_{\sigma s}^*(\mathbf{x}) \cdot \mathbf{u}_{\sigma' s'}(\mathbf{x}) = \delta_{\sigma \sigma'} \, \delta_{ss'}
\end{equation}
where $V$ is the cavity volume. In free space, we instead expand 
\begin{equation} \label{eq:free_A_expansion}
\mathbf{A}_q(\mathbf{x}) = \int \frac{d^3 \mathbf{k}}{(2\pi)^3} \frac{1}{\sqrt{2|\mathbf{k}|}} \sum_{\lambda = 1, 2} \left [\bm{\epsilon}_\lambda(\mathbf{k})e^{i \mathbf{k} \cdot \mathbf{x}} a_{\lambda \mathbf{k}} + {\rm h.c.} \right ]
\end{equation}
where $\bm{\epsilon}_\lambda(\mathbf{k})$ are the two transverse polarizations, for $\lambda \in \{+, -\}$, the mode functions $e^{i \mathbf{k} \cdot \mathbf{x}}$ are plane waves with angular frequency $ |\mathbf{k}|$. In both cases we are working in Coulomb gauge and dropping longitudinal parts of the field, as justified in Sec.~\ref{sec:general}. 

The mode functions $\mathbf{u}_{\sigma s}(\mathbf{x})$ and free-space polarization vectors $\bm{\epsilon}_\lambda(\mathbf{k})$ are given in App.~\ref{app:photon_modes}. Under the dipole approximation, it suffices to evaluate the mode functions at the position of the electron, which amounts to setting $e^{i \mathbf{k} \cdot \mathbf{x}}=1$ for the free space field. For spherical and cylindrical cavities, only a few families of modes are nonzero at the center of the cavity, and they can be written in the generic form
\begin{align}
\mathbf{u}_{\sigma,0\nu p}(\bm{0}) & \equiv u_{\sigma,\nu p}^{||} \hat{\mathbf{z}},\\
\mathbf{u}_{\sigma,\pm1\nu p}(\bm{0}) & \equiv \pm u_{\sigma,\nu p}^{\perp}(\hat{\mathbf{x}} \pm i \hat{\mathbf{y}}).
\end{align}
We call these modes ``axial'' and ``radial'' respectively. The first mode index is the azimuthal quantum number $m$, so axial modes have $m = 0$ and radial modes have $m = \pm 1$, and $\nu$ and $p$ are other numbers which index the modes. The coefficients are real-valued, with explicit expressions given in App.~\ref{app:photon_modes}. 

\subsection{Cavity and Free Space Results}
\label{sec:generic}

\paragraph{Cavity Result.} In this case, we need to compute the matrix element
\begin{equation}
\langle N; 0 | \mathbf{A}_q(\bm{0}) \cdot \bm{\pi} | N'; 1_{\sigma s} \rangle = \frac{1}{\sqrt{2\omega_{\sigma s}}}\langle nlq | \mathbf{u}_{\sigma s}(\bm{0}) \cdot \bpi |n'l'q' \rangle.
\end{equation}
The electron operators $\boldsymbol{\pi}$ are all linear in the creation and annihilation operators for various modes, and thus only change one quantum number at a time to leading order. The terms with cyclotron operators in $\pi_x$ and $\pi_y$, which couple to the radial modes, change the cyclotron level by 1 unit,
\begin{equation}
\label{eq:matrixelementcavity}
\langle n l q | \mathbf{u}^\perp_{\sigma,m\nu p}(\bm{0}) \cdot \bm{\pi} | n'l'q' \rangle \supset u_{\sigma,\nu p}^\perp \sqrt{2 m \omega_c} \, \delta_{ll'} \, \delta_{qq'} \times \begin{cases} \sqrt{n} \, \delta_{n',n-1} & m = +1, \\ \sqrt{n+1} \, \delta_{n',n+1} & m = -1, \end{cases}
\end{equation}
where we have approximated $\omega_c - \omega_m$ as $\omega_c$, accurate up to one part in $10^6$. Contributions which involve the axial and magnetron ladder operators are independent of the cyclotron level $n$, and thus do not affect the spacing between cyclotron levels. However, they do affect the spacing between axial and magnetron levels, which can affect the determination of $\omega_c^0$ through the invariance theorem Eq.~\eqref{eq:invariance}. For example, the axial electron operators couple to the axial modes and contribute 
\begin{equation}
\label{eq:matrixelementcavity_axial}
\langle n l q | \mathbf{u}^{||}_{\sigma,m\nu p}(\bm{0}) \cdot \bm{\pi} | n'l'q' \rangle = - i u_{\sigma,\nu p}^{||} \sqrt{\frac{m \omega_z}{2}} \langle n l q | (a_z - a_z^\dagger) | n'l'q' \rangle.
\end{equation}
This yields a shift of the axial level spacing of order $\Delta \omega_z \sim (\omega_z q / \omega_c) \Delta \omega_c$, where $\Delta \omega_c$ is the cavity shift due to Eq.~\eqref{eq:matrixelementcavity}. By the invariance theorem, this shift yields $\Delta \omega_c^0 \sim \omega_z \, \Delta \omega_z / \omega_c \sim (\omega_z^2 q / \omega_c^2) \Delta \omega_c$. Since $q \sim 100$ in the latest iteration of the experiment~\cite{Fan:2022oyb}, this contribution is $\sim 10^4$ times smaller than the cyclotron part of the cavity shift, and thus negligible. The contribution from magnetron modes is even more suppressed, since the typical magnetron number is $l \sim 100$ but $\omega_m/\omega_c \sim 10^{-6}$.

The dominant part of the cavity shift thus comes from the cyclotron operators in Eq.~(\ref{eq:matrixelementcavity}). Plugging this into Eq.~\eqref{eq:pert_expr} and suppressing the axial and magnetron numbers, which simply stay constant, we have 
\begin{align}
\delta E_n^{\mathrm{cav}} &= \frac{e^2 \omega_c}{m} \sum_{\sigma \nu p} \sum_{m = \pm 1} \sum_{n'} \frac{u_{\sigma,\nu p}^{\perp2}}{\omega_{\sigma,1\nu p}} \frac{1}{E_n - (E_{n'} + \omega_{\sigma, 1 \nu p})} \times \begin{cases} n \, \delta_{n',n-1} & m = +1, \\ (n+1) \, \delta_{n',n+1} & m = -1 \end{cases} \\
&= -\frac{e^2 \omega_c}{m} \sum_{\sigma \nu p} \frac{u_{\sigma,\nu p}^{\perp2}}{\omega_{\sigma,1\nu p}} \left ( \frac{n}{\omega_{\sigma,1\nu p} - \omega_c} + \frac{n+1}{\omega_{\sigma,1\nu p} + \omega_c} \right). \label{eq:nr_en_shift}
\end{align}
The cavity shift is the change in the spacing between cyclotron levels, $\Delta \omega_c^{\rm cav} = \delta E_{n+1}^{\rm cav} - \delta E_n^{\rm cav}$, which turns out to be independent of $n$. The result is 
\begin{equation} \label{eq:cav_shift_generic}
\frac{\Delta \omega_c^{\rm cav}}{\omega_c} = -\frac{8\pi \alpha}{m}\sum_{\sigma \nu p} \frac{u_{\sigma,\nu p}^{\perp2}}{\omega^2_{\sigma,1\nu p} - \omega_c^2} = -\frac{8\pi \alpha}{ma}\sum_{\sigma \nu p} \frac{u_{\sigma,\nu p}^{\perp2} a^3}{(\omega_{\sigma,1\nu p} a)^2 - z^2},
\end{equation}
where $u_{\sigma,\nu p}^\perp$ and $\omega_{\sigma,1\nu p}$ for spherical and cylindrical cavities are given in App.~\ref{app:photon_modes}, and in the second step we introduced a cavity length scale $a$ to nondimensionalize the sum, with $z = \omega_c a$.

For spherical cavities, only modes with $\sigma = \mathrm{TM}$ and $\nu = 1$ contribute, so only a single sum is required. Note that this result is essentially the same as Eq.~\eqref{eq:w_cav_classical}, which can be deduced by considering the particle's analogous classical trajectory. However, the quantum approach is more general, as it can give an answer even when a classical trajectory is not defined. 

\paragraph{Free Space Result.} In this case, the relevant matrix element is 
\begin{equation}
\langle N; 0 | \mathbf{A}_q(\bm{0}) \cdot \bm{\pi} | N'; 1_{\lambda \mathbf{k}} \rangle = \frac{1}{\sqrt{2 |\mathbf{k}|}} \, \langle nlq | \bm{\epsilon}_\lambda(\mathbf{k}) \cdot \bpi |n'l'q' \rangle.
\end{equation}
By the same logic as above, the dominant contribution comes from the cyclotron ladder operators. Suppressing the axial and magnetron numbers, we have
\begin{align}
\langle n | \bm{\epsilon}_\pm(\mathbf{k}) \cdot \bpi | n'\rangle &= \sqrt{\frac{m \omega_c}{2}} \, \langle n | \epsilon_\pm^x (a + a^\dagger) + i \epsilon_\pm^y (a - a^\dagger) | n' \rangle \\
&= \frac{\sqrt{m \omega_c}}{2} \Big ( \sqrt{n+1} e^{i \phi_k}(\hat{\mathbf{k}} \cdot \hat{\mathbf{z}} \mp 1) \delta_{n',n+1} + \sqrt{n} e^{-i \phi_k} (\hat{\mathbf{k}} \cdot \hat{\mathbf{z}} \pm 1) \delta_{n',n-1}\Big)
\end{align}
where we applied Eq.~\eqref{eq:polarization_freespace}. Squaring and summing over photon polarizations gives
\begin{equation}
\label{eq:squaredmatrixelementfree}
\sum_{\lambda = \pm} | \langle n | \bm{\epsilon}_\pm(\mathbf{k}) \cdot \bpi | n'\rangle |^2 = \frac{m \omega_c}{2} \, (1 + (\hat{\mathbf{k}} \cdot \hat{\mathbf{z}})^2) ((n+1) \, \delta_{n',n+1} + n \, \delta_{n',n-1} ).
\end{equation}
This corresponds to an energy level shift of 
\begin{equation}
\delta E_n^{\mathrm{free}} = -\frac{e^2 \omega_c}{2m} \int \frac{d^3 \mathbf{k}}{(2\pi)^3 (2|\mathbf{k}|)} \, (1 + (\hat{\mathbf{k}} \cdot \hat{\mathbf{z}})^2) \left ( \frac{n}{|\mathbf{k}| - \omega_c} + \frac{n+1}{|\mathbf{k}| + \omega_c} \right).
\end{equation}
The change in the spacing between cyclotron levels, $\Delta \omega_c^{\rm free} = \delta E_{n+1}^{\rm free} - \delta E_n^{\rm free}$, obeys
\begin{equation} \label{eq:general_free_result}
\frac{\Delta \omega_c^{\rm free}}{\omega_c} = -\frac{\alpha}{4 \pi^2 m} \int d^3 \mathbf{k} \, \frac{1 + (\hat{\mathbf{k}} \cdot \hat{\mathbf{z}})^2}{|\mathbf{k}|^2 - \omega_c^2}.
\end{equation}
Note that this integral is manifestly linearly divergent. As discussed below Eq.~\eqref{eq:pert_expr}, the integrand contains a pole at $|\mathbf{k}| = \omega_c$, and the integral should be regarded as a principal value. 

\section{Cavity Shifts in the Relativistic Quantum Theory}
\label{sec:qft_shift}

We now have a fully quantum expression for the cavity shift for a nonrelativistic electron, given by the difference between the linearly divergent sum Eq.~(\ref{eq:cav_shift_generic}) and integral Eq.~(\ref{eq:general_free_result}). Readers interested in the renormalization procedure may skip directly to the following section.

However, given that the analogous calculation of the Lamb shift requires a relativistic calculation to match low-energy and high-energy terms, we find it instructive to repeat the calculation in a fully relativistic formalism, taking the nonrelativistic limit only in the final step. Indeed, even the relativistic calculation of the cyclotron shift in free space (which, as discussed in Sec.~\ref{sec:general}, affects the spacing between cyclotron levels at $\mathcal{O}(B^3)$) requires a careful matching between low- and high-energy regimes~\cite{PhysRevD.8.3446}. It is also enlightening to see how the cyclotron shift arises from the same one-loop diagrams that can be used to compute $g-2$. 

As we will discuss in Sec.~\ref{sec:energyshiftsrel}, the relativistic calculation requires computing the difference of one-loop self-energy diagrams in Fig.~\ref{fig:1loop}, involving the in-cavity photon and free-space photon propagators. As for the electron propagator, there is no known exact solution in the full set of trap fields; instead, in Sec.~\ref{sec:Propagators} we will use Schwinger's result~\cite{PhysRev.82.664} for the electron propagator in a constant $B$-field alone, and include the effects of the quadrupole potential through the external-leg electron wavefunctions. As such, here we will not distinguish between $\omega_c^0$ and $\omega_c$. 

We will see that two new subtleties emerge in the relativistic calculation: 
\begin{itemize}
\item We must perform an additional renormalization to subtract the electron self-energy in the absence of an external magnetic field. This is exactly analogous to mass renormalization in the standard treatment of the Lamb shift~\cite{itzykson2012quantum}, and is absent in the nonrelativistic calculation where the electron mass is not treated as a contribution to the energy.
\item The contribution of the axial modes does not appear to be parametrically suppressed compared to the radial modes. This is an unphysical artifact arising from the fact that the Schwinger propagator does not know about the axial confinement of the electron, and thus the ``internal'' energy eigenstates from the propagator are not orthogonal to the external legs of the one-loop diagram.
\end{itemize}
After accounting for these issues, we show in Sec.~\ref{sec:rel_results} that taking the nonrelativistic limit $\omega_{\sigma s}, \omega_c \ll m$ recovers our previous results, with technical details relegated to App.~\ref{app:rel_details}.

\subsection{Energy Shifts from Self-Energy Diagrams}
\label{sec:energyshiftsrel}

First, we review standard arguments (e.g. see Ref.~\cite{weinberg}) for how self-energy diagrams compute energy-level shifts. The exact position-space propagator for a fermion $\Psi$ is defined as the time-ordered correlation function of fields
\begin{equation}
 i S'_{A}(X,X^\prime) = \langle T \{ \Psi(X), \overline{\Psi}(X^\prime) \} \rangle_A,
\end{equation}
where $X^\mu = (t, \mathbf{x})$, $X'^\mu = (t', \mathbf{x}')$, and the expectation value is taken with respect to the interacting vacuum in the presence of an external field $A_\mu$ which has both a classical background value and quantum fluctuations. Performing the time Fourier transform yields the mixed energy-position propagator
\begin{equation}
\label{eq:Smixed}
 S'_A(\mathbf{x}, \mathbf{x}^\prime; E) = \sum_n \frac{U_n(\mathbf{x}) \overline{U}_n(\mathbf{x}^\prime)}{E'_n - E + i\epsilon} - \sum_n \frac{V_n(\mathbf{x}) \overline{V}_n(\mathbf{x}^\prime)}{E'_n + E + i\epsilon}
\end{equation}
where $U_n$ and $V_n$ are the 4-component spinor eigenstates of the full Hamiltonian with energies $E'_n$ and $-E'_n$ respectively, and the energy denominators arise from the time-ordering and the time Fourier transform integral. In this form, it is clear that poles in the mixed propagator correspond to energy eigenvalues.

\begin{figure}
\centering
\includegraphics[width=0.56\paperwidth]{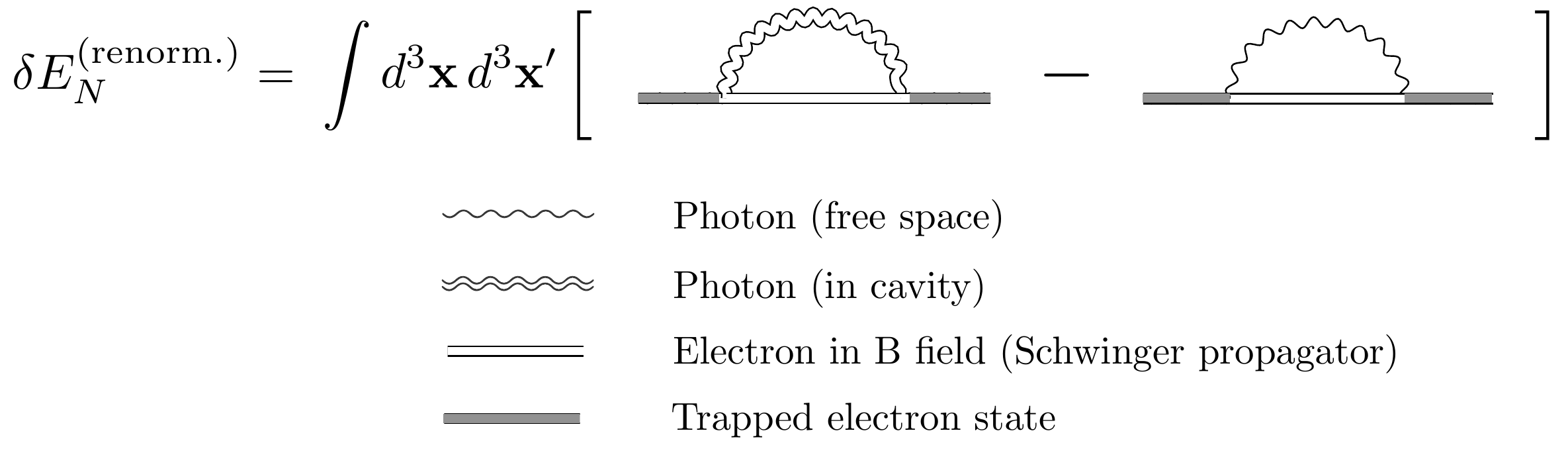}
\caption{One-loop contribution to the cavity shift in the relativistic quantum theory. The Schwinger propagator $S_A$ is the electron propagator in an external magnetic field, and the external electron states are approximate eigenstates of the full trapping potential, including the axial confinement.}
\label{fig:1loop}
\end{figure}

The exact electron propagator $S'_A$ may be expressed as a perturbative series in the electromagnetic coupling $e$,
\begin{equation}
\label{eq:Spert}
    S'_A(X,X^\prime) = S_A(X,X^\prime) + \delta S_A(X,X^\prime) + \dots,
\end{equation}
where $S_A$ is a Green's function for the Dirac equation with external \emph{classical} field $A_\mu$, which can be written in the mixed energy-position representation as 
\begin{equation}
\label{eq:Schwingermixed}
    S_A(\mathbf{x}, \mathbf{x}^\prime; E) = \sum_n \frac{u_n(\mathbf{x}) \overline{u}_n(\mathbf{x}^\prime)}{E_n - E + i\epsilon} - \sum_n \frac{v_n(\mathbf{x}) \overline{v}_n(\mathbf{x}^\prime)}{E_n + E + i\epsilon}.
\end{equation}
In the case where $A_\mu$ corresponds to a constant $B$-field, Schwinger derived a non-perturbative expression for $S_A$ as an integral over a Schwinger proper time variable, which we will use in detail in Sec.~\ref{sec:Propagators} below and refer to as the Schwinger propagator~\cite{PhysRev.82.664,kuznetsov2013electroweak}. The effect of $\delta S_A$ is to shift the energy eigenvalues away from $\pm E_n$, the poles of the Schwinger propagator (namely the relativistic Landau levels), and will also shift the spinor eigenstates $u_n, v_n$ of $S_A$. Taking $E'_n = E_n + \delta E_n$ and $U_n = u_n + \delta u_n$ in Eq.~\eqref{eq:Smixed} and expanding to first order in $\delta$'s, we have
\begin{equation}
    S'_A(\mathbf{x}, \mathbf{x}^\prime; E) = S_A(\mathbf{x}, \mathbf{x}^\prime; E) - \sum_n \frac{u_n(\mathbf{x}) \overline{u}_n(\mathbf{x}^\prime)}{(E_n - E)^2} \, \delta E_n + \mathcal{O}(\delta^2)
\end{equation}
so to find the first-order shift $\delta E_n$ of energy level $n$, we need to isolate the coefficient of $u_n(\mathbf{x}) \overline{u}_n(\mathbf{x}^\prime)/(E_n - E)^2$ in the exact propagator. 

The leading-order propagator correction $\delta S_A$ may be computed with a one-loop diagram, with the free-space correction subtracted from the cavity correction to obtain the physical cavity shift. In position space, the relevant diagram is an un-amputated diagram where the external legs contribute factors of the Schwinger propagator,
\begin{equation}
    \delta S_A(\mathbf{x}, \mathbf{x}^\prime; E) = \int d^3 \mathbf{w} \, d^3 \mathbf{w}^\prime \, S_A(\mathbf{x}, \mathbf{w}; E) \Sigma_A(\mathbf{w}, \mathbf{w}^\prime; E) S_A(\mathbf{w}^\prime, \mathbf{x}^\prime; E).
    \label{eq:deltaSA}
\end{equation}
Here, $-i \Sigma_A$ represents the amputated part of the one-loop diagram (also in the mixed energy-position representation), which involves two QED vertices and two propagators, and thus may be computed as
\begin{equation}
    -i\Sigma_A(\mathbf{x},\mathbf{x}^\prime;E)=(ie)^2 \int d\tau \, \gamma^\mu S_A(\mathbf{x},\mathbf{x}^\prime;\tau) \gamma^\nu D_{\mu \nu}(\mathbf{x},\mathbf{x}^\prime;\tau) e^{iE\tau},
    \label{eq:iSigmaANoPhase}
\end{equation}  
where $S_A$ is the electron propagator, $D_{\mu \nu}$ is the photon propagator, and $\tau = t -t'$. The two factors of $S_A$ in Eq.~(\ref{eq:deltaSA}) provide the required energy denominators, and taking inner products with a particular unperturbed eigenstate $u_n$ to isolate $E_n$, the energy shift is given by\footnote{This procedure has previously been used to calculate the energy level shift due to conducting plates~\cite{Bordag:1988yd}.}
\begin{equation}
\label{eq:energyshiftWeinberg}
    \delta E_n = \int d^3 \mathbf{x} \, d^3 \mathbf{x}^\prime \, \overline{u}_n(\mathbf{x}) \Sigma_A(\mathbf{x}, \mathbf{x}^\prime; E_n) u_n(\mathbf{x}^\prime).
\end{equation}
We illustrate this with Feynman diagrams in Fig.~\ref{fig:1loop}, where the electron-photon vertex has the familiar QED Feynman rule $i e \gamma^\mu$, but both the electron and the cavity photon propagators are modified from their free-space expressions, as we will describe in Sec.~\ref{sec:Propagators}.

The result for $\delta E_n$ makes intuitive sense, as it resembles a first-order energy shift in nonrelativistic quantum mechanics, with perturbing ``Hamiltonian'' given by the amputated one-loop diagram.\footnote{At one-loop, there is also a diagram corresponding to vacuum polarization, but as argued in Sec.~\ref{sec:general}, its effects are negligible here. Indeed, one can check that it vanishes identically in the limit of no quadrupole fields.} The setup of our calculation is very similar to Ref.~\cite{PhysRevD.8.3446}, which used the Schwinger propagator to compute free-space energy shifts to $\mathcal{O}(B^3)$ for a generic electron state; as a check, we have reproduced those results to $\mathcal{O}(B^2)$. In addition, Eq.~\eqref{eq:energyshiftWeinberg} can be used to compute the free-space one-loop correction to the spin-flip energy $\omega_s$, by taking the states $u_n$ to represent spin-down and spin-up electrons with the same cyclotron level index. As shown in Refs.~\cite{PhysRevD.8.3446,Suzuki:2005wra}, this recovers the usual one-loop anomalous magnetic moment ($g-2$), more commonly computed by considering the vertex renormalization. 

\subsection{Propagators and Wavefunctions}
\label{sec:Propagators}

Here we present the components necessary to evaluate Eq.~\eqref{eq:energyshiftWeinberg}. To avoid ambiguities with index heights for spatial indices, we write explicit spatial coordinates as $\mathbf{x} \equiv (x,y,z)$ and $\Delta \mathbf{x} \equiv \mathbf{x} - \mathbf{x}' = (\Delta x, \Delta y, \Delta z)$.

\paragraph{Electron Propagator.} The Schwinger propagator can be written in the form~\cite{kuznetsov2013electroweak}
\begin{equation}
S_A(X,X')=e^{-i\Phi(\mathbf{x},\mathbf{x}^\prime)}S(X-X'),
\end{equation}
which is the product of a gauge- and spacetime-translation-invariant piece with a gauge-dependent phase factor that breaks translation invariance. This phase arises because a constant magnetic field $\mathbf{B} = \nabla \times \mathbf{A}$ must come from a spatially-dependent vector potential, which breaks translation invariance.

In symmetric gauge, where the vector potential for a magnetic field $\mathbf{B} = B \hat{\mathbf{z}}$ is 
\begin{equation}
\label{eq:Asym}
\mathbf{A}=\frac{1}{2} \, \mathbf{B}\times\hat{\bm{\rho}}=\frac{1}{2}\begin{pmatrix}-By\\Bx\\0\end{pmatrix},
\end{equation} 
the Schwinger phase is~\cite{kuznetsov2013electroweak,Suzuki:2005wra}\footnote{Note that Ref.~\cite{kuznetsov2013electroweak} defines $\Phi$ with an additional minus sign and works in Landau gauge rather than symmetric gauge; by performing a gauge transformation to symmetric gauge, we have checked that our expressions for the wavefunctions and Schwinger phase agree with Ref.~\cite{Suzuki:2005wra} which uses symmetric gauge.}
\begin{equation}
\Phi(\mathbf{x},\mathbf{x}^\prime)=\frac{\beta}{4} \, [(x+x^\prime)(y-y^\prime)-(x-x^\prime)(y+y^\prime)],
\label{eq:SchwingerPhase}
\end{equation}
where $\beta = eB$.

The translation-invariant part of the Schwinger propagator, $S(X-X') \equiv S(\Delta\mathbf{x};\tau)$ with $\tau = t - t'$ can be expressed as an integral over a Schwinger proper time parameter $s$, 
\begin{multline}
S(\Delta\mathbf{x}; \tau)=-\frac{i\beta}{2(4\pi)^2}\int_0^\infty \frac{ds}{s \sin(\beta s)} \, e^{-i\left(m^2 s+\frac{1}{4s}(\tau^2-(\Delta z)^2)-\frac{\beta}{4\tan(\beta s)}((\Delta x)^2+(\Delta y)^2)\right)}\\
\quad \times\Big[\frac{\tau}{s}\left(\cos(\beta s)\gamma^0-i\sin(\beta s)\gamma^3\gamma^5\right)+2m\left(\cos(\beta s)+\sin(\beta s)\gamma^1\gamma^2\right)\\
-\frac{\Delta z}{s}\left(\cos(\beta s)\gamma^3-i\sin(\beta s)\gamma^0\gamma^5\right)-\frac{\beta}{\sin(\beta s)}(\Delta x\,\gamma^1+\Delta y\,\gamma^2)\Big].
\label{eq:SchwingerTranslationInvt}
\end{multline}
Exploiting the translation invariance of $S(X-X')$, we may Fourier transform to get a momentum-space propagator,
\begin{equation}
S_A(X, X') = e^{-i \Phi (X, X') } \int \frac{d^4 k}{(2\pi)^4} \, e^{-i k \cdot(X - X')}S(k),
\end{equation}
where
\begin{multline}
S(k)= \int_0^\infty \frac{ds}{\cos(\beta s)} \, e^{i s \left(k_0^2-k_3^2 - (k_1^2+k_2^2) \frac{\tan(\beta s)}{\beta s} - m^2 + i \epsilon\right)}\\
\times \left[ (k^0\gamma^0-k^3\gamma^3 + m)[\cos(\beta s) + \gamma^1 \gamma^2 \sin(\beta s)] - \frac{k^1\gamma^1+k^2\gamma^2}{\cos(\beta s)} \right].
\label{eq:SchwingerMomentumSpace}
\end{multline}

\paragraph{Electron Wavefunctions.} The electron wavefunctions which solve the Dirac equation in the presence of the vector potential $\mathbf{A}$ in Eq.~\eqref{eq:Asym} are plane waves in $z$. There are no known exact solutions of the Dirac equation once the confining quadrupole potential is included, but the axial confinement of the electron is crucial in order to apply the dipole approximation. As a result, we will take our electron wavefunctions to be the \emph{approximate} energy eigenstates of the Dirac equation in symmetric gauge~\cite{kuznetsov2013electroweak,Suzuki:2005wra},
\begin{equation}
u_n(\mathbf{x})=\sqrt{\frac{\beta}{2\pi\sigma_z}}\left(\frac{2}{\pi}\right)^{1/4}e^{-\beta (x^2+y^2)/4}e^{-(z-\bar{z})^2/\sigma_z^2} \hat{u}_n(\mathbf{x}),
\label{eq:electronwavefunction}
\end{equation}
where $\bar{z}$ is the $z$-coordinate of the cavity center. By assumption, the electron is localized on the cavity axis at $x = y = 0$. We have put in a Gaussian axial confinement, parameterized by $\sigma_z$, by hand. However, we will show that the final result is independent of $\sigma_z$, indicating that the exact form of the confining wavefunction is irrelevant, as long as the axial confinement is sufficiently weak. The condition we obtain on $\sigma_z$ is parametrically identical to the one obtained in Ref.~\cite{Bordag:1988yd} for the parallel-plate geometry.

To compute the cyclotron shift, we only need the ground state and first excited state, $n = 0$ and $n = 1$, with energy eigenvalues $E_0 = m$ and $E_1 = \sqrt{m^2 + 2 \beta}$, and spinor wavefunctions
\begin{equation} \label{eq:spinors}
\hat{u}_0(\mathbf{x}) =\begin{pmatrix}0\\1\\0\\0\end{pmatrix}, \qquad \hat{u}_1(\mathbf{x}) =\sqrt{\frac{\beta}{E_1(E_1+m)}}\begin{pmatrix}0 \\ (E_1 + m)(x+iy)/2 \\ -i \\ 0\end{pmatrix}.
\end{equation}
In the weak-field limit $\beta \ll m^2$, we have $E_1 \approx E_0 + \frac{eB}{m} = E_0 + \omega_c$ as expected. As in the nonrelativistic calculation, we have assumed that the contribution to the energy from the axial motion is negligible, and thus the energy does not depend on $\sigma_z$, which is correct to an accuracy of $\mathcal{O}(\ell^2/\sigma_z^2) \sim 10^{-3}$.

\paragraph{Photon Propagator.} As in Sec.~\ref{sec:nrqm_shift}, we work in Coulomb gauge. Then the mode expansion of $\mathbf{A}$ is the same as in Eq.~\eqref{eq:cavity_A_expansion} and Eq.~\eqref{eq:free_A_expansion}, though we now work in the interaction picture, where we include the time dependence of the operators $a_{\sigma s} \to e^{-i \omega_{\sigma s} t} a_{\sigma s}$. We can obtain the transverse component of the propagator from the time-ordered correlation function
\begin{align}
D^{ij}_{\rm cav}(\mathbf{x},\mathbf{x}';\tau) &\equiv \langle 0|TA^i(t,\mathbf{x})A^j(t^\prime,\mathbf{x}')|0 \rangle \\
&=\frac{1}{2\pi i}\sum_{\sigma s}\frac{1}{2\omega_{\sigma s}}\int_{-\infty}^\infty d\omega \, e^{- i\omega\tau}\left(\frac{u_{\sigma s}^{i*}(\mathbf{x})u_{\sigma s}^{j}(\mathbf{x}')}{\omega+(\omega_{\sigma s}-i\epsilon)}-\frac{u_{\sigma s}^i(\mathbf{x})u_{\sigma s}^{*j}(\mathbf{x}')}{\omega-(\omega_{\sigma s}-i\epsilon)}\right),
\end{align}
where $\tau = t - t'$ as before. This propagator is time-translation-invariant, but spatial translation invariance is broken by the cavity. Note that the assumed axial symmetry implies that all modes have azimuthal dependence $e^{im \phi}$ for integer $m$, and that modes with $m = 0$ are real, while modes with $m > 0$ are complex conjugates of modes with $m < 0$, with the same frequency eigenvalues $\omega_{\sigma s}$. Thus, when we sum over all modes in the dipole approximation, we can reindex the sum to make the numerators of the two terms agree,
\begin{align}
\sum_{\sigma s}\frac{1}{2\omega_{\sigma s}}\left(\frac{u_{\sigma s}^{i*}(\mathbf{0})u_{\sigma s}^{j}(\mathbf{0})}{\omega+(\omega_{\sigma s}-i\epsilon)}-\frac{u_{\sigma s}^i(\mathbf{0})u_{\sigma s}^{*j}(\mathbf{0})}{\omega-(\omega_{\sigma s}-i\epsilon)}\right) = & \sum_{\sigma s}\frac{1}{2\omega_{\sigma s}}u_{\sigma s}^i(\bm{0})u_{\sigma s}^{*j}(\bm{0}) \nonumber \\ & \times \left(\frac{1}{\omega+(\omega_{\sigma s}-i\epsilon)}-\frac{1}{\omega-(\omega_{\sigma s}-i\epsilon)}\right).
\end{align}
Thus, the position-space propagator in the dipole approximation simplifies to
\begin{equation}
\label{eq:PhotonPropDipolePosition}
D^{ij}_{\rm cav}(\bm{0}, \bm{0};\tau) = \sum_{\sigma s}u_{\sigma s}^i(\bm{0})u_{\sigma s}^{*j}(\bm{0})\int \frac{d\omega}{2\pi}e^{-i\omega\tau}\frac{i}{\omega^2-\omega_{\sigma s}^2+i\epsilon}.
\end{equation}
Because the mode function values $u_{\sigma s}^i(\mathbf{0})$ are now constants, this propagator is trivially spatially translation-independent because it is spatially constant. Using the time-translation invariance as well, we can Fourier transform to 4-momentum-space,
\begin{equation}
\label{eq:PhotonPropDipoleMomentum}
D^{ij}(k) =\sum_{\sigma s}u_{\sigma s}^i(\bm{0})u_{\sigma s}^{*j}(\bm{0}) \, \frac{(2\pi)^3\delta^{(3)}(\mathbf{k})}{k_0^2-\omega_{\sigma s}^2+i\epsilon}.
\end{equation}
For the spherical and cylindrical cavity geometries we consider, the mode functions obey
\begin{equation}
u^i_{\sigma s}(\bm{0})u^{*j}_{\sigma s}(\bm{0})=
\begin{cases}
u_{\sigma,\nu p}^{\perp2}, & i=j=\{1,2\}\\
u_{\sigma,\nu p}^{||2}, & i=j=3,
\end{cases}
\label{eq:relativisticmodes}
\end{equation}
where the values of $u_{\sigma,\nu p}^{\perp2}$ and $u_{\sigma,\nu p}^{||2}$ are given in App.~\ref{app:photon_modes}. 

QED calculations in Coulomb gauge typically also require the use of the longitudinal part of the propagator, $D_{00}$, which arises from the static Coulomb interaction between charges. In our case, because this component of the propagator is time-independent, it will shift all energy levels uniformly and therefore not contribute to the shift of the cyclotron frequency~\cite{PhysRevD.32.729}. Therefore, we are free to ignore it and use only the transverse propagator $D^{ij}$.

In free space, rather than using the Feynman-gauge propagator as is typical in relativistic QED calculations, we will use the same Coulomb gauge as we did for the cavity propagator. The Coulomb-gauge propagator has transverse part
\begin{equation}
D^{ij}_{\rm free}(\mathbf{x},\mathbf{x}';\tau) \equiv \langle 0|TA^i(t,\mathbf{x})A^j(t^\prime,\mathbf{x}')|0 \rangle =i\int \frac{d^4k}{(2\pi)^4}\left(\delta^{ij}-\frac{k^i k^j}{|\mathbf{k}|^2}\right)\frac{e^{-ik_0 \tau+i\mathbf{k}\cdot(\mathbf{x}-\mathbf{x}')}}{k_0^2-|\mathbf{k}|^2+i\epsilon}.
\end{equation}
By the dipole approximation, we may drop the spatially-dependent phase factor, giving
\begin{equation}
\label{eq:PhotonPropDipolePositionFree}
D^{ij}_{\rm free}(\bm{0},\bm{0};\tau)=i\int \frac{d^4k}{(2\pi)^4}\left(\delta^{ij}-\frac{k^i k^j}{|\mathbf{k}|^2}\right)\frac{e^{-ik_0 \tau}}{k_0^2-|\mathbf{k}|^2+i\epsilon}.
\end{equation} 

\subsection{Cavity and Free Space Results}
\label{sec:rel_results}

We can now put the pieces of the one-loop diagram together to compute the amputated diagram $-i\Sigma_A$. In position space, this diagram contains the usual factor $i e \gamma^\mu$ at each vertex, a factor of the Schwinger propagator~\eqref{eq:SchwingerTranslationInvt} connecting $\mathbf{x}$ to $\mathbf{x}'$, the Schwinger phase~\eqref{eq:SchwingerPhase} and the dipole-approximation photon propagator~\eqref{eq:PhotonPropDipolePosition}. Both propagators depend on $\tau = t - t'$, so we may Fourier transform in $\tau$ to get the mixed energy-position representation 
\begin{equation}
-i\Sigma_A(\mathbf{x},\mathbf{x}^\prime;E_n)=(ie)^2 e^{-i \Phi(\mathbf{x},\mathbf{x}')}\int d\tau \, \gamma^\mu S(\mathbf{x}-\mathbf{x}^\prime;\tau) \gamma^\nu D_{\mu \nu}(\mathbf{0},\mathbf{0};\tau) e^{iE_n\tau},
\label{eq:positionselfenergy}
\end{equation}
which identical to Eq.~(\ref{eq:iSigmaANoPhase}) but with the Schwinger phase explicitly included and the dipole approximation used for the photon propagator. 

Equivalently, we may use the momentum-space propagators to write
\begin{equation}
-i \Sigma_A(X, X') = (ie)^2 \gamma^\mu e^{-i \Phi(X,X')} \int \frac{d^4 k}{(2\pi)^4}e^{-i k \cdot(X - X')}S(k) \gamma^\nu \int \frac{d^4 q}{(2\pi)^4} e^{-i q \cdot(X-X')} D_{\mu \nu}(q).
\end{equation}
The only piece of this expression which breaks translation invariance is the Schwinger phase, so we may factor it out and implicitly define an amputated momentum-space self-energy as
\begin{equation}
     -i \Sigma_A(X, X') = e^{-i \Phi(X,X')} \int \frac{d^4 p}{(2\pi)^4} \, e^{-i p \cdot (X-X')} \Sigma_A(p),
    \end{equation}
where
\begin{equation}
-i \Sigma_A(p) = (ie)^2 \int \frac{d^4 k}{(2\pi)^4} \, \gamma^\mu S(k) \gamma^\nu D_{\mu \nu}(p-k).
\label{eq:allmomentumselfenergy}
\end{equation}
Performing the time Fourier transform gives
\begin{equation}
\Sigma_A(\mathbf{x},\mathbf{x}'; E_n) = \int d \tau \, e^{i E_n \tau} \, \Sigma_A(X, X') = e^{-i \Phi(\mathbf{x}, \mathbf{x}')} \int \frac{d^3 \mathbf{p}}{(2\pi)^3}e^{i \mathbf{p} \cdot (\mathbf{x} - \mathbf{x}')}\, \Sigma_A(p^0 = E_n; \mathbf{p}).
\label{eq:momentumselfenergy}
\end{equation}

\paragraph{Cavity Result.} The details of the evaluation of the energy shift Eq.~\eqref{eq:energyshiftWeinberg} using position space \eqref{eq:positionselfenergy} and momentum space \eqref{eq:momentumselfenergy} are given in App.~\ref{sec:position} and App.~\ref{sec:momentum}, respectively. Because the spatial dependence of the wavefunctions are Hermite polynomials times a Gaussian weight, and the Schwinger propagator is linear in both position and momentum multiplied by a quadratic phase, the evaluation of the self-energy simply boils down to a large number of complex-valued Gaussian integrals. Depending on the representation, it is not necessarily more convenient to evaluate $\Sigma_A$ before evaluating the energy shift, as some of the Gaussian integrals simplify after first performing the spatial integrals over the external-state wavefunctions. 

After all position and momentum integrals have been performed, both approaches yield an energy shift in terms of a single Schwinger proper time integral and a photon mode sum,
\begin{multline}
\delta E_n=-\frac{ie^2}{4}\sum_{\sigma s}\frac{1}{\omega_{\sigma s}}\int_0^\infty ds \, \sqrt{\frac{\sigma_z^2}{\sigma_z^2+2is}} \, e^{-i(m^2+\beta)s}\\
\hspace{-5mm}\times\Big[\Big(m\,f_n(s)+(E_n+\omega_{\sigma s})g_n(s)+h_n(s)\Big)\text{erfc}\Big(e^{i\pi/4}\sqrt{s}(E_n+\omega_{\sigma s})\Big)e^{is(E_n+\omega_{\sigma s})^2}\\
\quad+\Big(m\,f_n(s)+(E_n-\omega_{\sigma s})g_n(s)+h_n(s)\Big)\text{erfc}\Big(-e^{i\pi/4}\sqrt{s}(E_n-\omega_{\sigma s})\Big)e^{is(E_n-\omega_{\sigma s})^2}\Big].
\label{eq:rel_main_start}
\end{multline}
For the ground state, $E_0=m$ and
\begin{equation} \label{eq:ground_results}
g_0(s)=-f_0(s)=2e^{-i\beta s}u_{\sigma,\nu p}^{\perp2}+e^{i\beta s}u_{\sigma,\nu p}^{\parallel2},\qquad h_0(s)=0.
\end{equation}
For the first excited state, $E_1=\sqrt{m^2 + 2\beta}$, and analogous expressions for $f_1(s)$, $g_1(s)$, and $h_1(s)$ are derived in App.~\ref{sec:position}.

Recall that $\sigma_z$ was introduced by hand in Eq.~\eqref{eq:electronwavefunction} to confine the electron in the $z$-direction. The external-state wavefunctions are consequently not exact eigenstates of the Dirac operator for which the Schwinger propagator is a Green's function. The result of this setup is that the energy shifts induced from the axial modes $u_{\sigma,\nu p}^\parallel$ of the photon are not suppressed compared to the radial modes $u_{\sigma,\nu p}^\perp$. Based on the analysis of Sec.~\ref{sec:nrqm_shift}, we assume that the axial modes \textit{should} be suppressed and drop them for the remainder of the calculation.\footnote{One might wonder if one could simply leave the electron unconfined along the $z$-direction. For a spherical cavity, this is impossible to reconcile with the cavity geometry, but it seems viable for a cylindrical cavity. However, besides invalidating the dipole approximation for the axial motion, this approach either leads to inconsistencies from the fact that both the electron and photon wavefunctions must satisfy boundary conditions at the cavity endcaps, or leads to spurious infrared divergences for an infinitely-long cylinder with no endcaps.}

At this stage, the energy shift still appears to depend explicitly on $\sigma_z$. However, since $E_n \sim m \gg \omega_{\sigma s}$ in the infrared, and $m^2 \gg \beta$ in the weak-field limit, the dominant behavior of the oscillating exponential in the integrand for large $s$ is
\begin{align}
    e^{-i m^2 s}{\rm erfc}\left(e^{i\pi/4}\sqrt{s}(m + \omega_{\sigma s})\right)e^{i (m + \omega_{\sigma s})^2 s} & \sim A e^{-i m^2 s} \\
     e^{-i m^2 s}{\rm erfc}\left(-e^{i\pi/4}\sqrt{s}(m - \omega_{\sigma s})\right)e^{i (m - \omega_{\sigma s})^2 s} & \sim B e^{-i m^2 s} + C e^{- 2 i m \omega_{\sigma s} s},
\end{align}
where $A$, $B$, and $C$ are non-oscillating functions of $s$, and we have used the expansion
\begin{equation}
    {\rm erfc}\left(e^{i\pi/4}z\right) \sim \frac{e^{-i z^2}}{z}\left(\frac{1-i}{\sqrt{2\pi}} + \mathcal{O}\left(\frac{1}{z}\right)\right)
\end{equation}
for large $z$. The fast oscillations of the exponent will damp the $A$ and $B$ terms unless $s < 1/m^2$, and the $C$ term is damped unless $s < 1/(2m \omega_{s\sigma})$. Since $\omega_{s\sigma} < m$ as long as the cavity shift is infrared-dominated, the first condition is automatically satisfied if the second is. The largest undamped region of integration for both terms is thus
\begin{equation}
0 < \frac{2s}{\sigma_z^2} < \frac{1}{ m \omega_{s\sigma} \sigma_z^2}.
\label{eq:sintegralregion}
\end{equation}
The $\sigma_z$ dependence cancels in the integrand, $\sqrt{\frac{\sigma_z^2}{\sigma_z^2+2is}} \sim 1$, as long as the upper bound of the inequality in~(\ref{eq:sintegralregion}) is much less than 1. For a cavity of size $R$, the lowest-lying cavity mode is of order $\omega_{\sigma s}^0 \sim 1/R$, so the energy shift is independent of $\sigma_z$ as long as $\sigma_z \gg \sqrt{R/m} \sim 4 \times 10^{-5} \ {\rm mm}$, where we took $R \sim 4 \ {\rm mm}$. (This condition was first noted in~\cite{Bordag:1988yd}, where $R$ was the separation between parallel plates.) This is satisfied in typical experimental setups, where $\sigma_z \sim 4 \times 10^{-4} \ {\rm mm}$. 

Using this approximation to set the $\sigma_z$-dependent factor to unity, and performing the trivial sum over $m=\pm1$, the remaining integral over $s$ can be performed analytically to obtain a ground state energy shift of
\begin{align}
\label{eq:deltaE0perpRel}
\delta E_0^\perp &= e^2\sum_{\sigma \nu p}u_{\sigma,\nu p}^{\perp2}\left[\frac{1-\frac{E_{0;\sigma,\nu p}^+}{\sqrt{m^2+2\beta}}}{\omega_{\sigma,1\nu p}^2+2m\omega_{\sigma,1\nu p}-2\beta}-\frac{1+\frac{E_{0;\sigma,\nu p}^-}{\sqrt{m^2+2\beta}}}{\omega_{\sigma,1\nu p}^2-2m\omega_{\sigma,1\nu p}-2\beta}\right]\\
&\approx \frac{e^2}{m}\sum_{\sigma \nu p}\frac{u_{\sigma,\nu p}^{\perp2}}{\omega_{\sigma,1\nu p}+\omega_c},
\end{align}
and, using Eqs.~\eqref{eq:f1_eq},~\eqref{eq:g1_eq}, and~\eqref{eq:h1_eq}, an excited state energy shift of
\begin{align}
\delta E_1^\perp &= \frac{e^2\beta}{E_1(E_1+m)}\sum_{\sigma \nu p}\frac{u_{\sigma,\nu p}^{\perp2}}{\omega_{\sigma,1\nu p}} \label{eq:rel_E1_shift} \\
&\times\Bigg[\frac{m+E_{1;\sigma,\nu p}^+}{\omega_{\sigma,1\nu p}^2+2(\omega_{\sigma,1\nu p}E_1+\beta)}\bigg(1-\frac{E_{1;\sigma,\nu p}^+}{\sqrt{\left(E_{1;\sigma,\nu p}^+\right)^2-\omega_{\sigma,1\nu p}^2-2(\omega_{\sigma,1\nu p}E_1+\beta)}}\bigg) \nonumber \\
&~~-\frac{}{}\frac{(E_1+m)^2\left(m-E_{1;\sigma,\nu p}^+\right)}{2\beta\left(\omega_{\sigma,1\nu p}^2+2(\omega_{\sigma,1\nu p}E_1-\beta)\right)}\bigg(1-\frac{E_{1;\sigma,\nu p}^+}{\sqrt{\left(E_{1;\sigma,\nu p}^+\right)^2-\omega_{\sigma,1\nu p}^2-2(\omega_{\sigma,1\nu p}E_1-\beta)}}\bigg) \nonumber \\
&~~+\frac{m+E_{1;\sigma,\nu p}^-}{\omega_{\sigma,1\nu p}^2-2(\omega_{\sigma,1\nu p}E_1-\beta)}\bigg(1+\frac{E_{1;\sigma,\nu p}^-}{\sqrt{\left(E_{1;\sigma,\nu p}^-\right)^2-\omega_{\sigma,1\nu p}^2+2(\omega_{\sigma,1\nu p}E_1-\beta)}}\bigg) \nonumber \\
&~~-\frac{(E_1+m)^2\left(m-E_{1;\sigma,\nu p}^-\right)}{2\beta\left(\omega_{\sigma,1\nu p}^2-2(\omega_{\sigma,1\nu p}E_1+\beta)\right)}\bigg(1+\frac{E_{1;\sigma,\nu p}^-}{\sqrt{\left(E_{1;\sigma,\nu p}^-\right)^2-\omega_{\sigma,1\nu p}^2+2(\omega_{\sigma,1\nu p}E_1+\beta)}}\bigg)\Bigg] \nonumber \\
&\approx \frac{e^2}{m}\sum_{\sigma \nu p}\frac{u_{\sigma,\nu p}^{\perp2}}{\omega_{\sigma,1\nu p}}\left(\frac{\omega_{\sigma,1\nu p}-\omega_c}{\omega_{\sigma,1\nu p}+\omega_c}-\frac{\omega_c}{\omega_{\sigma,1\nu p}-\omega_c}\right),
\end{align}
where we defined $E^\pm_{n;\sigma, \nu p}=E_n\pm\omega_{\sigma,1\nu p}$ and substituted $\beta=m\omega_c$. In both cases, we have finally taken the nonrelativistic limit by making the approximation $\omega_c, \omega_{\sigma,1\nu p} \ll m$ in the last line. As discussed in Sec.~\ref{sec:general}, these approximations are good because $\omega_c/m \sim 10^{-9}$, and the highest cavity modes are below the plasma frequency, $\omega_{\sigma,1\nu p}/m \lesssim \omega_p/m \sim 10^{-4}$.

Finally, because this is a relativistic calculation, we must perform a mass renormalization analogous to the one needed for the Lamb shift. The physical electron mass is defined by setting the full self-energy equal to $m$ in the absence of any external fields. Thus, we obtain the physical energy level shift by subtracting off the $B=0$ (i.e.~$\omega_c = 0$) result from $\delta E_i^\perp$, so
\begin{align}
\delta E_0^\perp &= -\frac{e^2}{m}\sum_{\sigma \nu p}\frac{u_{\sigma,\nu p}^{\perp2}}{\omega_{\sigma,1\nu p}}\frac{\omega_c}{\omega_{\sigma,1\nu p}+\omega_c}, \\
\delta E_1^\perp &=-\frac{e^2}{m}\sum_{\sigma \nu p}\frac{u_{\sigma,\nu p}^{\perp2}}{\omega_{\sigma,1\nu p}}\left(\frac{\omega_c}{\omega_{\sigma,1\nu p}-\omega_c}+\frac{2\omega_c}{\omega_{\sigma,1\nu p}+\omega_c}\right).
\end{align}
This now agrees with the energy level shifts Eq.~\eqref{eq:nr_en_shift} in the nonrelativistic calculation. Defining the cavity shift $\Delta \omega_c^{\text{cav}} = \delta E_1^\perp - \delta E_0^\perp$ as before, we recover 
\begin{equation}
\frac{\Delta \omega_c^{\rm cav}}{\omega_c} = -\frac{8 \pi \alpha}{m}\sum_{\sigma \nu p} \frac{u_{\sigma,\nu p}^{\perp2}}{\omega_{\sigma,1\nu p}^2-\omega_c^2},
\label{eq:cav_shift_Schwinger}
\end{equation}
in agreement with Eq.~\eqref{eq:cav_shift_generic}.

\paragraph{Free Space Result.} Our derivation implies that the cavity shift induced from radial modes is proportional to the quantity $D^{xx}_{\rm cav}(\bm{0}, \bm{0};\tau) + D^{yy}_{\rm cav}(\bm{0}, \bm{0};\tau)$, which is consistent with the cylindrical symmetry of the problem. By comparing Eq.~\eqref{eq:PhotonPropDipolePosition} and Eq.~\eqref{eq:PhotonPropDipolePositionFree}, we see that we can convert the cavity value of this quantity to the free space value by substituting
\begin{equation}
\sum_{\sigma s} \to \frac{d^3 \mathbf{k}}{(2 \pi)^3}, \qquad \omega_{\sigma s} \to |\mathbf{k}|, \qquad 2 u_{\sigma,\nu p}^{\perp2} \to (1 - (\hat{\mathbf{k}} \cdot \hat{\mathbf{x}})^2) + (1 - (\hat{\mathbf{k}} \cdot \hat{\mathbf{y}})^2)
\end{equation}
Applying these transformations to Eq.~\eqref{eq:cav_shift_Schwinger}, and noting that a sum over $s$ is equivalent to a sum over $m =\pm 1$ and over $\nu$ and $p$, we read off
\begin{equation}
\frac{\Delta \omega_c^{\rm free}}{\omega_c} = -\frac{\alpha}{4 \pi^2 m} \int d^3 \mathbf{k} \, \frac{1 + (\hat{\mathbf{k}} \cdot \hat{\mathbf{z}})^2}{|\mathbf{k}|^2 - \omega_c^2}.
\end{equation}
which agrees with Eq.~\eqref{eq:general_free_result}. We have thus successfully recovered the nonrelativistic results. 

\section{Spherical Cavity}
\label{sec:sphere_result}

Now we explicitly compute the cavity shift for a spherical cavity. We express it in terms of the difference of a divergent sum and divergent integral, appropriately regularized, and show that the difference can be evaluated by considering an appropriate contour integral. The final result precisely matches the classical result of Ref.~\cite{PhysRevA.34.2638}. 

\subsection{Setup}

\paragraph{Divergent Sum and Integral.} As shown in App.~\ref{app:photon_modes}, applying the cavity shift result Eq.~\eqref{eq:cav_shift_generic} to a spherical cavity of radius $a$ yields 
\begin{equation} \label{eq:sum_def}
\frac{\Delta \omega_c^{\text{cav}}}{\omega_c} = - \frac{8}{3 \pi} \frac{\alpha}{m a} \, S
\end{equation}
where
\begin{equation} \label{eq:spherical_sum}
S \equiv \sum_{p=1}^\infty f(p) = \sum_{p=1}^\infty \frac{\pi}{4} \left( \int_0^1 dx \, x^2 j_1^2(c_p x) \right)^{-1} \frac{1}{c_p^2 - z^2}. 
\end{equation}
Here, $j_1$ is a spherical Bessel function, and the dimensionless cyclotron angular frequency is $z = a \omega_c$. The dimensionless angular frequencies of the relevant modes are $c_p = \zeta'_{1p}$, the $p^{\text{th}}$ zero of the derivative of $x j_1(x) = (\sin x)/x - \cos x$. This implies $c_p \approx \pi p$ for large $p$, so that the sum is linearly divergent. 

To get a physical result, we must subtract $\Delta \omega_c^{\text{cav}}$ against the free space result Eq.~\eqref{eq:general_free_result}. Motivated by the spherical symmetry of the problem, we write the $\mathbf{k}$ integral in spherical coordinates and perform the angular integral, leaving an integral over the magnitude $k = |\mathbf{k}|$,
\begin{equation} \label{eq:sphere_int_def}
\frac{\Delta \omega_c^{\text{free}}}{\omega_c} = - \frac{8}{3 \pi} \frac{\alpha}{m a} \, I
\end{equation}
written in terms of the dimensionless integration variable $c = ka$,
\begin{equation} \label{eq:spherical_int}
I = \frac{1}{2} \int_0^{\infty} dc \, \frac{c^2}{c^2 - z^2}
\end{equation}
which should be evaluated as a principal value. This integral is also linearly divergent.

To make the subtraction well-defined, the sum and the integral must be regulated in a compatible way. Note that the mode angular frequencies in the sum and integral are $\omega_p = c_p/a$ and $\omega = k = c/a$, respectively. Thus, a frequency-dependent regulator should treat $c_p$ and $c$ in the same way. The contour integration method below will allow us to keep this regulator implicit, while automatically ensuring it is applied consistently. 

\paragraph{Rewriting the Sum.} The sum is written in terms of a $p^{\text{th}}$ Bessel zero, so to relate it to an integral we must render it well-defined and well-behaved for non-integer $p$.\footnote{The following technique is due to Olver~\cite{Olver_1950}, and is further discussed in Ref.~\cite{10.1063/1.1665021,abramowitz1965handbook}.} To start, note that the definition of $c_p$ is equivalent to $\tan c_p = - c_p / (c_p^2 - 1)$. Inverting this expression gives 
\begin{equation} \label{eq:analytic_ext}
c_p + \arctan \left( \frac{c_p}{c_p^2 - 1} \right) = \pi p
\end{equation}
where the principal branch of the arctangent is implied. This is an analytic relation between $c_p$ and $p$, which is smooth and monotonic in the entire relevant range $p \geq 1$.

Some useful identities follow straightforwardly from Eq.~\eqref{eq:analytic_ext}. For integer $p$, we have
\begin{equation} \label{eq:trig_results}
\cos(2 c_p) = \frac{c_p^4 - 3 c_p^2 + 1}{c_p^4 - c_p^2 + 1}, \qquad \sin(2 c_p) = -\frac{2 c_p (c_p^2 - 1)}{c_p^4 - c_p^2 + 1}.
\end{equation}
In addition, for arbitrary $p$, we have
\begin{align}
\frac{1}{e^{- 2 \pi i p} - 1} &= \frac12 \left(-1 + \frac{(c_p^2 - 1) \cos c_p - c_p \sin c_p}{c_p \cos c_p + (c_p^2 - 1)\sin c_p} \, i \right), \label{eq:exp_denom_result} \\
\frac{dc_p}{dp} &= \pi \, \frac{c_p^4 - c_p^2 + 1}{c_p^2 (c_p^2 - 2)}, \label{eq:deriv_result}
\end{align}
where the second line follows from differentiating Eq.~\eqref{eq:analytic_ext}. These results allow us to substantially simplify the summand. Carrying out the normalization integral in Eq.~\eqref{eq:spherical_sum} by writing $j_1$ in terms of trigonometric functions, we have 
\begin{align}
f(p) &= \frac{\pi}{2} \left( 1 + \frac{-1 + \cos(2 c_p) + c_p\sin(2 c_p)/2}{c_p^2} \right)^{-1} \frac{c_p^2}{c_p^2 - z^2} \\
&= \frac12 \frac{d c_p}{dp} \frac{c_p^2}{c_p^2 - z^2} \label{eq:final_fp}
\end{align}
by using Eq.~\eqref{eq:trig_results} and Eq.~\eqref{eq:deriv_result}. This form closely resembles the desired integrand. From this point onward, we drop the subscript on $c_p$ and view it as a continuous-valued quantity $c(p)$. 

\begin{figure}
\centering
\includegraphics[width=0.32\columnwidth]{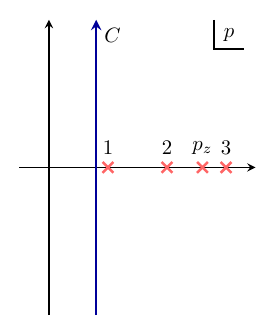}
\includegraphics[width=0.32\columnwidth]{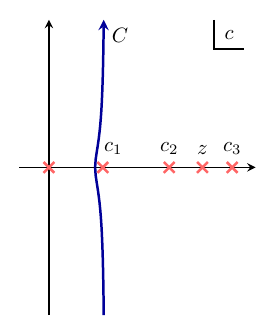}
\includegraphics[width=0.32\columnwidth]{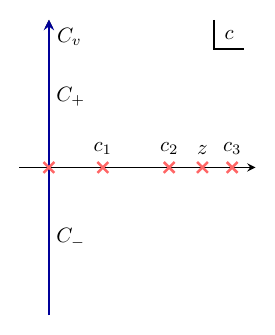}
\caption{Integration contours, with poles marked with red crosses. \textbf{Left:} the original integration contour, in terms of $p$.\textbf{ Center:} the same integration contour after changing variables to $c$. \textbf{Right:} moving the contour to the imaginary axis $C_v$ picks up half the residue of the pole at $c = 0$, and yields a pair of integrals, over $C_+$ and $C_-$, that can be related by symmetry.}
\label{fig:sphere}
\end{figure}

\subsection{Contour Integration}

\paragraph{Sum From Contour Integral.} Our next step is to consider the contour integral
\begin{equation}
A = \int_C \frac{f(p)}{e^{- 2 \pi i p} - 1} \, dp
\end{equation}
where $C$ runs up the imaginary $p$ axis with $\text{Re}(p) = 1 - i \epsilon$, as shown at left in Fig.~\ref{fig:sphere}. We will evaluate $A$ in two different ways, giving a relation between the desired sum and integral. 

First, we can evaluate $A$ by closing the contour in the right half of the complex plane. Here we assume the implicit regulator falls off quickly enough so that the contour at infinity contributes nothing, and that the regulator introduces no additional singularities within the contour. The implicit function $c(p)$ is also analytic within this contour, so the only singularities are poles at $p = 1, 2, \ldots$, and at the point where $c = z$, which we call $p = p_z$. 

Thus, the residue theorem implies 
\begin{equation}
A = - 2 \pi i \left( \text{Res}(p = 1) + \text{Res}(p = 2) + \ldots + \text{Res}(p = p_z) \right).
\end{equation}
We have $\text{Res}(p = n) = f(n)/(- 2 \pi i)$ for any integer $n$, so $A$ contains the regularized sum $S$. As for the additional pole at $p = p_z$, we have
\begin{equation}
f(p) = \frac12 \frac{d c}{dp} \frac{c^2}{c + z} \frac{1}{c - z} \simeq \frac12 \frac{d c}{dp} \frac{c^2}{c + z} \frac{dp}{d c} \frac{1}{p - p_z} \bigg|_{c = z} = \frac{z}{4} \frac{1}{p - p_z},
\end{equation}
which implies 
\begin{equation}
\text{Res}(p = p_z) = \frac{z}{4} \frac{1}{e^{- 2 \pi i p_z} - 1}.
\end{equation}
Applying the identity Eq.~\eqref{eq:exp_denom_result}, we conclude 
\begin{equation} \label{eq:A_first}
A = S + \frac{\pi z}{4} \left( \frac{(1 - z^2) \cos z + z \sin z}{(1-z^2) \sin z -z \cos z} + i \right). 
\end{equation}

\paragraph{Integral From Contour Integral.} We can also evaluate $A$ by first changing variables to $c$. This cancels the Jacobian factor in $f(p)$, and applying Eq.~\eqref{eq:exp_denom_result} gives
\begin{equation}
A = \int_{C} \frac12 \frac{c^2}{c^2 - z^2} \, g(c) \, dc, \qquad g(c) = \frac12 \left(-1 + \frac{(c^2 - 1) \cos c - c \sin c}{c \cos c + (c^2 - 1)\sin c} \, i \right).
\end{equation}
Now we are free to deform $C$ in the $c$-plane, though we must recall that since $c(p)$ is not bijective, there can be additional singularities which were not visible in the $p$-plane. In particular, there is a pole at $c = 0$ with residue $3i/(8 z^2)$.

On the imaginary axis, the function $g(c)$ obeys the useful symmetry property
\begin{equation} \label{eq:g_trick}
g(-iy) + 1 = - g(iy)
\end{equation}
for real $y > 0$. Furthermore, $g(iy)$ exponentially decays for large $y$, while $g(-iy)$ approaches $-1$. This motivates us to move $C$ to a contour $C_v$ which passes straight up along the imaginary axis, as shown at right in Fig.~\ref{fig:sphere}. Accounting for the pole at $c = 0$,
\begin{equation}
A = \frac{3 i}{8 z^2} (\pi i) + I_v, \qquad I_v = \int_{C_v} \frac12 \frac{c^2}{c^2 - z^2} \, g(c) \, dc.
\end{equation}
To evaluate $I_v$, we break the contour into $C_+$ and $C_-$, corresponding to the positive and negative imaginary axis respectively. Then by Eq.~\eqref{eq:g_trick}, the exponentially damped contribution from $C_+$ cancels with part of that from $C_-$. The remaining integral is
\begin{equation}
I_v = -\int_{C_-} \frac12 \frac{c^2}{c^2 - z^2} \, dc = \int_0^\infty \frac12 \frac{c^2}{c^2-z^2} \bigg|_{c = x - i \epsilon}
\end{equation}
where in the second equality, we rotated the contour through the fourth quadrant, giving an integral passing just below the positive real axis. This is very close to the integral $I$ in Eq.~\eqref{eq:spherical_int}, but in $I_v$ the integration contour passes below the pole at $c = z$, which has residue $z/4$, rather than straight through. This implies $I_v = I + \pi i z/4$, so that
\begin{equation} \label{eq:A_second}
A = - \frac{3 \pi}{8z^2} + I + \frac{\pi i z}{4}.
\end{equation}

Equating our results Eq.~\eqref{eq:A_first} and Eq.~\eqref{eq:A_second}, we conclude that
\begin{equation} \label{eq:sphere_final}
I - S = \frac{\pi z}{4} \left( \frac{(1 - z^2) \cos z + z \sin z}{(1-z^2) \sin z -z \cos z} + \frac{3}{2z^3} \right).
\end{equation}
Note that in this section, we implicitly assumed $z > c_1$. If $z < c_1$, then the term $\pi i z / 4$ is subtracted from both Eq.~\eqref{eq:A_first} and Eq.~\eqref{eq:A_second}, giving the same result for $I-S$. 

\subsection{Discussion}

Our result Eq.~\eqref{eq:sphere_final} implies that the physical cavity shift $\Delta \omega_c = \Delta \omega_c^{\text{cav}} - \Delta \omega_c^{\text{free}}$ satisfies
\begin{equation} \label{eq:sphere_final_2}
\frac{\Delta \omega_c}{\omega_c} = \frac{\alpha}{ma} \left( \frac23 \frac{(1 - z^2) \cos z + z \sin z}{(1-z^2) \sin z - z \cos z} \, z + \frac{1}{z^2} \right).
\end{equation}
Ref.~\cite{PhysRevA.34.2638} computed the cavity shift classically; taking its Eq.~(12) and plugging in the free-space decay rate (in the cgs units used in that work) gives exactly the same answer. 

Throughout this derivation, we have been keeping the regulator implicit. We argued earlier that the sum and integral are regulated consistently if the integrand's integration variable is regulated in the same way as $c_p$. This is indeed the case, because the integral above arose from changing variables from $p$ to $c$. We have also assumed the regulator is approximately equal to $1$ for low-lying modes, but falls to zero sufficiently quickly in the right halves of the $p$-plane and $c$-plane to close contours at infinity, while also being analytic in these regions. These properties are straightforward to satisfy, e.g.~one could regulate the integral by multiplying the integrand by $1 / (1 + c/c_0)^n$ for a large positive $c_0$ and sufficiently high $n$. 

Assuming these properties, our final answer is regulator-independent because it expresses the difference of the regulated sum and integral in terms of residues of poles at low values of $c$, specifically $c = 0$ and $c = z$. One subtlety is that above, we used the symmetry property Eq.~\eqref{eq:g_trick} to cancel off two integrals; however the symmetry property is generically modified at large $y$ by the regulator. This manipulation was legitimate because the integrals were exponentially damped; that is, their support is at small $y$, making them regulator-independent. 

The method we have used is related to the Abel--Plana formula,
\begin{equation}
\sum_{n=0}^\infty F(n) - \int_0^\infty F(x) \, dx = \frac{F(0)}{2} + i \int_0^\infty \frac{F(it) - F(-it)}{e^{2 \pi t} - 1} \, dt
\end{equation}
which holds for $F$ analytic in the right half-plane. In both cases one evaluates the difference of a regulated sum and integral by considering an appropriate contour integral. The difference is that our sum and integral are naturally written in terms of different variables ($p$ and $c$, respectively), and that additional singularities appear. 

\section{Cylindrical Cavity}
\label{sec:cylinder_result}

For a cylindrical cavity, our strategy will be qualitatively similar, but several new complications will arise. The mode sum will be significantly more complicated, depending on both the length $L$ and radius $a$ of the cylinder, and involving both TE and TM modes. In addition, the contour integrals we consider will contain both poles and branch cuts. 

\subsection{Setup}

\paragraph{Divergent Sum and Integral.}  As shown in App.~\ref{app:photon_modes}, applying the cavity shift result Eq.~\eqref{eq:cav_shift_generic} to a cylindrical cavity and performing the sum over the longitudinal index $p$ yields
\begin{equation}
\frac{\Delta \omega_c^{\text{cav}}}{\omega_c} = - \frac{\alpha}{ma} \, S
\end{equation}
where the sum contains contributions from TM and TE modes respectively, 
\begin{equation}
S = \sum_{n=1}^\infty \tanh\bigg( \frac{L c_n}{2a} \bigg) f_{\mathrm{TM}}^{(1)}(n) + \tanh\bigg( \frac{L \sqrt{c_n^2-z^2}}{2a} \bigg) f_{\mathrm{TM}}^{(2)}(n) + \tanh\bigg( \frac{L \sqrt{\bar{c}_n^2 - z^2}}{2a} \bigg) f_{\mathrm{TE}}(n)
\end{equation}
where
\begin{align}
f_{\mathrm{TM}}(n) &= \frac{c_n - \sqrt{c_n^2 - z^2}}{z^2 J_2^2(c_n)} \equiv f_{\mathrm{TM}}^{(1)}(n) + f_{\mathrm{TM}}^{(2)}(n) \\
f_{\mathrm{TE}}(n) &= \frac{1}{(J_1^2(\bar{c}_n) - J_2^2(\bar{c}_n)) \sqrt{\bar{c}_n^2 - z^2}}
\end{align}
and $f_{\mathrm{TM}}^{(1)}(n)$ and $f_{\mathrm{TM}}^{(2)}(n)$ are the parts of $f_{\mathrm{TM}}(n)$ with and without a square root, respectively. 

We have again defined $z = a \omega_c$, relabeled the mode index $\nu$ to the more conventional $n$ since there is now no possibility of confusion with the electron state index, and let $c_n$ and $\bar{c}_n$ be the $n^{\text{th}}$ zeroes of the cylindrical Bessel function $J_1$ and its derivative $J_1'$, respectively. Note that the arguments of the square roots can be negative; here and for the rest of this section, the principal branch of the square root is implied. For large $n$, we have $c_n \approx (n+1/4) \, \pi$ and $\bar{c}_n \approx (n-1/4) \, \pi$, so that the sum is linearly divergent. 

To compare this to the free space result Eq.~\eqref{eq:general_free_result}, we note that $n$ indexes the radial dependence in cylindrical coordinates. Motivated by the cylindrical symmetry of the problem, we write the $\mathbf{k}$ integral in cylindrical coordinates and integrate over the azimuthal angle and the axial momentum $k_z$, giving
\begin{equation}
\frac{\Delta \omega_c^{\text{free}}}{\omega_c} = - \frac{\alpha}{ma} \, I
\end{equation}
where, defining the integration variable $c = k_\perp a$, we have
\begin{align}
I &= \frac{a}{2 \pi} \int_0^{\infty} dk_\perp \, k_\perp \int_{-\infty}^{\infty}dk_z \left(1 + \frac{k_z^2}{k_\perp^2 + k_z^2}\right)\left(\frac{1}{k_\perp^2 + k_z^2 -\omega_c^2}\right) \\
&= \frac{1}{2z^2} \left(\int_0^\infty dc \, c^2 - \int_{z}^\infty dc \, \frac{c (c^2 - 2 z^2)}{\sqrt{c^2 - z^2}}\right). \label{eq:cyl_full_int}
\end{align}
It will be useful to decompose the integral as $I = I_1 + I_2 + I_3$, where
\begin{align}
I_1 &= \int_0^\infty dc \, \frac{c^2}{2z^2}, \\
I_2 &= \int_z^\infty dc \, \frac{-c \sqrt{c^2-z^2}}{2z^2}, \\
I_3 &= \int_z^\infty dc \, \frac{c}{2 \sqrt{c^2-z^2}}.
\end{align}
Some of these integrals are cubically divergent, but the full integral is only linearly divergent.

To match the sum and integral, we again demand the regulator depend only on frequency, and note that the TM modes have squared angular frequency $(c_n/a)^2 + (p \pi/L)^2$, while the plane wave modes have $\omega^2 = k_\perp^2 + k_z^2$. The axial momentum $k_z$ just corresponds to $p \pi / L$, and neither the $p$ sum nor the $k_z$ integral have to be regulated, as their results are finite. Thus, for the TM modes it suffices to identify $c_n$ with $k_\perp a = c$. 

However, for the TE modes, a similar argument shows that we should identify $\bar{c}_n$ with $c$, and asymptotically $c_n$ and $\bar{c}_n$ differ by $\pi/2$. This difference leads to an order-one ambiguity in $S-I$, so at this point it would not be possible to evaluate it using an explicit regulator. Instead, we must find which parts of the integral correspond to the TE and TM modes. 

\paragraph{Rewriting the Sum.} Since the integral does not depend on the cylinder length $L$, it is useful to decompose the sum as $S = S_0 + S'$, where 
\begin{equation} \label{eq:S0_def}
S_0 \equiv \sum_{n=1}^\infty f_{\mathrm{TM}}(n) + f_{\mathrm{TE}}(n) 
\end{equation}
is also independent of $L$, and the remainder term $S'$ depends only on the combination $1 - \tanh(x)$ for various arguments $x$. Though the form of $S'$ is messy, it converges exponentially because $1 - \tanh(x) \approx 2 e^{-2x}$ for large $x$, so that it can be quickly evaluated with high numeric precision. 

Next, we rewrite the summand in a way that is well-defined and well-behaved for non-integer $n$. First, for $f_{\mathrm{TM}}(n)$, we start with Olver's trick (see section 9.5 of Ref.~\cite{abramowitz1965handbook}), defining
\begin{equation} \label{eq:cn_trick}
\tan(\pi n) + \frac{J_1(c_n)}{Y_1(c_n)} = 0,
\end{equation}
where $Y_1$ is a Bessel function of the second kind. Similar to Eq.~\eqref{eq:analytic_ext}, this can be regarded as an analytic equation relating $c_n$ and $n$. By differentiating with respect to $n$, we can show that 
\begin{equation}
\frac{dc_n}{dn} = \frac{\pi^2 c_n}{2} \, (Y_1(c_n)^2 + J_1(c_n)^2).
\end{equation}
The summand is only evaluated at integer $n$, where $J_1(c_n)$ vanishes. Using the Wronskian $J_2(x) Y_1(x) - J_1(x) Y_2(x) = 2 / (\pi x)$, we find 
\begin{equation}
f_{\mathrm{TM}}(n) = \frac{c_n}{2} \frac{c_n - \sqrt{c_n^2 - z^2}}{z^2} \frac{dc_n}{dn}.
\end{equation}
Up to the Jacobian factor, this is just the integrand of $I_1 + I_2$.

As for $f_{\mathrm{TE}}(n)$, we can extend $\bar{c}_n$ to real $n$ using 
\begin{equation}
\tan(\pi n) + \frac{J_1'(\bar{c}_n)}{Y_1'(\bar{c}_n)} = 0.
\end{equation}
Differentiating with respect to $n$ and rewriting the result in terms of Bessel derivatives gives
\begin{equation}
\frac{d\bar{c}_n}{dn} = \frac{\pi^2}{2} \frac{\bar{c}_n^3}{\bar{c}_n^2 - 1} \, (J_1'(\bar{c}_n)^2 + Y_1'(\bar{c}_n)^2).
\end{equation}
The summand is only evaluated at integer $n$, where $J_1'(\bar{c}_n)$ vanishes. Using the recursion relation $J_1'(x) + J_2(x) = J_1(x)/x$ and the Wronskian $J_1'(x) Y_1(x) - J_1(x) Y_1'(x) = 2 / (\pi x)$,
\begin{equation}
f_{\mathrm{TE}}(n) = \frac{1}{2} \frac{\bar{c}_n}{\sqrt{\bar{c}_n^2 - z^2}} \frac{d\bar{c}_n}{dn}.
\end{equation}
Up to the Jacobian factor, this is just the integrand of $I_3$. From this point on, we will drop subscripts on $c_n$ and $\bar{c}_n$. 

Having identified which parts of the integral correspond to the TE and TM sums, we could now choose a regulator and evaluate $S_0 - I$ numerically. We will consider this in Sec.~\ref{sec:conclusion}, but here we will continue with an analytic argument, as it will yield a compact explicit result. 

\subsection{Contour Integration}

Again, we will relate the sum and integral with appropriately chosen contour integrals. The above argument shows that the TM mode summand can be decomposed as
\begin{equation}
f_{\mathrm{TM}}^{(1)}(n) = \frac{c^2}{2 z^2} \frac{dc}{dn}, \qquad f_{\mathrm{TM}}^{(2)}(n) = - \frac{c \sqrt{c^2 - z^2}}{2z^2} \, \frac{dc}{dn}.
\end{equation}
The first term can be treated similarly to the spherical cavity, while the second term will require considering the branch cut from the square root.

\paragraph{TM Modes, No Branch Cut.} 

Here we consider the contour integral 
\begin{equation}
A = \int_C \frac{f_{\mathrm{TM}}^{(1)}(n)}{e^{- 2 \pi i n} - 1} \, dn 
\end{equation}
where the contour $C$ runs up along the imaginary $n$ axis with $\text{Re}(n) = 1 - i \epsilon$. Closing the contour in the right half-plane, there are no poles besides those at integer $n$, so that
\begin{equation}
A = \sum_{n=1}^\infty f_{\mathrm{TM}}^{(1)}(n). 
\end{equation}

Next, we evaluate the integral again, first changing variables from $n$ to $c$, which yields
\begin{equation} \label{eq:gc_expr}
A = \int_{C'} \frac{c^2}{2 z^2} \, g(c) \, dc, \qquad g(c) = - \frac12 \left( 1 + \frac{Y_1(c)}{J_1(c)} \, i \right)
\end{equation}
where $g(c)$ was evaluated using Eq.~\eqref{eq:cn_trick}. Here we need to check for additional poles and branch cuts, in the region of $c$-space not covered by $p$. First, $Y_1(c)$ has a branch point at $c = 0$, with a branch cut along the negative real axis. Second, near the origin $g(c) \sim 1/c^2$, but this cancels against the factor of $c^2$ in the integrand.

Thus, there is no obstruction to deforming the contour $C'$ to the imaginary $c$ axis. Along this axis, $g(c)$ obeys the symmetry property
\begin{equation} \label{eq:g_sym_prop}
g(-iy) + 1 = g(iy) 
\end{equation}
for real $y > 0$. Thus, the top half of the contour gives a well-behaved, exponentially damped integral, while the bottom half contains the same exponentially damped integral, with the same sign, plus an integral which becomes $I_1$ after rotating the contour to the real axis. Thus,
\begin{equation}
A = I_1 - \int_0^\infty \frac{y^2}{z^2} \, g(iy) \, (i \, dy).
\end{equation}
Comparing our results, we conclude that 
\begin{equation} \label{eq:cyl_result_1}
\sum_{n=1}^\infty f_{\mathrm{TM}}^{(1)}(n) - I_1 = - \frac{1}{\pi} \int_0^\infty \frac{K_1(y)}{I_1(y)} \, \frac{y^2}{z^2} \, dy.
\end{equation}
Here we have used $g(c) = K_1(-ic)/ (\pi i \, I_1(-ic))$, valid everywhere that the integrand is evaluated, to make the exponential damping of the integrand manifest. 

\paragraph{TM Modes, With Branch Cut.} 

\begin{figure}
\centering
\includegraphics[width=0.32\columnwidth]{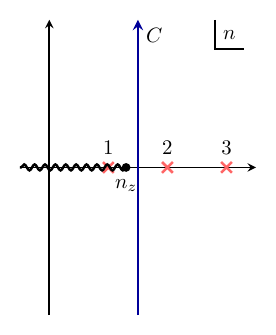}
\includegraphics[width=0.32\columnwidth]{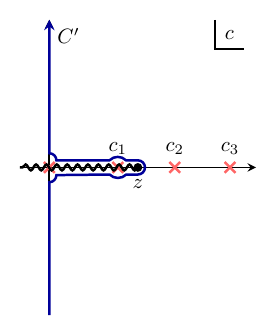}
\includegraphics[width=0.32\columnwidth]{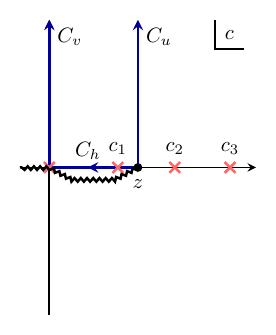}
\caption{Integration contours, with poles marked with red crosses. We show the result for the part of the TM mode integral with a branch cut, though the TE mode integral is qualitatively similar. \textbf{Left:} the original integration contour, in terms of $n$. \textbf{Center:} an equivalent integration contour in terms of $c$, which diverts around the branch cut. \textbf{Right:} using symmetry, the two halves of the contour integral can be combined into the single contour $C_h+C_v$, which passes straight through poles on the real axis.  The branch cut has been pushed downward for clarity. Alternatively, deforming the contour to $C_u$ yields an integral which is easier to numerically evaluate.}
\label{fig:cylinder}
\end{figure}

In this case, $f_{\mathrm{TM}}^{(2)}(c)$ has branch points at $c = \pm z$, and we take the branch cut to go straight between them. This implies $f_{\mathrm{TM}}^{(2)}(n)$ has a branch point at $n = n_z$, the point corresponding to $c = z$, with a branch cut going straight to the left. 

Now, we consider the contour integral
\begin{equation}
A = \int_C \frac{f_{\mathrm{TM}}^{(2)}(n)}{e^{- 2 \pi i n} - 1} \, dn
\end{equation}
where $C$ goes straight up, crossing the real axis immediately to the right of the $n = n_z$ branch point, as shown at left in Fig.~\ref{fig:cylinder}. Closing the contour in the right-half plane gives
\begin{equation}
A = \sum_{n = n^*+1}^\infty f_{\mathrm{TM}}^{(2)}(n)
\end{equation}
where $n^*$ is the highest $n$ to the left of the branch point. 

Next, we evaluate $A$ by changing variables to $c$, where 
\begin{equation} \label{eq:A_TM_expr}
A = - \int_{C'} \frac{c \, \sqrt{c^2-z^2}}{2 z^2} \, g(c) \, dc
\end{equation}
and $g(c)$ is as defined in Eq.~\eqref{eq:gc_expr}. There is now a simple pole at $c = 0$, and a branch cut extending to the left of $c = z$. We thus deform the contour so that it goes up along the imaginary $c$ axis, with a rightward detour around the branch cut, as shown in Fig.~\ref{fig:cylinder}. 

Letting the curved arcs have radius $\epsilon$, we can decompose $A = A_h + A_v + A_c$, containing the contributions from the horizontal, vertical, and curved parts of the contour respectively. Without the branch cut, $A_h$ would vanish, while $A_c$ would contain contributions from enclosed poles. However, since crossing the branch cut flips the sign, here $A_c$ cancels while the two contributions to $A_h$ have the same sign. That is, we have 
\begin{equation}
A_h = -\int_{C_h} \frac{c \sqrt{c^2 - z^2}}{z^2} \, g(c) \, dc
\end{equation}
where the principal branch of the square root is implied, and the contour $C_h$ passes straight along the real axis from $c = z$ to $c = \epsilon$, directly through a number of poles. It will be useful to extract out the imaginary part, 
\begin{equation}
A_h = \text{Re}(A_h) - \frac{i}{2} \int_0^z \frac{x \sqrt{z^2 - x^2}}{z^2} \, dx.
\end{equation}

Next, we decompose $A_v = A_+ + A_-$, from the top and bottom parts of the contour respectively, so that $A_+$ is
\begin{equation}
A_+ = - \int_{C_v} \frac{c \, \sqrt{c^2-z^2}}{2 z^2} \, g(c) \, dc
\end{equation}
where $C_v$ passes straight along the imaginary axis from $c = i \epsilon$ to $c = i \infty$. Since $g(iy) = K_1(y) / (\pi i I_1(y))$, $A_+$ is a real, exponentially damped integral. As for $A_-$, decomposing $g(-iy) = g(iy) - 1$, the first term yields another copy of the integral $A_+$. As for the other term, we rotate the contour through the fourth quadrant so that it passes from positive infinity to $- i \epsilon$, just below the real axis. This yields 
\begin{equation}
A_- = A_+ - \int_z^\infty \frac{x \sqrt{x^2-z^2}}{2 z^2} \, dx + \frac{i}{2} \int_0^z \frac{x \sqrt{z^2-x^2}}{z^2} \, dx
\end{equation}
where the sign on the last term is positive since the contour passes just below the branch cut. 

Combining our results, the second term of $A_-$ is the desired regulated integral $I_2$, while the third term cancels with the imaginary part of $A_h$, giving 
\begin{equation}
A = I_2 - \text{Re} \int_{C_h + C_v} \frac{c \sqrt{c^2 - z^2}}{z^2} \, g(c) \, dc.
\end{equation}
Parametrizing the contour by $c = \sqrt{z^2-u^2}$ gives 
\begin{equation} \label{eq:cyl_result_2}
\sum_{n = n^* + 1}^{\infty} f_{\mathrm{TM}}^{(2)}(n) - I_2 = \frac{1}{\pi} \, \text{Re} \int_0^\infty \frac{u^2}{z^2} \frac{K_1(\sqrt{u^2-z^2})}{I_1(\sqrt{u^2-z^2})} \, du
\end{equation}
where a principal value is implied for the pole at $u = z$. Alternatively, we can deform the contour to $C_u$, which passes directly upward from $c = z$, as shown at right in Fig.~\ref{fig:cylinder}. This picks up residues at the poles at $c = 0$ and $c = c_n$ for $n \leq n^*$, but these contributions are purely imaginary and do not affect the real part. Parametrizing by $c = z + iy$ gives 
\begin{equation} \label{eq:cyl_result_2_alt}
\sum_{n = n^* + 1}^{\infty} f_{\mathrm{TM}}^{(2)}(n) - I_2 = \frac{1}{\pi} \, \text{Re} \int_0^\infty \frac{(y-iz) \sqrt{y^2-2iyz}}{z^2} \frac{K_1(y-iz)}{I_1(y-iz)} \, dy.
\end{equation}
As we will see, these two expressions are both useful in different cases. 

\paragraph{TE Modes.} This is very similar to the previous case. Taking the analogous initial integral $\bar{A}$ and changing variables from $p$ to $\bar{c}$ gives 
\begin{equation} \label{eq:A_TE_expr}
\bar{A} = \int_{C'} \frac12 \frac{\bar{c}}{\sqrt{\bar{c}^2 - z^2}} \, \bar{g}(\bar{c}) \, d\bar{c} = \sum_{n = n^*+1}^\infty f_{\mathrm{TE}}(n)
\end{equation}
where $\bar{n}^*$ is the highest integer to the left of the branch point at $\bar{c} = z$, and 
\begin{equation}
\bar{g}(\bar{c}) = - \frac12 \left( 1 + \frac{Y_1'(\bar{c})}{J_1'(\bar{c})} \, i \right)
\end{equation}
which, on the imaginary axis, obeys the symmetry property
\begin{equation}
\bar{g}(-iy) + 1 = \bar{g}(iy).
\end{equation}
By repeating the reasoning above, we have 
\begin{equation}
\bar{A} = I_3 + \text{Re} \int_{C_h + C_v} \frac{\bar{c}}{\sqrt{\bar{c}^2 - z^2}} \, \bar{g}(\bar{c}) \, d\bar{c}.
\end{equation}
We have $\bar{g}(\bar{c}) = K_1'(-i\bar{c})/ (\pi i \, I_1'(-i\bar{c}))$ everywhere the integrand is evaluated, which implies
\begin{equation} \label{eq:cyl_result_3}
\sum_{n = \bar{n}^*+1}^\infty f_{\mathrm{TE}}(n) - I_3 = \frac{1}{\pi} \, \text{Re} \int_0^\infty \frac{K_1'(\sqrt{u^2-z^2})}{I_1'(\sqrt{u^2-z^2})} \, du.
\end{equation}
Again, we may deform the contour to $C_u$, which goes directly upward. This picks up purely imaginary contributions from the poles at $\bar{c} = 0$ and $\bar{c} = \bar{c}_n$ for $n \leq \bar{n}^*$, which do not affect the real part of the integral. Then we conclude 
\begin{equation} \label{eq:cyl_result_3_alt}
\sum_{n = \bar{n}^*+1}^\infty f_{\mathrm{TE}}(n) - I_3 = \frac{1}{\pi} \, \text{Re} \int_0^\infty \frac{y-iz}{\sqrt{y^2-2iyz}} \, \frac{K_1'(y-iz)}{I_1'(y-iz)} \, dy.
\end{equation}

\subsection{Discussion}

\paragraph{Final Result.} By summing~\eqref{eq:cyl_result_1},~\eqref{eq:cyl_result_2}, and~\eqref{eq:cyl_result_3}, we have
\begin{equation} \label{eq:cyl_answer}
S - I = J + S' + \sum_{n = 1}^{n^*} f_{\mathrm{TM}}^{(2)}(n) + \sum_{n = 1}^{\bar{n}^*} f_{\mathrm{TE}}(n) \equiv J + \Delta S,
\end{equation}
where $S' = S - S_0$ contains all the dependence on the cavity length $L$. We can write $\Delta S$ as
\begin{align} \label{eq:Delta_S_result}
\Delta S &= \sum_{n=1}^\infty \Big(\tanh\Big( \frac{L c_n}{2a} \Big) - 1\Big) \, f_{\mathrm{TM}}^{(1)}(n) \nonumber \\ 
&\quad + \sum_{n = n^* + 1}^\infty \Big(\tanh\Big( \frac{L\sqrt{c_n^2 - z^2}}{2a} \Big) - 1\Big) \, f_{\mathrm{TM}}^{(2)}(n) + \sum_{n=1}^{n^*} \tan \Big( \frac{L \sqrt{z^2-c_n^2}}{2a} \Big) \, i f_{\mathrm{TM}}^{(2)}(n) \nonumber \\
&\quad + \sum_{n = \bar{n}^* + 1}^\infty \Big(\tanh\Big( \frac{L\sqrt{\bar{c}_n^2 - z^2}}{2a} \Big) - 1 \Big) \, f_{\mathrm{TE}}(n) \, + \sum_{n=1}^{\bar{n}^*} \tan \Big( \frac{L \sqrt{z^2 - \bar{c}_n^2}}{2a} \Big) \, i f_{\mathrm{TE}}(n),
\end{align}
in which all terms are manifestly real. The remaining integral is
\begin{equation} \label{eq:J_int_def}
J = \frac{1}{\pi} \, \text{Re} \int_0^\infty \left [ \frac{K_1'(\sqrt{u^2-z^2})}{I_1'(\sqrt{u^2-z^2})} + \frac{u^2}{z^2} \left(\frac{K_1(\sqrt{u^2-z^2})}{I_1(\sqrt{u^2-z^2})} - \frac{K_1(u)}{I_1(u)} \right) \right ]\, du
\end{equation}
where a principal value is implied for all poles of the integrand. Alternatively, by summing~\eqref{eq:cyl_result_1},~\eqref{eq:cyl_result_2_alt}, and~\eqref{eq:cyl_result_3_alt}, we have
\begin{equation} \label{eq:J_alt_def}
J = \frac{1}{\pi} \, \text{Re} \int_0^\infty\left [ \frac{(y-iz) \sqrt{y^2-2iyz}}{z^2} \frac{K_1(y-iz)}{I_1(y-iz)} - \frac{K_1(y)}{I_1(y)} \, \frac{y^2}{z^2} + \frac{y-iz}{\sqrt{y^2-2iyz}} \, \frac{K_1'(y-iz)}{I_1'(y-iz)}\right ] \, dy.
\end{equation}
In both cases, the answer is in terms of an exponentially damped integral $J$, finite sums, and exponentially damped infinite sums. As for the sphere, the exponential damping implies that the cavity shift is regulator-independent, and dominantly due to low-lying modes. 

We expect the cavity shift to become large when the cyclotron frequency crosses a resonant cavity mode; this behavior is captured within $S'$. The first form Eq.~\eqref{eq:J_int_def} is useful because it is more closely related to the classical result, while the second form Eq.~\eqref{eq:J_alt_def} has an integrand with no poles, making it easier to numerically evaluate.

\begin{figure}
\centering
\includegraphics[width=0.8\columnwidth]{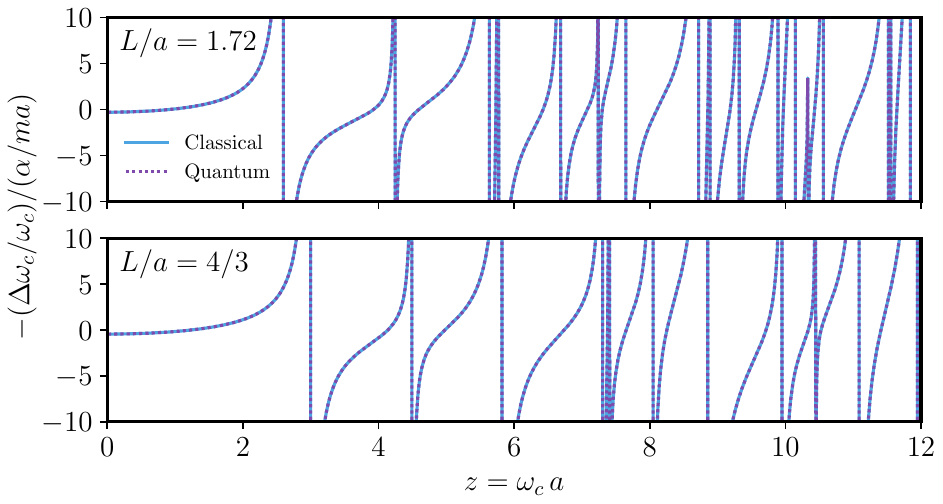}
\caption{Numeric values of the cavity shift for the cylinder, computed using our renormalized mode sum result Eq.~\eqref{eq:cyl_answer}, and using the existing classical Green's function result Eq.~\eqref{eq:classical_cylinder}. The two agree perfectly, and cannot be distinguished on the plot. The sums are evaluated up to $n = 50$, but summing to $n = 10$ already gives a match to $\sim 1\%$ accuracy. The top value of $L/a$ was chosen to match the most recent measurement~\cite{Fan:2022oyb}, and the bottom value was chosen to match that considered in Ref.~\cite{PhysRevA.32.3204}.}
\label{fig:cylinder_result}
\end{figure}

\paragraph{Comparison to Classical Result.} The physical fractional cavity shift is
\begin{equation}
\frac{\Delta \omega_c}{\omega_c} = - \frac{\alpha}{m a} (S - I).
\end{equation}
The classical result, given by taking the real part of Eq.~(4.23) of Ref.~\cite{PhysRevA.32.3204}, is 
\begin{multline} \label{eq:classical_cylinder}
\frac{\Delta \omega_c}{\omega_c} = \frac{2 \alpha}{m L} \Bigg[ \frac12 \log(4 \cos^2(\xi/2)) + \sum_{n=1}^\infty (-1)^n \left(\frac{\sin(n\xi)}{n^2 \xi} + \frac{\cos(n \xi) - 1}{n^3 \xi^2} \right) \\ - \text{Re} \sum_{p=0}^\infty \left( \frac{K_1'(\mu_p a)}{I_1'(\mu_p a)} + \frac{k_p^2}{\omega_c^2} \left( \frac{K_1(\mu_p a)}{I_1(\mu_p a)} - \frac{K_1(k_p a)}{I_1(k_p a)} \right) \right) \Bigg]
\end{multline}
where $\xi = \omega_c L$, $k_p = (2p+1) \pi / L$ and $\mu_p = \sqrt{k_p^2 - \omega_c^2}$. (We note that here we are taking $L$ to be the cylinder's total length, but in Ref.~\cite{PhysRevA.32.3204} it was the half of the cylinder's length.) The first line of Eq.~\eqref{eq:classical_cylinder} depends only on $L$, and represents the effect of the cavity endcaps. The second line is the correction due to the curved sides of the cavity, and contains poles when the cyclotron frequency is equal to a cavity mode frequency. 

As $L \to \infty$, for fixed $a$ and $\omega_c$, it is straightforward to see why these answers match. Then in Eq.~\eqref{eq:Delta_S_result}, the infinite sums go to zero, and the finite sums oscillate rapidly. For simplicity, we suppose the limit $L \to \infty$ is taken with some small fractional smearing in $L$, so that the tangent terms average to zero, giving $\Delta \omega_c / \omega_c \simeq -\alpha J/ma$. Similarly, in the classical result Eq.~\eqref{eq:classical_cylinder}, the first term averages to zero, the sum over $n$ goes to zero, and the sum over $p$ becomes a discrete approximation of the integral $\pi J$ in Eq.~\eqref{eq:J_int_def} with bin spacing $2 \pi a / L$, giving $\Delta \omega_c / \omega_c \simeq - (2 \alpha / m L)(L / 2 \pi a) (\pi J)$, precisely the same result. 

More generally, it is straightforward to check using Eq.~\eqref{eq:J_alt_def} that our result matches the classical result numerically; we show this comparison in Fig.~\ref{fig:cylinder_result}.

\section{Conclusion}
\label{sec:conclusion}

We have shown that the cavity shift for an electron undergoing cyclotron motion in a perfectly conducting cavity can be computed quantum-mechanically, treating the electron either nonrelativistically or relativistically. For a spherical cavity, the result is a perfect analytic match to the result derived using classical Green's functions, while for a cylindrical cavity, the two expressions appear quite different but match numerically. To conclude, we place our results in a broader context, and look toward future work. 

\paragraph{Relation to Other Calculations.} The renormalization of the cavity shift is qualitatively similar to that of the Casimir effect, reviewed in Refs.~\cite{Milton:2004ya,Jaffe:2005vp,Lamoreaux:2005gf,Klimchitskaya:2009cw}. Unlike the Lamb shift, which contains an $\mathcal{O}(1)$ high-energy contribution that must be calculated in the relativistic theory, both the cavity shift and Casimir effect can be computed accurately using nonrelativistic electrons. In fact, they both arise far in the infrared, from modes with energies on the scale $1/R$, where $R$ is the length scale of the conductors. Furthermore, we have seen that earlier calculations of the cavity shift using the (self)-field of the electron give the same results as our calculation which directly uses cavity modes. This is reminiscent of how the Casimir effect can be thought of as due to the fields of fluctuating electrons in the cavity walls, as was shown by Lifshitz and Schwinger~\cite{lifshitz1956theory,schwinger_1978}, or equivalently as due to the vacuum energy of modes.

On a more technical level, a great deal of formalism has been developed to calculate the Casimir effect in various geometries, which may be of use for future cavity shift calculations. In particular, our contour integration method is similar in spirit to the treatment of the Casimir effect for spherical shells given in Ref.~\cite{10.1063/1.1666141}, and it may be possible to apply the generalizations of the Abel--Plana formula derived in Refs.~\cite{inui2003generalized,Saharian:2007ph} to more general cavity shapes. 

\paragraph{Higher-Order Corrections.} Since we have shown that the one-loop self energy gives precisely the same cavity shift as a classical calculation, it is natural to ask whether this agreement continues at higher order. Though we do not have a definitive answer to this question, we can give some heuristic arguments. A two-loop diagram only contributes to the cavity shift in regions where at least one of the loop momenta is at the infrared scale $\sim 1/R$. 

First, suppose only one of the loop momenta is at the low scale. By dimensional analysis, one would expect a contribution to the cavity shift scaling as $\Delta \omega_c / \omega_c \sim \alpha^2 /( mR)$, which is potentially relevant. Restoring factors of $\hbar$, we have $\alpha \propto 1/\hbar$, and $1/m$ becomes the Compton scale $\hbar/m$, so that this contribution to the cavity shift is not $\hbar$-independent. In this sense, it is not classical. However, from the perspective of the low-energy loop, one can absorb the high-energy loop's effect into a renormalized mass and charge for the electron. Then our one-loop calculation would already include this effect if one uses the physical electron mass and charge. 

There may be a residual contribution from the region where both loop momenta are at the low scale, which would scale roughly as $\Delta \omega_c / \omega_c \sim (\alpha / mR)^2$. This is $\hbar$-independent, and thus could potentially arise from a higher-order classical calculation. In the leading-order classical calculation, one considers the backreaction of the electron's cavity field on the electron's motion. At second-order, one could account for how this backreaction modifies the electron's cavity field, which then acts back on the electron again. We do not know if this contribution would agree with the two-loop calculation, but it will be negligible for the foreseeable future.  

\paragraph{Longitudinal Cavity Shift.} As shown in Eq.~\eqref{eq:shift_scaling}, the fractional cavity shifts due to transverse and longitudinal (i.e.~electrostatic) fields are proportional to $1/m$ and $m$ respectively. As a result of this scaling, only the transverse field is relevant for trapped electrons, while only the longitudinal field is relevant for trapped ions. In the context of trapped ions, this effect is called the ``image charge shift.'' It was first measured in 1989~\cite{PhysRevA.40.6308}, and is routinely accounted for in precision measurements, e.g.~see Refs.~\cite{vandyck2006231,BASE:2014drs,BASE:2016yuo,Schneider:2017lff,PhysRevLett.119.033001}.

Exact solutions for the image charge shift exist in idealized geometries~\cite{tinkle_2001,Winters_2006}, while in cylindrically symmetric geometries a semianalytic solution is available~\cite{PhysRevA.64.023403}. More generally, it can be determined numerically using the measured cavity geometry to $\sim 1\%$ precision~\cite{PhysRevA.100.023411}. This approach also applies to the cavities used for trapped electrons, allowing the longitudinal cavity shift to be found with accuracy $\Delta \omega_c / \omega_c \sim 10^{-16}$, sufficient for the near future. 

These results and methods do not seem to be applicable to the transverse cavity shift, which must be separately accounted for. The longitudinal cavity shift is substantially simpler because it can be computed without renormalization, e.g.~by considering the Coulomb fields of the static charges induced on the cavity walls. Furthermore, as already noted in Ref.~\cite{PhysRevA.64.023403}, the transverse cavity shift depends on the cavity's full mode structure, and is therefore much more sensitive to the details of the cavity geometry and boundary conditions.

\begin{figure}
\centering
\includegraphics[width=0.75\columnwidth]{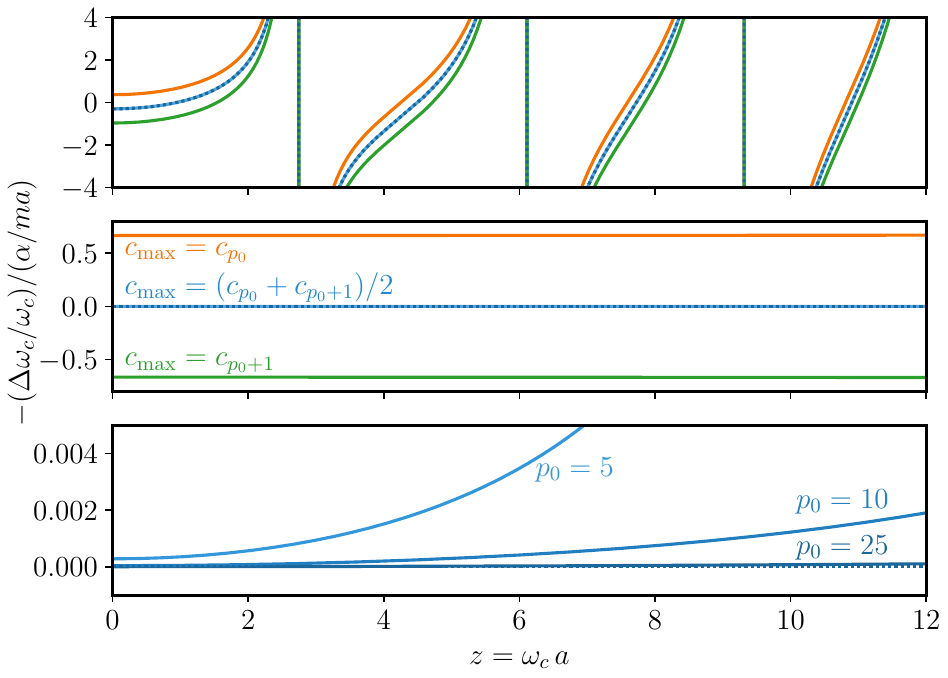}
\caption{Evaluation of the cavity shift in a spherical cavity using hard cutoffs, and its deviation from the exact result. The correct cutoff for the integral is midway between two frequencies in the sum. The result is accurate even for a small number of sum elements $p_0$.}
\label{fig:sph_cutoff}
\end{figure}

\begin{figure}
\centering
\includegraphics[width=0.75\columnwidth]{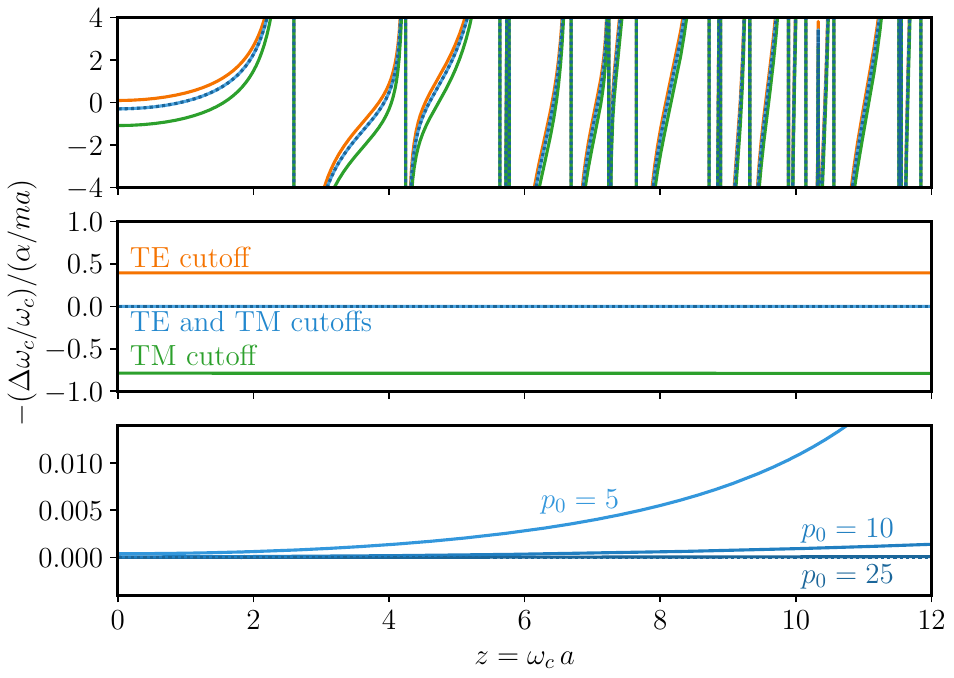}
\caption{Evaluation of the cavity shift in a cylindrical cavity with $L/a=1.72$ using hard cutoffs, and its deviation from the exact result. In addition to the shift of $c_{\mathrm{max}}$ noted in Fig.~\ref{fig:sph_cutoff}, one must use different cutoffs for the parts of the integral corresponding to TE and TM modes, as discussed in the text. The result is accurate even for a small number of sum elements $p_0$.}
\label{fig:cyl_cutoff}
\end{figure}

\paragraph{Evaluation with Explicit Regulators.} In a more general cavity geometry, it may be intractable to compute the renormalized cavity shift analytically. However, the cavity shift can also be readily evaluated using explicit regulators. As a proof of concept, we show that this can be done by applying appropriate hard cutoffs to the sum and integral. 

First, for the spherical case, we have argued that the regulator must treat $c_p$ in the sum in the same way as $c$ in the integral. Then we would get the correct result for their difference by, e.g.~multiplying the integrand by a smooth regulator such as $e^{-c/c_0}$ and the summand by $e^{-c_p / c_0}$ for some large $c_0$, and evaluating both numerically. 

Here we will consider the even simpler case of a hard cutoff, evaluating the sum up to $p = p_0$ and the integral up to $c = c_{\mathrm{max}}$. This is not a smooth regulator, and one could reasonably take $c_{\mathrm{max}} \in [c_{p_0}, c_{p_0+1}]$. Since $c_p \approx \pi p$ for large $p$, this yields fractional cavity shifts that vary by $\sim \alpha / ma$. However, it was shown in Ref.~\cite{Barton_1981} that for linearly divergent sums, one will obtain the correct answer (i.e.~the one obtained for a smooth regulator) if one takes the integral cutoff to lie halfway between sum elements, $c_{\mathrm{max}} = (c_{p_0} + c_{p_0 + 1})/2$, as shown in Fig.~\ref{fig:sph_cutoff}. Furthermore, a close match to the exact answer is obtained even for relatively low $p_0$. 

For the cylindrical cavity, the selection of cutoffs is somewhat more subtle, because the TE and TM mode frequencies differ. Cutting off the entire integral $I = I_1 + I_2 + I_3$ in Eq.~\eqref{eq:cyl_full_int} at $c_{\mathrm{max}} = (c_{n_0} + c_{n_0+1})/2$, appropriate for TM modes, or at $c_{\mathrm{max}} = (\bar{c}_{n_0} + \bar{c}_{n_0+1})/2$, appropriate for TE modes, yields fractional cavity shifts that are incorrect by $\sim \alpha / ma$, as shown in Fig.~\ref{fig:cyl_cutoff}. As discussed in Sec.~\ref{sec:cylinder_result}, the integral $I_1 + I_2$ corresponds to the TM mode sum, while $I_3$ corresponds to the TE mode sum. Cutting off these integrals accordingly gives a highly accurate result, even for low $p_0$. (Alternatively, one could naively average the TE and TM cutoff results with a $2{:}1$ weighting, on the grounds that the linear divergence in the TE mode sum is twice as large. This would also yield a good, albeit slightly less accurate result.)

\paragraph{Implications for Experiment.} The cavity shift has long been one of the dominant systematic uncertainties of electron $g-2$ measurements. In fact, one of the main reasons a cylindrical cavity is used is that the cavity shift in an ideal cylinder has been computed~\cite{cylinder_1989}. 

In practice, the real cavity differs from an ideal cylinder, due to geometrical imperfections and the presence of the trapping electrodes. Thus, in the most recent measurement~\cite{Fan:2022oyb}, several dozen low-lying cavity modes were measured and identified with specific cylindrical cavity mode numbers. Compared to the ideal modes, the measured modes have slightly different frequencies and coupling strengths, as well as finite quality factors. The result for an ideal cylinder is then modified by subtracting out what an ideal mode would have contributed, and adding back the contribution from the corresponding measured mode. 

We have shown that one can indeed compute the cavity shift from a renormalized mode sum, which implies that this ``add and subtract'' method is fundamentally sound. In addition, our numeric results with hard cutoffs indicate that a very accurate result can be obtained by considering only a small number of low-lying modes. However, one must ensure there is no asymptotic shift between the real and ideal frequencies of the higher-frequency, unmeasured modes, which could contribute systematic error. 

Future work can extend our results to a realistic Penning trap, which involves both shifts of mode properties, and an off-center electron position. It would be useful to explore a wider variety of prescriptions for subtracting the sum and integral, beyond hard cutoffs, investigate how they should be generalized to account for imperfections, and quantify their convergence rates and associated uncertainties. Such developments may help enable the next order of magnitude in sensitivity for future measurements of the electron $g$-factor. 

\acknowledgments

We are indebted to Gerald Gabrielse for proposing this problem and valuable discussions. We thank Thomas Myers, Banibrato Sinha, Lillian Soucy, Benedict Sukra, and especially Xing Fan for detailed discussions about the electron magnetic moment measurements. We thank Asher Berlin for collaboration in the early stages of this project, and Asimina Arvanitaki, Carl Beadle, Zach Bogorad, Cliff Burgess, David Curtin, Patrick Draper, and Jorge Noronha for enlightening discussions. KZ was supported by the Office of High Energy Physics of the U.S. Department of Energy under contract DE-AC02-05CH11231. 
RH's  work was primarily supported by the Office of High Energy Physics of the U.S. Department of Energy QuantISED program. 
HD, SP, and YK are supported in part by DOE grant DE-SC0015655. HD, RH, SP, and YK are also partially supported by the Department of Energy, Office of Science, National Quantum Information Science Research Centers,
Superconducting Quantum Materials and Systems Center (SQMS) under contract number DE-AC02-07CH11359. SQMS's support was specifically critical for framing and launching this project, and supporting numerical studies. 
YK acknowledges the support of a Discovery Grant from the Natural Sciences and Engineering Research Council of Canada (NSERC).

\newpage
\appendix
\renewcommand\thefigure{\thesection.\arabic{figure}}
\setcounter{figure}{0}

\section{Photon Mode Functions}
\label{app:photon_modes}

Here we review some well-known properties of mode functions, for convenience and clarity, and show how they are used to derive the cavity shift for spherical and cylindrical cavities.

\paragraph{Spherical Cavity.} For a cavity of radius $a$, the mode functions are~\cite{hill2009electromagnetic}
\begin{align}
    \mathbf{u}_{{\rm TE},m\nu p} & = \frac{c^{(\rm TE)}_{m\nu p}}{\omega_{{\rm TE},\nu p} r} \, \hat{J}_\nu (\omega_{{\rm TE},\nu p} r) e^{i m \phi} \left [ -\frac{im}{\sin \theta}P_\nu ^m(\cos \theta) \hat{\bm{\theta}} + \frac{d}{d\theta} P_\nu^m(\cos \theta) \hat{\bm{\phi}} \right] \\
    \mathbf{u}_{{\rm TM},m\nu p} & = \frac{c^{(\rm TM)}_{m\nu p}}{\omega_{{\rm TM},\nu p} r} \, e^{i m \phi} \left [ \frac{\nu (\nu +1)}{\omega_{{\rm TM},\nu p} r}\hat{J}_\nu (\omega_{{\rm TM},\nu p} r)P_\nu^m(\cos \theta)\hat{\mathbf{r}} \nonumber \right. \\
    & \left. \qquad \qquad \qquad + \hat{J}'_\nu (\omega_{{\rm TM},\nu p} r) \frac{d}{d\theta} P_\nu^m(\cos \theta) \hat{\bm{\theta}} + \frac{im}{\sin \theta} \hat{J}'_\nu (\omega_{{\rm TM},\nu p} r) P_\nu^m(\cos \theta) \hat{\bm{\phi}} \right]
\end{align}
where $\nu $ and $p$ are positive integers, $m$ is an integer with $|m| \leq \nu $, $\hat{J}_\nu (x) \equiv x j_\nu (x)$, $j_\nu (x)$ is a spherical Bessel function, $\hat{J}'_\nu (x) \equiv \frac{d}{dx}\hat{J}_\nu (x)$, and the resonant angular frequencies are
\begin{equation}
\omega_{{\rm TE},m\nu p} = \frac{\zeta_{\nu p}}{a}, \qquad \omega_{{\rm TM},m\nu p} = \frac{\zeta'_{\nu p}}{a}
\end{equation}
where $\zeta_{\nu p}$ and $\zeta'_{\nu p}$ are the $p^{\text{th}}$ roots of $\hat{J}_\nu (x)$ and $\hat{J}'_\nu (x)$, respectively. The $m$ index is a dummy index added to match the notation of the cylindrical case, because the spherical cavity eigenfrequencies are degenerate in $m$.

The only modes that are nonzero at the center of the cavity are the $\nu = 1$ TM modes. Thus, in the dipole approximation, we only need to consider
\begin{align}
\mathbf{u}_{{\rm TM},01p}(\bm{0}) &= \frac{2}{3}c^{\rm (TM)}_{01p} \, \hat{\mathbf{z}} \equiv u_{\text{TM},1p}^{||,\text{sphere}} \, \hat{\mathbf{z}}, \\
\mathbf{u}_{{\rm TM},\pm1,1p}(\bm{0}) &= -\frac{2}{3}c^{\rm (TM)}_{11p} \,(\hat{\mathbf{x}} \pm i \hat{\mathbf{y}}) \equiv u_{\text{TM},1p}^{\perp,\text{sphere}} \, (\hat{\mathbf{x}} \pm i \hat{\mathbf{y}}).
\end{align}
As shown in the main text, only $c_{11p}^{(\mathrm{TM})}$ is relevant for the cavity shift. There are many equivalent expressions for it, but the one that will be most useful comes from considering the normalization of the mode's magnetic field. Suppressing the $\mathrm{TM}$ and $11$ subscripts, it is~\cite{hill2009electromagnetic}
\begin{equation}
    \mathbf{B}_{p} = c^{\rm (TM)}_{11p} \frac{\hat{J}_1(\omega_p r)}{\omega_p r} e^{i\phi} [-i \hat{\bm{\theta}} + \cos \theta \, \hat{\bm{\phi}}].
\end{equation}
Squaring and integrating, the normalization condition is 
\begin{equation}
\int d^3 \mathbf{r}\, |\mathbf{B}_{p}|^2 = \frac{16 \pi}{3} (c^{\rm (TM)}_{11p})^2 \int_0^a dr \, r^2 j_1^2(\omega_p r) = 1
\end{equation}
which yields the form of $c^{\rm (TM)}_{11p}$ we will use in our analytic argument,
\begin{equation}
(c^{\rm (TM)}_{11p})^2 = \frac{3}{16 \pi a^3} \left( \int_0^1 dx \, x^2 j_1^2(\zeta_{1p}' x) \right)^{-1}.
\end{equation}
Using the generic result Eq.~\eqref{eq:cav_shift_generic} for the cavity shift, we have
\begin{equation} 
\frac{\Delta \omega_c^{\rm cav}}{\omega_c} = -\frac{8\pi \alpha}{m a}\sum_{p=1}^\infty \frac{(u_{\text{TM},1p}^{\perp,\text{sphere}})^2 a^3}{(\omega_{{\rm TM},11 p} a)^2 - z^2} = - \frac23 \frac{\alpha}{ma} \sum_{p=1}^\infty \frac{\left( \int_0^1 dx \, x^2 j_1^2(\zeta_{1p}' x) \right)^{-1}}{(\zeta_{1p}')^2- z^2}.
\end{equation}
This is the starting point of Sec.~\ref{sec:sphere_result}. We also note that an equivalent form of $(c^{\rm (TM)}_{11p})^2$ is
\begin{equation}
(c^{\rm (TM)}_{11p})^2 = \frac{3}{8\pi a^3}\frac{(\zeta'_{1p})^2}{(\zeta'_{1p})^2-2} \frac{1}{j_1(\zeta_{1p}')^2}.
\end{equation}
Using the asymptotic behavior of $j_1$ and $\zeta'_{1p}$, one can easily see from this form of the normalization factor that the mode sum is linearly divergent. 

\paragraph{Cylindrical Cavity.} For a cavity of radius $a$ and length $L$, with the cavity endcaps at $z = 0$ and $z = L$, the mode functions are \cite{Kakazu:1995va,hill2009electromagnetic}
\begin{align}
\mathbf{u}_{{\rm TE},m\nu p} &= c^{(\rm TE)}_{m\nu p} \left [\frac{m}{\rho}J_m \left(\frac{\chi'_{m\nu }\rho}{a}\right)\hat{\bm{\rho}} + \frac{i \chi'_{m\nu }}{a}J'_m\left(\frac{\chi'_{m\nu }\rho}{a}\right)\hat{\bm{\phi}}\right]\omega_{{\rm TE},m\nu p}\sin\left(\frac{p\pi z}{L}\right) e^{i m \phi} \\
\mathbf{u}_{{\rm TM},m\nu p} &= c^{(\rm TM)}_{m\nu p} \left [ -\left( \frac{\chi_{m\nu }}{a}J'_m \left(\frac{\chi_{m\nu }\rho}{a}\right)\hat{\bm{\rho}} + \frac{im}{\rho}J_m\left(\frac{\chi_{m\nu }\rho}{a}\right)\hat{\bm{\phi}} \right)\frac{p\pi}{L}\sin\left(\frac{p\pi z}{L}\right) \right. \nonumber\\
& \qquad \qquad \left. + \frac{\chi_{m\nu }^2}{a^2}J_m\left(\frac{\chi_{m\nu }\rho}{a}\right)\cos\left(\frac{p\pi z}{L}\right) \hat{\mathbf{z}} \right]e^{i m \phi}
\end{align}
where $\chi_{m\nu }$ is the $\nu ^{\text{th}}$ zero of the Bessel function $J_m$, and $\chi'_{m\nu }$ is the $\nu ^{\text{th}}$ zero of its derivative $J_m^\prime$. Here the integer $m$ is the azimuthal index, the positive integer $\nu$ is the radial index, and the nonnegative integer $p$ is the longitudinal index, and $J_{-m}(x) = (-1)^m J_m(x)$. Note that here the Bessel order is indexed by $m$ instead of $\nu$. Also, the radial index is usually denoted by $n$, but here we use $\nu$ to avoid confusion with the electron Landau level. 

The normalization coefficients are
\begin{align}
c^{({\rm TM})}_{m\nu p} &= \sqrt{\frac{2 - \delta_{p 0}}{\pi L}}\frac{1}{J_{m+1}(\chi_{m\nu })\,\chi_{m\nu }\,\omega_{{\rm TM},m\nu p}} \\
c^{({\rm TE})}_{m\nu p} &= \sqrt{\frac{2}{\pi L}}\frac{1}{\sqrt{J_m^2(\chi'_{m\nu})-J_{m+1}^2(\chi'_{m\nu})}\,\chi'_{m\nu}\,\omega_{{\rm TE},m\nu p}}
\end{align}
where the resonant angular frequencies are
\begin{equation}
\omega_{{\rm TM},m\nu p} = \sqrt{\frac{\chi_{m\nu}^2}{a^2}+ \frac{p^2\pi^2}{L^2}}, \qquad \omega_{{\rm TE},m\nu p} = \sqrt{\frac{{\chi'}_{m\nu}^2}{a^2}+ \frac{p^2\pi^2}{L^2}}.
\end{equation}

Under the dipole approximation, we only evaluate the mode functions at the center of the cavity, $\mathbf{r} = \bm{0}$, which here stands for $\rho = 0$ and $z = L/2$. For small arguments, we have $J_m(x) \sim x^{|m|}$ and $J_m'(x) \sim x^{|m-1|}$, which implies that only a few families of modes can be nonzero at $\rho = 0$. These are the ``axial'' ($m = 0$) TM modes, 
\begin{equation} 
\label{eq:ModeFunctionsDipoleAxial}
\mathbf{u}_{{\rm TM},0\nu p}(\bm{0}) = c_{0\nu p}^{\text{(TM)}}\frac{\chi_{0\nu}}{a^2}\cos\left(\frac{p\pi}{2}\right)\hat{\mathbf{z}} \equiv u_{\text{TM},\nu p}^{||,\text{cyl.}} \, \hat{\mathbf{z}}
\end{equation}
and the ``radial'' ($m = \pm 1$) TM and TE modes, 
\begin{align}
\label{eq:ModeFunctionsDipoleRadialTM}
\mathbf{u}_{{\rm TM},\pm1\nu p}(\bm{0}) &= -c_{1\nu p}^{\text{(TM)}} \, \frac{\chi_{1\nu}}{2a}\frac{p\pi}{L}\sin\left(\frac{p\pi}{2}\right)\left(\hat{\mathbf{x}} \pm i\hat{\mathbf{y}}\right) \equiv -u_{\text{TM},\nu p}^{\perp,\text{cyl.}} \, (\hat{\mathbf{x}} \pm i \hat{\mathbf{y}}),\\
\label{eq:ModeFunctionsDipoleRadialTE}
\mathbf{u}_{{\rm TE},\pm1\nu p}(\bm{0}) &= \pm c_{1\nu p}^{(\text{TE})} \, \omega_{\text{TE},1\nu p}\frac{\chi'_{1\nu}}{2a}\sin\left(\frac{p\pi}{2}\right)\left(\hat{\mathbf{x}} \pm i\hat{\mathbf{y}}\right) \equiv \pm u_{\text{TE},\nu p}^{\perp,\text{cyl.}} \, (\hat{\mathbf{x}} \pm i \hat{\mathbf{y}}).
\end{align}
The signs on the right-hand sides of Eqs.~\eqref{eq:ModeFunctionsDipoleRadialTM} and~\eqref{eq:ModeFunctionsDipoleRadialTE} are purely due to the mode conventions, and will drop out when squared to compute physical quantities. 

Again, only the radial modes will contribute, and the generic result Eq.~\eqref{eq:cav_shift_generic} is 
\begin{equation}
\frac{\Delta \omega_c^{\rm cav}}{\omega_c} = -\frac{\alpha}{ma}\sum_{\sigma \nu p} \frac{8 \pi (u_{\sigma,\nu p}^{\perp, \text{cyl.}})^2 a^3}{(\omega_{\sigma,1\nu p} a)^2 - (\omega_c a)^2} = -\frac{\alpha}{ma} (S_{\mathrm{TM}} + S_{\mathrm{TE}}).
\end{equation}
We will show the evaluation of $S_{\mathrm{TM}}$ explicitly. Using the above results,  
\begin{align}
S_{\mathrm{TM}} &= \sum_{\nu=1}^\infty \frac{4a}{L} \frac{1}{J_{2}^2(\chi_{1\nu })} \sum_{p=1}^\infty \frac{\sin^2(p \pi / 2)}{\omega^2_{{\rm TM},1\nu p} a^2} \frac{(p \pi a / L)^2}{\chi_{1\nu}^2 - z^2 + (p \pi a / L)^2} \\
&= \sum_{\nu=1}^\infty \frac{4a}{L} \frac{1}{J_{2}^2(\chi_{1\nu })} \sum_{q=0}^\infty \frac{((2q+1) r)^2}{(\chi_{1\nu}^2 + ((2q+1) r)^2) (\chi_{1\nu}^2 - z^2 + ((2q+1) r)^2)}
\end{align}
where we let $r = \pi a / L$ and reindexed the sum. The inner sum can be rewritten as 
\begin{equation}
\frac{1}{z^2} \sum_{q=0}^\infty \left ( \frac{1}{1 + ((2q+1) r)^2 / \chi_{1\nu}^2} - \frac{1}{1 + ((2q+1) r)^2 / (\chi_{1\nu}^2 - z^2)} \right)
\end{equation}
and then can be evaluated using the identity 
\begin{equation}
\sum_{q=0}^\infty \frac{1}{1 + (2 q + 1)^2/A} = \frac{\pi}{4} \sqrt{A} \tanh\left(\frac{\pi}{2} \sqrt{A} \right)
\end{equation}
which can be derived with the Poisson summation formula. The result is 
\begin{equation}
S_{\mathrm{TM}} = \sum_{\nu=1}^\infty \frac{1}{z^2 J_{2}^2(\chi_{1\nu})} \left(\chi_{1\nu} \tanh\Big(\frac{L \chi_{1\nu}}{2a}\Big) - \sqrt{\chi_{1\nu}^2-z^2} \tanh\Big(\frac{L \sqrt{\chi_{1\nu}^2-z^2}}{2a} \Big) \right),
\end{equation}
and $S_{\mathrm{TE}}$ is derived similarly. This is the starting point of Sec.~\ref{sec:cylinder_result}. 

\paragraph{Free Space.} If we quantize in a box of side $D$, the continuum limit is
\begin{equation}
\frac{1}{D^3} \sum_{\mathbf{k}}\to \int \frac{d^3 \mathbf{k}}{(2\pi)^3}
\label{eq:continuumlimit}
\end{equation}
and the wavevectors become a continuous variable. In the continuum limit, the photon mode functions are the usual transverse polarization vectors times a plane wave $e^{i \mathbf{k} \cdot \mathbf{x}}$. For a wavevector $\mathbf{k} = (k \sin \theta_k \cos \phi_k, k \sin \theta_k \sin \phi_k, k \cos \phi_k)$, the normalized polarization vectors in Cartesian coordinates are
\begin{equation}
\bm{\epsilon}_{\pm}(\mathbf{k}) = \frac{1}{\sqrt{2}}\left [ (\cos \theta_k \cos \phi_k \mp i \sin \phi_k)\hat{\mathbf{x}} + (\cos \theta_k \sin \phi_k \pm i \cos \phi_k)\hat{\mathbf{y}}- \sin \theta_k \hat{\mathbf{z}}\right],
\label{eq:polarization_freespace}
\end{equation}
satisfying $\bm{\epsilon}_\pm \cdot \bm{\epsilon}^*_{\pm} = 1$, $\bm{\epsilon}_\pm \cdot \bm{\epsilon}^*_{\mp} = 0$, and $\bm{\epsilon}_\pm \cdot \mathbf{k} = 0$. 

\section{Details of the Relativistic Calculation}
\label{app:rel_details}

In this appendix, we give additional technical details on the evaluation of the energy shift in relativistic quantum field theory from Sec.~\ref{sec:qft_shift}. To avoid ambiguity, all spatial coordinates will be denoted without index heights, $(x^1, x^2, x^3) \equiv (x,y,z)$.

\subsection{Position Space Calculation}
\label{sec:position}

Using the definitions in Sec.~\ref{sec:Propagators}, Eq.~\eqref{eq:energyshiftWeinberg} with Eq.~\eqref{eq:positionselfenergy} becomes
\begin{align}
\delta E_n&=-ie^2\int d^3 \mathbf{x}\, d^3 \mathbf{x}^\prime \int d\tau \, \bar{u}_n(\mathbf{x})\gamma^i S_A(\mathbf{x},\mathbf{x}^\prime;\tau) \gamma^j D_{ij}(\mathbf{0};\tau) e^{iE_n\tau} u_n(\mathbf{x}^\prime)\\
&=-\frac{i\beta^2e^2}{4(2\pi)^{\frac{9}{2}}\sigma_z}\sum_{\sigma, s}\int_0^\infty \frac{ds\,e^{-im^2 s}}{s \sin(\beta s)}\int \frac{d\omega}{\omega^2-\omega_{\sigma s}^2+i\epsilon}\int d\tau\, e^{i\left((E_n-\omega)\tau-\frac{\tau^2}{4s}\right)} \nonumber \\
&\qquad\times \int d^3 \mathbf{x}\, d^3\mathbf{x}'e^{\varphi(\mathbf{x},\mathbf{x'})}\left(2m\,f_n(s,\mathbf{x},\mathbf{x'})+\frac{\tau}{s} \, g_n(s,\mathbf{x},\mathbf{x'})+h_n(s,\mathbf{x},\mathbf{x'})\right),
\end{align}
where
\begin{multline}
\varphi(\mathbf{x},\mathbf{x'})=i\left(\frac{1}{4s}(z-z')^2+\frac{\beta}{4\tan(\beta s)}((x-x')^2+(y-y')^2)\right)\\
-i\frac{\beta}{4}\big((x+x')(y-y')-(x-x')(y+y')\big)\\
-\frac{\beta}{4}(x^2+y^2+x'^2+y'^2)-\frac{1}{\sigma_z^2}\left(z-\bar{z}\right)^2-\frac{1}{\sigma_z^2}\left(z'-\bar{z}\right)^2
\end{multline}
and 
\begin{align}
f_n(s,\mathbf{x},\mathbf{x'})&=u_{\sigma s}^i(\bm{0})u_{\sigma s}^{*j}(\bm{0})\hat{u}_n^*(\mathbf{x})\gamma^0\gamma^i\left(\cos(\beta s)+\sin(\beta s)\gamma^1\gamma^2\right)\gamma^j\hat{u}_n(\mathbf{x}'),\\
g_n(s,\mathbf{x},\mathbf{x'})&=u_{\sigma s}^i(\bm{0})u_{\sigma s}^{*j}(\bm{0})\hat{u}_n^*(\mathbf{x})\gamma^0\gamma^i\left(\cos(\beta s)\gamma^0-i\sin(\beta s)\gamma^3\gamma^5\right)\gamma^j\hat{u}_n(\mathbf{x}'),\\
h_n(s,\mathbf{x},\mathbf{x'})&=u_{\sigma s}^i(\bm{0})u_{\sigma s}^{*j}(\bm{0})\hat{u}_n^*(\mathbf{x})\gamma^0\gamma^i\left(\tfrac{\beta}{\sin(\beta s)}((x-x')\gamma^1+(y-y')\gamma^2)\right. \nonumber \\
&\hspace{135pt}\left.+\tfrac{z-z'}{s}\big(\cos(\beta s)\gamma^3-i\sin(\beta s)\gamma^0\gamma^5\big)\right)\gamma^j\hat{u}_n(\mathbf{x}')
\end{align}
with 
\begin{equation}
u^i_{\sigma s}(\bm{0})u^{*j}_{\sigma s}(\bm{0})=
\begin{cases}
u_{\sigma,\nu p}^{\perp2} & i=j=\{1,2\},\\
u_{\sigma,\nu p}^{||2} & i=j=3.
\end{cases}
\end{equation}
We now need to separate the $\omega$ fraction and use the Schwinger trick,
\begin{align} \label{eq:Schwinger_trick}
\int \frac{d\omega}{\omega^2-\omega_{\sigma s}^2+i\epsilon}&=\frac{1}{2\omega_{\sigma s}}\int d\omega \left(\frac{1}{\omega-(\omega_{\sigma s}-i\epsilon)}-\frac{1}{\omega+(\omega_{\sigma s}-i\epsilon)}\right)\\
&=\frac{1}{2\omega_{\sigma s}}\int d\omega \, (-i)\int_0^\infty d\lambda\left(e^{i\lambda(\omega-\omega_{\sigma s}+i\epsilon)}+e^{-i\lambda(\omega+\omega_{\sigma s}-i\epsilon)}\right)\\
&=\frac{1}{2i\omega_{\sigma s}}\int_0^\infty d\lambda \, e^{-i\lambda(\omega_{\sigma s}-i\epsilon)}\int d\omega \left(e^{i\lambda\omega}+e^{-i\lambda\omega}\right),
\end{align}
to get
\begin{multline}
\delta E_n=-\frac{i\beta^2e^2}{4(2\pi)^{\frac{9}{2}}\sigma_z}\sum_{\sigma, s}\frac{1}{2i\omega_{\sigma s}}\int_0^\infty \frac{ds\,e^{-im^2 s}}{s \sin(\beta s)}\int_0^\infty d\lambda \, e^{-i\lambda(\omega_{\sigma s}-i\epsilon)}\\
\times \int d\omega \left(e^{i\lambda\omega}+e^{-i\lambda\omega}\right)\int d\tau\, e^{i\left((E_n-\omega)\tau-\frac{\tau^2}{4s}\right)}\\
\times\int d^3\mathbf{x} \, d^3\mathbf{x}' \, e^{\varphi(\mathbf{x},\mathbf{x'})}\left(2m\,f_n(s,\mathbf{x},\mathbf{x'})+\frac{\tau}{s}g_n(s,\mathbf{x},\mathbf{x'})+h_n(s,\mathbf{x},\mathbf{x'})\right).
\end{multline}
The $\tau$ and $\omega$ integrals are essentially Gaussian, and thus can be evaluated as
\begin{align}
\int d\omega & \left(e^{i\lambda\omega}+e^{-i\lambda\omega}\right) \int d\tau\, e^{i\left((E_n-\omega)\tau-\frac{\tau^2}{4s}\right)}\left(A+\frac{\tau}{s}B\right) \nonumber\\
&=\int d\omega \left(e^{i\lambda\omega}+e^{-i\lambda\omega}\right)e^{is(E_n-\omega)^2}\sqrt{4\pi s}e^{-i\pi/4}\left(A+2(E_n-\omega)B\right)\\
&=\sqrt{4\pi^2} \, e^{-i\frac{\lambda^2}{4s}}\left[e^{iE_n\lambda}\left(A+\frac{\lambda}{s}B\right)+(\lambda\rightarrow-\lambda)\right],
\end{align}
where for the first equality we used $(E_n-\omega)\tau-\frac{\tau^2}{4s}=s(E_n-\omega)^2-\frac{(\tau-2s(E_n-\omega))^2}{4s}$, and for the second equality we used $\pm\lambda\omega+s(E_n-\omega)^2=\pm E_n\lambda-\frac{\lambda^2}{4s}+s\left(\omega-(E_n\mp\frac{\lambda}{2s})\right)^2$, to obtain
\begin{multline}
\delta E_n=-\frac{i\beta^2e^2}{4(2\pi)^{\frac{7}{2}}\sigma_z}\sum_{\sigma, s}\frac{1}{2i\omega_{\sigma s}}\int_0^\infty \frac{ds\,e^{-im^2 s}}{s \sin(\beta s)}\int_0^\infty d\lambda \, e^{-i\left(\frac{\lambda^2}{4s}+\lambda\omega_{\sigma s}\right)}\int d^3\mathbf{x} \, d^3\mathbf{x}' \, e^{\varphi(\mathbf{x},\mathbf{x'})}\\
\times\left[e^{iE_n\lambda}\left(2m\,f_n(s,\mathbf{x},\mathbf{x'})+\frac{\lambda}{s}g_n(s,\mathbf{x},\mathbf{x'})+h_n(s,\mathbf{x},\mathbf{x'})\right)+(\lambda\rightarrow-\lambda)\right].
\label{eq:mom_coeff_struct}
\end{multline}
The $\lambda$ integral then evaluates to
\begin{equation}
\int_0^\infty d\lambda \, e^{-i\frac{\lambda^2}{4s}-i\omega_{\sigma s}\lambda\pm i E_n\lambda}=-e^{-i\pi/4}\sqrt{\pi s}\left(1\pm\text{erf}(e^{i\pi/4}\sqrt{s}E^\mp_{n;\sigma s})\right)e^{is(E^\mp_{n;\sigma s})^2}
\label{eq:lambda0}
\end{equation}
and
\begin{equation}
\frac{1}{s}\int_0^\infty d\lambda \, \lambda e^{-i\frac{\lambda^2}{4s}-i\omega_{\sigma s}\lambda\pm i E_n\lambda}=\mp2e^{-i\pi/4}\sqrt{\pi s}E^\mp_{n;\sigma s}\left(1\pm\text{erf}(e^{i\pi/4}\sqrt{s}E^\mp_{n;\sigma s})\right)e^{is(E^\mp_{n;\sigma s})^2}-2i,
\label{eq:lambda1}
\end{equation}
where we defined $E^\pm_{n;\sigma s}=E_n\pm\omega_{\sigma s}$, so
\bea
\delta E_n&=-\frac{i\pi^{3/2}\beta^2e^2}{(2\pi)^{\frac{9}{2}}\sigma_z}e^{-i\pi/4}\sum_{\sigma, s}\frac{1}{2i\omega_{\sigma s}}\int_0^\infty \frac{ds\,e^{-im^2 s}}{\sqrt{s} \sin(\beta s)}\int d^3\mathbf{x}\, d^3\mathbf{x}'\,e^{\varphi(\mathbf{x},\mathbf{x'})}\\
&\times\left[\left(m\,f_n(s,\mathbf{x},\mathbf{x'})+E^+_{n;\sigma s}g_n(s,\mathbf{x},\mathbf{x'})+\frac{1}{2}h_n(s,\mathbf{x},\mathbf{x'})\right)\right.\\
&\hspace{150pt}\times\left.\left(1-\text{erf}(e^{i\pi/4}\sqrt{s}E^+_{n;\sigma s})\right)e^{is(E^+_{n;\sigma s})^2}\right.\\
&~~+\left.\left(m\,f_n(s,\mathbf{x},\mathbf{x'})+E^-_{n;\sigma s}g_n(s,\mathbf{x},\mathbf{x'})+\frac{1}{2}h_n(s,\mathbf{x},\mathbf{x'})\right)\right.\\
&\hspace{150pt}\times\left.\left(1+\text{erf}(e^{i\pi/4}\sqrt{s}E^-_{n;\sigma s})\right)e^{is(E^-_{n;\sigma s})^2}\right].
\eae
We now need to perform the gamma contractions, for both the ground and excited state.

\paragraph{Ground State.} Using the spinor $\hat{u}_0(\mathbf{x})$ from Eq.~\eqref{eq:spinors}, we have
\bea
f_0(s,\mathbf{x},\mathbf{x'})&=u_{\sigma s}^i(\bm{0})u_{\sigma s}^{*j}(\bm{0})\hat{u}_0^*(\mathbf{x})\gamma^0\gamma^i\left(\cos(\beta s)+\sin(\beta s)\gamma^1\gamma^2\right)\gamma^j\hat{u}_0(\mathbf{x}')\\
&=2u_{\sigma,\nu p}^{\perp2}(-\cos(\beta s)+i\sin(\beta s))+u_{\sigma,\nu p}^{||2}(-\cos(\beta s)-i\sin(\beta s))\\
&=-2u_{\sigma,\nu p}^{\perp2}e^{-i\beta s}-u_{\sigma,\nu p}^{||2}e^{i\beta s}\equiv f_0(s),
\eae
\bea
g_0(s,\mathbf{x},\mathbf{x'})&=u_{\sigma s}^i(\bm{0})u_{\sigma s}^{*j}(\bm{0})\hat{u}_0^*(\mathbf{x})\gamma^0\gamma^i\left(\cos(\beta s)\gamma^0-i\sin(\beta s)\gamma^3\gamma^5\right)\gamma^j\hat{u}_0(\mathbf{x}')\\
&=2u_{\sigma,\nu p}^{\perp2}(\cos(\beta s)-i\sin(\beta s))+u_{\sigma,\nu p}^{||2}(\cos(\beta s)+i\sin(\beta s))\\
&=2u_{\sigma,\nu p}^{\perp2}e^{-i\beta s}+u_{\sigma,\nu p}^{||2}e^{i\beta s}\equiv g_0(s),
\eae
\bea
h_0(s,\mathbf{x},\mathbf{x'})&=u_{\sigma s}^i(\bm{0})u_{\sigma s}^{*j}(\bm{0})\hat{u}_0^*(\mathbf{x})\gamma^0\gamma^i\left(\tfrac{\beta}{\sin(\beta s)}((x-x')\gamma^1+(y-y')\gamma^2)\right.\\
&\hspace{130pt}\left.+\tfrac{z-z'}{s}\big(\cos(\beta s)\gamma^3-i\sin(\beta s)\gamma^0\gamma^5\big)\right)\gamma^j\hat{u}_0(\mathbf{x}')\\
&=0\equiv h_0(s).
\eae
Note that none of these terms depend on position. Finally, we need to perform the position integrals, which are all Gaussian because $\varphi$ is quadratic in the spatial coordinates,
\begin{equation} \label{eq:gaussian_int}
I_0 \equiv \int d^3\mathbf{x}\, d^3\mathbf{x}'\,e^{\varphi(\mathbf{x},\mathbf{x'})}=i\frac{8\pi^2}{\beta^2}e^{-i\beta s}\sin(\beta s)\pi\sigma_z^2\sqrt{\frac{2s}{2s-i\sigma_z^2}}.
\end{equation}
Putting everything together, we get
\begin{multline}
\delta E_0=-\frac{ie^2}{4}\sum_{\sigma s}\frac{1}{\omega_{\sigma s}}\int_0^\infty ds \, \sqrt{\frac{\sigma_z^2}{\sigma_z^2+2is}} \, e^{-i(m^2+\beta)s}\\
\hspace{-5mm}\times\Big[\Big(m\,f_0(s)+E^+_{0;\sigma s}g_0(s)+\frac{1}{2}h_0(s)\Big)\left(1-\text{erf}(e^{i\pi/4}\sqrt{s}E^+_{0;\sigma s})\right)e^{is(E^+_{0;\sigma s})^2}\\
\quad+\Big(m\,f_0(s)+E^-_{0;\sigma s}g_0(s)+\frac{1}{2}h_0(s)\Big)\left(1+\text{erf}(e^{i\pi/4}\sqrt{s}E^-_{0;\sigma s})\right)e^{is(E^-_{0;\sigma s})^2}\Big], \label{eq:ground_shift}
\end{multline}
with $f_0(s)$, $g_0(s)$ and $h_0(s)$ defined above, as stated in Eq.~\eqref{eq:ground_results} of the main text.

\paragraph{Excited State.} Here we use $E_1 = \sqrt{m^2 + 2\beta}$, and the spinor $\hat{u}_1(\mathbf{x})$ from Eq.~\eqref{eq:spinors}. Defining the variables $w = x+iy$ and $w' = x'+iy'$ for convenience, we have
\bea
f_1(s,\mathbf{x},\mathbf{x'})&=u_{\sigma s}^i(\bm{0})u_{\sigma s}^{*j}(\bm{0})\hat{u}_1^*(\mathbf{x})\gamma^0\gamma^i\left(\cos(\beta s)+\sin(\beta s)\gamma^1\gamma^2\right)\gamma^j\hat{u}_1(\mathbf{x}')\\
&=\tfrac{\beta}{E_1(E_1+m)}\left\{2u_{\sigma,\nu p}^{\perp2}\left[\left(1-\tfrac{(E_1+m)^2}{4}w^*w'\right)\cos(\beta s)+i\left(1+\tfrac{(E_1+m)^2}{4}w^*w'\right)\sin(\beta s)\right]\right.\\
&\left.\hspace{60pt}+u_{\sigma,\nu p}^{||2}\left[\left(1-\tfrac{(E_1+m)^2}{4}w^*w'\right)\cos(\beta s)-i\left(1+\tfrac{(E_1+m)^2}{4}w^*w'\right)\sin(\beta s)\right]\right\}\\
&=\tfrac{\beta}{E_1(E_1+m)}\left[2u_{\sigma,\nu p}^{\perp2}\left(e^{i\beta s}-\tfrac{(E_1+m)^2}{4}w^*w'e^{-i\beta s}\right)\right.\\
&\left.\hspace{60pt}+u_{\sigma,\nu p}^{||2}\left(e^{-i\beta s}-\tfrac{(E_1+m)^2}{4}w^*w'e^{i\beta s}\right)\right],
\label{eq:f1_unevaluated}
\eae
\bea
g_1(s,\mathbf{x},\mathbf{x'})&=u_{\sigma s}^i(\bm{0})u_{\sigma s}^{*j}(\bm{0})\hat{u}_1^*(\mathbf{x})\gamma^0\gamma^i\left(\cos(\beta s)\gamma^0-i\sin(\beta s)\gamma^3\gamma^5\right)\gamma^j\hat{u}_1(\mathbf{x}')\\
&=\tfrac{\beta}{E_1(E_1+m)}\left\{2u_{\sigma,\nu p}^{\perp2}\left[\left(1+\tfrac{(E_1+m)^2}{4}w^*w'\right)\cos(\beta s)+i\left(1-\tfrac{(E_1+m)^2}{4}w^*w'\right)\sin(\beta s)\right]\right.\\
&\left.\hspace{60pt}+u_{\sigma,\nu p}^{||2}\left[\left(1+\tfrac{(E_1+m)^2}{4}w^*w'\right)\cos(\beta s)-i\left(1-\tfrac{(E_1+m)^2}{4}w^*w'\right)\sin(\beta s)\right]\right\}\\
&=\tfrac{\beta}{E_1(E_1+m)}\left[2u_{\sigma,\nu p}^{\perp2}\left(e^{i\beta s}+\tfrac{(E_1+m)^2}{4}w^*w'e^{-i\beta s}\right)\right.\\
&\left.\hspace{60pt}+u_{\sigma,\nu p}^{||2}\left(e^{-i\beta s}+\tfrac{(E_1+m)^2}{4}w^*w'e^{i\beta s}\right)\right],
\label{eq:g1_unevaluated}
\eae
\bea
h_1(s,\mathbf{x},\mathbf{x'})&=u_{\sigma s}^i(\bm{0})u_{\sigma s}^{*j}(\bm{0})\hat{u}_1^*(\mathbf{x})\gamma^0\gamma^i\left(\tfrac{\beta}{\sin(\beta s)}((x-x')\gamma^1+(y-y')\gamma^2)\right.\\
&\hspace{116pt}\left.+\tfrac{z-z'}{s}\big(\cos(\beta s)\gamma^3-i\sin(\beta s)\gamma^0\gamma^5\big)\right)\gamma^j\hat{u}_1(\mathbf{x}')\\
&=\tfrac{\beta}{E_1(E_1+m)}\tfrac{\beta(E_1+m)}{2\sin(\beta s)}u_{\sigma,\nu p}^{||2}\{[-iw^*+iw'](x-x') +[w^*+w'](y-y')\}\\
&=-\tfrac{\beta^2}{2E_1\sin(\beta s)}u_{\sigma,\nu p}^{||2}[i(x-x')^2+i(y-y')^2 +(x-x')(y+y')-(x+x')(y-y')].
\label{eq:h1_unevaluated}
\eae
Finally, we need to perform the position integrals, which now are integrals of quadratic functions of position against the same Gaussian weight as before. We can thus pull out the appropriate second moment to find
\begin{align}
&\int d^3\mathbf{x}\, d^3\mathbf{x}'\,e^{\varphi(\mathbf{x},\mathbf{x'})}(x-iy)(x'+iy')=\frac{2}{\beta}e^{-2i\beta s} I_0,\\
&\int d^3\mathbf{x}\, d^3\mathbf{x}'\,e^{\varphi(\mathbf{x},\mathbf{x'})}(x\pm x')(y\mp y')=\pm \frac{2}{\beta}\sin(\beta s)e^{-i\beta s} I_0, \\
&\int d^3\mathbf{x}\, d^3\mathbf{x}'\,e^{\varphi(\mathbf{x},\mathbf{x'})}(x-x')^2=\int d^3\mathbf{x}\, d^3\mathbf{x}'\,e^{\varphi(\mathbf{x},\mathbf{x'})}(y-y')^2=\frac{2i}{\beta}\sin(\beta s)e^{-i\beta s} I_0,
\end{align}
where the remaining Gaussian integral $I_0$ is given by Eq.~\eqref{eq:gaussian_int}. 

Putting everything together, we get
\begin{multline}
\delta E_1=-\frac{ie^2}{4}\sum_{\sigma s}\frac{1}{\omega_{\sigma s}}\int_0^\infty ds \, \sqrt{\frac{\sigma_z^2}{\sigma_z^2+2is}} \, e^{-i(m^2+\beta)s}\\
\hspace{-5mm}\times\Big[\Big(m\,f_1(s)+E^+_{1;\sigma s}g_1(s)+\frac{1}{2}h_1(s)\Big)\left(1-\text{erf}(e^{i\pi/4}\sqrt{s}E^+_{1;\sigma s})\right)e^{is(E^+_{1;\sigma s})^2}\\
\quad+\Big(m\,f_1(s)+E^-_{1;\sigma s}g_1(s)+\frac{1}{2}h_1(s)\Big)\left(1+\text{erf}(e^{i\pi/4}\sqrt{s}E^-_{1;\sigma s})\right)e^{is(E^-_{1;\sigma s})^2}\Big], \label{eq:excited_shift}
\end{multline}
where 
\begin{align}
f_1(s)&\equiv\tfrac{\beta}{E_1(E_1+m)}\left[2\left(e^{i\beta s}-\tfrac{(E_1+m)^2}{2\beta}e^{-3i\beta s}\right)u_{\sigma,\nu p}^{\perp2}+\left(1-\tfrac{(E_1+m)^2}{2\beta}\right)e^{-i\beta s}u_{\sigma,\nu p}^{\parallel2}\right], \label{eq:f1_eq}\\
g_1(s)&\equiv\tfrac{\beta}{E_1(E_1+m)}\left[2\left(e^{i\beta s}+\tfrac{(E_1+m)^2}{2\beta}e^{-3i\beta s}\right)u_{\sigma,\nu p}^{\perp2}+\left(1+\tfrac{(E_1+m)^2}{2\beta}\right)e^{-i\beta s}u_{\sigma,\nu p}^{\parallel2}\right], \label{eq:g1_eq}\\
h_1(s)&\equiv\tfrac{4\beta}{E_1} \, e^{-i\beta s}u_{\sigma,\nu p}^{\parallel2}. \label{eq:h1_eq}
\end{align}
These are the results used to obtain Eq.~\eqref{eq:rel_E1_shift} in the main text. 

\subsection{Momentum Space Calculation}
\label{sec:momentum}
Using the definitions in Sec.~\ref{sec:Propagators}, Eq.~\eqref{eq:allmomentumselfenergy} becomes
\bea
-i \Sigma(p) &= (ie)^2 \int \frac{d^4 k}{(2\pi)^4}\gamma^\mu S(k) \gamma^\nu D_{\mu \nu}(p-k).\\
&= -i e^2 \sum_{\sigma s} u^{i*}_{\sigma s}(\bm{0}) u^j_{\sigma s}(\bm{0})\int \frac{d^4 k}{(2\pi)^4} (2\pi)^3 \delta^{(3)}(\mathbf{p}-\mathbf{k})\frac{\gamma^i S(k) \gamma^j}{(p_0-k_0)^2 - \omega_{\sigma s}^2 + i\epsilon}\\
&=-ie^2 \sum_{\sigma s} u^{i*}_{\sigma s}(\bm{0}) u^j_{\sigma s}(\bm{0})\int \frac{d k_0}{2\pi}\frac{\gamma^i S(k^0,\mathbf{p}) \gamma^j}{(p_0-k_0)^2 - \omega_{\sigma s}^2 + i\epsilon}
\eae
with
\bea
S(k^0,\mathbf{p}) &= \int_0^\infty \frac{ds}{\cos(\beta s)} \, e^{i s \left(k_0^2-p_3^2 - (p_1^2+p_2^2) \frac{\tan(\beta s)}{\beta s} - m^2 + i \epsilon\right)}\\
&\qquad \qquad \times \left[ (k^0\gamma^0-p^3\gamma^3 + m)[\cos(\beta s) + \gamma^1 \gamma^2 \sin(\beta s)] - \frac{p^1\gamma^1+p^2\gamma^2}{\cos(\beta s)} \right]
\eae
and 
\begin{equation}
u^i_{\sigma s}(\bm{0})u^{*j}_{\sigma s}(\bm{0})=
\begin{cases}
u_{\sigma,\nu p}^{\perp2} & i=j=\{1,2\},\\
u_{\sigma,\nu p}^{||2} & i=j=3.
\end{cases}
\end{equation}
Using the Schwinger parametrization analogously to Eq.~\eqref{eq:Schwinger_trick}, we can write
\bea
\frac{1}{(p_0-k_0)^2 -\omega_{\sigma s}^2+ i\epsilon} &= \frac{1}{2 \omega_{\sigma s}} \left(\frac{1}{(p_0-k_0) -\omega_{\sigma s}+ i\epsilon} + \frac{1}{-(p_0-k_0) -\omega_{\sigma s} +i\epsilon}\right) \\
&=\frac{-i}{2 \omega_{\sigma s}} \int_0^\infty d\lambda\left(e^{i \lambda[(p_0-k_0)-\omega_{\sigma s}+ i\epsilon]} + e^{-i \lambda[(p_0-k_0)+\omega_{\sigma s} - i\epsilon]} \right).
\eae
The integral over $k^0$ can now be evaluated as a Gaussian, 
\bea
\int dk^0 (A+k^0B)e^{i(sk_0^2\pm \lambda k^0)}=\sqrt{\frac{\pi}{s}} \, e^{i\pi/4}e^{-i\frac{\lambda^2}{4s}}\left(A\mp\frac{\lambda}{2s}B\right),
\eae
so that
\begin{align} \label{eq:sigmap}
\Sigma(p) &= -e^{i\pi/4}\frac{i e^2}{2\sqrt{\pi}} \sum_{\sigma s} \frac{1}{2\omega_{\sigma s}}\int_0^{\infty} d\lambda \int_0^{\infty} \frac{ds}{\sqrt{s}\cos(\beta s)} \\
&\times e^{-i\left[\frac{\lambda^2}{4s}+s\left( p_3^2 + (p_1^2+p_2^2) \frac{\tan(\beta s)}{\beta s} + m^2 \right)+\lambda\omega_{\sigma s} - i\epsilon \right]}\left(e^{i\lambda p^0}M(\mathbf{p},\lambda,s)+e^{-i\lambda p^0}M(\mathbf{p},-\lambda,s)\right) \nonumber
\end{align}
with
\beq
M(\mathbf{p},\lambda,s)\equiv u^i_{\sigma s}(\bm{0})u^{*j}_{\sigma s}(\bm{0})\gamma^i \Big( \Big(\tfrac{\lambda}{2s}\gamma^0 -\gamma^3 p^3+ m\Big)\left(\cos(\beta s) + \gamma^1 \gamma^2 \sin(\beta s)\right) - \tfrac{p^1 \gamma^1+p^2\gamma^2}{\cos(\beta s)} \Big)\gamma^j.
\eeq

We can now evaluate the energy shift Eq.~\eqref{eq:energyshiftWeinberg} with the amputated self-energy diagram Eq.~\eqref{eq:momentumselfenergy},
\begin{equation}
\delta E_n = \int d^3 \mathbf{x} \, d^3 \mathbf{x}' \, \bar{u}_n(\mathbf{x}) \left [ e^{i \Phi(\mathbf{x},\mathbf{x}')} \int \frac{d^3 \mathbf{p}}{(2\pi)^3} e^{i \mathbf{p} \cdot (\mathbf{x} - \mathbf{x}')} \Sigma(p^0 = E_n, \mathbf{p})\right] u_n(\mathbf{x}').
\end{equation}
Since the spatial components of the momentum are now being treated as distinct from the time component, we will use $(p^1,p^2,p^2)\equiv(p_x,p_y,p_z)$ for the remainder of the section. Using the electron wavefunctions in Eq.~\eqref{eq:spinors}, we find the energy shift for the ground and the excited state to be 
\begin{multline}
\delta E_n = \frac{\beta}{2\pi \sigma_z}\sqrt{\frac{2}{\pi}}\int \frac{d^3 \mathbf{p}}{(2\pi)^3} \int d^3 \mathbf{x} \, d^3 \mathbf{x}' \: e^{i\psi(\mathbf{x},\mathbf{x}')}\\
\quad \times
\begin{cases}
\Sigma_{22}(E_0, \mathbf{p}) & n=0\\
\begin{aligned}
\tfrac{\beta}{E_1(E_1+m)}&\Big[ \tfrac{(E_1+m)^2}{4}(x-iy)(x'+iy')\Sigma_{22}(E_1, \mathbf{p}) - \Sigma_{33}(E_1, \mathbf{p}) \\
&+  i\tfrac{(E_1+m)}{2}\Big((x-iy)\Sigma_{23}(E_1, \mathbf{p}) +(x'+iy')\Sigma_{32}(E_1, \mathbf{p})\Big)\Big]
\end{aligned}
& n=1,
\end{cases}
\end{multline}
where
\begin{align}
\psi(\mathbf{x},\mathbf{x}')& =i \, \frac{\beta (x^2+y^2+x'^2+y'^2)}{4} +i \, \frac{(z-\bar{z})^2 + (z'-\bar{z}')^2}{\sigma_z^2} \nonumber \\
& \qquad - \frac{\beta}{4}((x+x')(y-y')-(x-x')(y+y'))+\mathbf{p} \cdot (\mathbf{x}-\mathbf{x}')
\end{align}
and $\Sigma_{ij}(E_n, \mathbf{p})$ is the $ij^\text{th}$ component of $\Sigma(E_n, \mathbf{p})$. The spatial integrals are Gaussian and can be evaluated as
\begin{align}
\int d^3\mathbf{x}\, d^3\mathbf{x'}\, e^{i\psi(\mathbf{x},\mathbf{x'})}&= \frac{8\pi^3 \sigma_z^2}{\beta^2} \, e^{-\frac{(p_x^2+p_y^2)}{\beta} - \frac{\sigma_z^2 p_z^2}{2}} \equiv \bar{I}_0 \\
\int d^3\mathbf{x}\, d^3\mathbf{x'}\, e^{i\psi(\mathbf{x},\mathbf{x'})}(x-iy)(x'+iy')&=\frac{2}{\beta^2}(2(p_x^2+p_y^2)-\beta) \bar{I}_0 \\
\int d^3\mathbf{x}\, d^3\mathbf{x'}\, e^{i\psi(\mathbf{x},\mathbf{x'})}(x-iy)&=\frac{2i}{\beta}(p_x- ip_y) \bar{I}_0 \\
\int d^3\mathbf{x}\, d^3\mathbf{x'}\, e^{i\psi(\mathbf{x},\mathbf{x'})}(x'+iy')&=-\frac{2i}{\beta}(p_x+ip_y) \bar{I}_0.
\end{align}
to obtain
\begin{multline}
\delta E_n = \frac{2\sigma_z}{\beta\sqrt{2\pi}} \int \frac{d^3 \mathbf{p}}{2\pi} \exp \left( - \frac{p_x^2 + p_y^2}{\beta} - \frac{\sigma_z^2 p_z^2}{2} \right) \\
\times
\begin{cases}
\Sigma_{22}(E_0, \mathbf{p}) & n=0\\
\begin{aligned}
\tfrac{1}{2E_1(E_1+m)} &\big[ (E_1+m)^2 \left(\tfrac{2}{\beta}(p_x^2+p_y^2)-1\right)\Sigma_{22}(E_1, \mathbf{p}) - 2 \beta \Sigma_{33}(E_1, \mathbf{p})\\
&- 2(E_1+m)\big((p_x - ip_y)\Sigma_{23}(E_1, \mathbf{p}) - (p_x + ip_y)\Sigma_{32}(E_1, \mathbf{p})\big) \big]
\end{aligned}
& n=1
\end{cases}
\end{multline}
\begin{multline}
\hspace{15pt}= -e^{i\pi/4}\frac{i e^2}{\beta\sqrt{\pi}} \frac{ \sigma_z}{\sqrt{2\pi}} \sum_{\sigma s} \frac{1}{2\omega_{\sigma s}}\int_0^{\infty} d\lambda \int_0^{\infty} \frac{ds}{\sqrt{s}\cos(\beta s)}e^{-i\left(\frac{\lambda^2}{4s}+sm^2+\lambda\omega_{\sigma s} - i\epsilon \right)} \\
\times \int \frac{d^3 \mathbf{p}}{2\pi}\exp\left[- \frac{(1+ i \tan(\beta s))}{\beta} (
p_x^2+p_y^2) - \frac{(\sigma_z^2 + 2is )}{2} p_z^2\right] \\
\times
\begin{cases}
\left(e^{i\lambda E_0}M_{22}(\mathbf{p},\lambda,s)+(\lambda\rightarrow-\lambda)\right) & n=0\\
\begin{aligned}
&\tfrac{e^{i\lambda E_1}}{2E_1(E_1+m)} \Big[(E_1+m)^2 \left(\tfrac{2}{\beta}(p_x^2+p_y^2)-1\right)M_{22}(\mathbf{p},\lambda,s) - 2 \beta M_{33}(\mathbf{p},\lambda,s)\\
&\hspace{50pt}- 2(E_1+m)\big((p_x - ip_y)M_{23}(\mathbf{p},\lambda,s) - (p_x + ip_y)M_{32}(\mathbf{p},\lambda,s)\big) \Big]\\
&\hspace{20pt}+(\lambda\rightarrow-\lambda)
\end{aligned}
& n=1.
\end{cases}
\label{eq:int_with_coeffs}
\end{multline}
We now need to consider the gamma matrix structure of the self-energy. Noting that only even powers of $p$ contribute to the integrals, the relevant terms are 
\begin{align}
M_{22}(\mathbf{p},\lambda,s)&\supset u^{i*}_{\sigma s}(\bm{0}) u^j_{\sigma s}(\bm{0})\left(\gamma^i\left(\tfrac{\lambda}{2s}\gamma^0 + m\right)\left(\cos(\beta s) + \gamma^1 \gamma^2 \sin(\beta s)\right)\gamma^j\right)_{22} \nonumber \\
&= i\left(m-\tfrac{\lambda}{2s}\right)u^{i*}_{\sigma s}(\bm{0}) u^j_{\sigma s}(\bm{0})\cos(\beta s) \nonumber \\
&\quad \times\left[ (i+\tan(\beta s))\big(\delta^{i1}\delta^{j1}+\delta^{i2}\delta^{j2}- i\delta^{i1}\delta^{j2}+i\delta^{i2}\delta^{j1} \big) +(i -\tan(\beta s))\delta^{i3}\delta^{j3}\right] \nonumber \\
&=\left(-m+\tfrac{\lambda}{2s}\right)\left(2e^{-i\beta s}u_{\sigma,\nu p}^{\perp2}+e^{i\beta s}u_{\sigma,\nu p}^{||2}\right), \\
M_{33}(\mathbf{p},\lambda,s)&\supset u^{i*}_{\sigma s}(\bm{0}) u^j_{\sigma s}(\bm{0})\left(\gamma^i\left(\tfrac{\lambda}{2s}\gamma^0 + m\right)\left(\cos(\beta s) + \gamma^1 \gamma^2 \sin(\beta s)\right)\gamma^j\right)_{33} \nonumber \\
&= -\left(m+\tfrac{\lambda}{2s} \right)u^{i*}_{\sigma s}(\bm{0}) u^j_{\sigma s}(\bm{0})\cos(\beta s) \nonumber \\
&\quad \times\left[ (1+ i\tan(\beta s)) (\delta^{i1}\delta^{j1}+\delta^{i2}\delta^{j2}+i\delta^{i1}\delta^{j2} -i\delta^{i2}\delta^{j1}) +(1-i\tan(\beta s)) \delta^{i3}\delta^{j3} \right] \nonumber \\
&=-\left(m+\tfrac{\lambda}{2s}\right)\left(2e^{i\beta s}u_{\sigma,\nu p}^{\perp2}+e^{-i\beta s}u_{\sigma,\nu p}^{||2}\right),
\end{align}
and the combinations
\begin{align}
&M_{23}(\mathbf{p},\lambda,s)-M_{32}(\mathbf{p},\lambda,s)\supset -\frac{u^{i*}_{\sigma s}(\bm{0}) u^j_{\sigma s}(\bm{0})}{\cos(\beta s)}\left((\gamma^i \slashed{p}_\perp\gamma^j)_{23}-(\gamma^i \slashed{p}_\perp\gamma^j)_{32}\right) \nonumber \\
&\quad= -2\frac{u^{i*}_{\sigma s}(\bm{0}) u^j_{\sigma s}(\bm{0})}{\cos(\beta s)}\left(p_x  (\delta^{i1}\delta^{j1}-\delta^{i2}\delta^{j2}-\delta^{i3}\delta^{j3})+p_y(\delta^{i1}\delta^{j2}+\delta^{i2}\delta^{j1})\right) \nonumber \\ 
&\quad= -\frac{2p_x u_{\sigma,\nu p}^{||2}}{\cos(\beta s)}, \\
&M_{23}(\mathbf{p},\lambda,s)+M_{32}(\mathbf{p},\lambda,s)\supset -\frac{u^{i*}_{\sigma s}(\bm{0}) u^j_{\sigma s}(\bm{0})}{\cos(\beta s)}\left((\gamma^i \slashed{p}_\perp\gamma^j)_{23}+(\gamma^i \slashed{p}_\perp\gamma^j)_{32}\right) \nonumber \\
&\quad= -2i\frac{u^{i*}_{\sigma s}(\bm{0}) u^j_{\sigma s}(\bm{0})}{\cos(\beta s)}\left(p_y (\delta^{i2}\delta^{j2}-\delta^{i1}\delta^{j1}-\delta^{i3}\delta^{j3})+p_x(\delta^{i1}\delta^{j2}+\delta^{i2}\delta^{j1})\right) \nonumber \\
&\quad= -\frac{2ip_y u_{\sigma,\nu p}^{||2}}{\cos(\beta s)},
\end{align}
where $\slashed{p}_\perp=p_x \gamma^1+p_y \gamma^2$, and we dropped the off-diagonal components $u^{1*}_{\sigma s}(0)u^2_{\sigma s}(0)$ and $u^{2*}_{\sigma s}(0)u^1_{\sigma s}(0)$ since they vanish once summed over the polarizations. The remaining momentum integrals can be performed using
\begin{align}
&\int d^3 \mathbf{p} \: e^{- \big(\frac{(1+ i \tan(\beta s))}{\beta} (
p_x^2+p_y^2) + \frac{(\sigma_z^2 + 2is )}{2} p_z^2\big)} = \frac{\pi \beta}{ 1+i \tan(\beta s)} \sqrt{\frac{2\pi}{\sigma_z^2+2is}} \equiv J_0, \label{eq:mom_ints_1} \\
&\int d^3 \mathbf{p} \: p_{x,y}^2 e^{-\big(\frac{(1+ i \tan(\beta s))}{\beta} (
p_x^2+p_y^2) + \frac{(\sigma_z^2 + 2is )}{2} p_z^2 \big)} = \frac{ \beta}{ 2(1+i \tan(\beta s))} \, J_0. \label{eq:mom_ints_2}
\end{align}

At this point we can connect to the position space derivation. In Eq.~\eqref{eq:int_with_coeffs}, the combinations of $M_{ij}(\mathbf{p},\lambda,s)$ will form combinations of the $f_n(s)$, $g_n(s)$, and $h_n(s)$ defined in App.~\ref{sec:position}. For the ground state case in Eq.~\eqref{eq:int_with_coeffs}, 
\bea
M_{22}(\mathbf{p},\lambda,s) &= \left(-m+\frac{\lambda}{2s}\right)\left(2e^{-i\beta s}u_{\sigma,\nu p}^{\perp2}+e^{i\beta s}u_{\sigma,\nu p}^{||2}\right) \\
&= mf_0(s)+\frac{\lambda}{2s} \, g_0(s).
\eae
The first line of the excited state case in Eq.~\eqref{eq:int_with_coeffs} becomes
\begin{align}
\phantom{}&\frac{1}{2E_1(E_1+m)}\left[(E_1+m)^2 \left(\tfrac{2}{\beta}(p_x^2+p_y^2)-1\right)M_{22}(\mathbf{p},\lambda,s) - 2 \beta M_{33}(\mathbf{p},\lambda,s)\right] \nonumber \\
&\rightarrow\frac{1}{2E_1(E_1+m)}\left[(E_1+m)^2\left(\frac{2}{1+i\tan(\beta s)}-1\right)\left(-m+\frac{\lambda}{2s}\right)\left(2e^{-i\beta s}u_{\sigma,\nu p}^{\perp2}+e^{i\beta s}u_{\sigma,\nu p}^{||2}\right)\right. \nonumber \\
&\hspace{200pt}\left.+2\beta \left(m+\frac{\lambda}{2s}\right)\left(2e^{i\beta s}u_{\sigma,\nu p}^{\perp2}+e^{-i\beta s}u_{\sigma,\nu p}^{||2}\right)\right] \nonumber \\
&=\frac{\beta}{E_1(E_1+m)}\left[\frac{(E_1+m)^2}{2\beta}e^{-2i\beta s}\left(\frac{\lambda}{2s}-m\right)\left(2e^{-i\beta s}u_{\sigma,\nu p}^{\perp2}+e^{i\beta s}u_{\sigma,\nu p}^{||2}\right)\right. \nonumber \\
&\hspace{200pt}\left.+\left(\frac{\lambda}{2s}+m\right)\left(2e^{i\beta s}u_{\sigma,\nu p}^{\perp2}+e^{-i\beta s}u_{\sigma,\nu p}^{||2}\right)\right] \nonumber \\
&= mf_1(s)+\frac{\lambda}{2s} \, g_1(s),
\end{align}
where we replaced $p_{x,y}^2\rightarrow\frac{\beta}{2(1+i\tan(\beta s))}$ using Eq.~\eqref{eq:mom_ints_2}. The second line of the excited state case in Eq.~\eqref{eq:int_with_coeffs} becomes
\begin{align}
\phantom{}&\frac{-2(E_1+m)}{2E_1(E_1+m)}\left[(p_x-ip_y)M_{23}(\mathbf{p},\lambda,s)-(p_x+ip_y)M_{32}(\mathbf{p},\lambda,s)\right] \nonumber \\
&\quad=-\frac{1}{E_1}\left[p_x(M_{23}(\mathbf{p},\lambda,s)-M_{32}(\mathbf{p},\lambda,s))-ip_y(M_{23}(\mathbf{p},\lambda,s)+M_{32}(\mathbf{p},\lambda,s))\right] \nonumber \\
&\quad=-\frac{1}{E_1}\left[-\frac{2p_x^2 u_{\sigma,\nu p}^{||2}}{\cos(\beta s)}-\frac{2p_y^2 u_{\sigma,\nu p}^{||2}}{\cos(\beta s)}\right] \nonumber  \\
&\quad\rightarrow \frac{2\beta}{E_1}e^{-i\beta s} u_{\sigma,\nu p}^{||2} \nonumber \\
&\quad=\frac{1}{2} h_1(s),
\end{align}
where we again replaced $p_{x,y}^2\rightarrow\frac{\beta}{2(1+i\tan(\beta s))}$ using Eq.~\eqref{eq:mom_ints_2}. Putting everything together,
\begin{multline}
\delta E_n = -e^{i\pi/4}\frac{ ie^2}{2\sqrt{\pi}} \sum_{\sigma s} \frac{1}{\omega_{\sigma s}} \int_{0}^{\infty} d\lambda \int_0^{\infty} \frac{ds}{\sqrt{s}} \sqrt{\frac{\sigma^2_z}{\sigma^2_z+2is}} \, e^{-i\left(\frac{\lambda^2}{4s}+s(m^2+\beta)+\lambda\omega_{\sigma s} - i\epsilon \right)} \\
\times
\begin{cases}
e^{i\lambda E_0}\left(mf_0(s)+\frac{\lambda}{2s}g_0(s)\right) +(\lambda\rightarrow-\lambda) & n=0,\\
e^{i\lambda E_1} \left(mf_1(s)+\frac{\lambda}{2s}g_1(s)+\frac{1}{2}h_1(s)\right)+(\lambda\rightarrow-\lambda) & n=1.
\end{cases}
\end{multline}
Finally, performing the $\lambda$ integrals using Eqs.~\eqref{eq:lambda0} and~\eqref{eq:lambda1} recovers Eqs.~\eqref{eq:ground_shift} and~\eqref{eq:excited_shift}, derived in the previous section. 

\bibliographystyle{utphys3}
\bibliography{refs}

\end{document}